\documentclass[aps,prd,twocolumn,showpacs,amsmath,amssymb,nofootinbib,superscriptaddress]{revtex4-2}
\usepackage{graphicx}  
\usepackage{color}
\usepackage{times}
\usepackage[utf8]{inputenc} 
\usepackage{float}
\usepackage{bm}
\usepackage{ulem}
\usepackage{multirow}
\usepackage{makecell}
\usepackage{url}
\usepackage{natbib}
\usepackage{mathrsfs}
\usepackage{physics}
\usepackage{comment}
\usepackage[table,xcdraw]{xcolor}
\usepackage{booktabs,makecell}
\usepackage{amsmath}
\usepackage{pifont}
\usepackage[colorlinks=true,citecolor=blue,urlcolor=blue,linkcolor=blue]{hyperref}
\usepackage{microtype}
\usepackage[none]{hyphenat}

\DeclareUnicodeCharacter{03BB}{\ensuremath{\lambda}} 
\DeclareUnicodeCharacter{039B}{\ensuremath{\Lambda}} 

\begin{document}

\title{Multi-Messenger and Cosmological Constraints on Dark Matter through Two-Fluid Neutron Star Modeling}

\author{Ankit Kumar}
\email{ankitlatiyan25@gmail.com \\ankit.k@iopb.res.in}
\affiliation{Department of Mathematics and Physics, Kochi University, Kochi, 780-8520, Japan}

\author{Sudhakantha Girmohanta}
\affiliation{Tsung-Dao Lee Institute, Shanghai Jiao Tong University, No. 1 Lisuo Road, Pudong New Area, Shanghai, 201210, China}
\affiliation{School of Physics and Astronomy, Shanghai Jiao Tong University,
800 Dongchuan Road, Shanghai, 200240, China}

\author{Hajime Sotani}
\affiliation{Department of Mathematics and Physics, Kochi University, Kochi, 780-8520, Japan}
\affiliation{Interdisciplinary Theoretical \& Mathematical Science Program (iTHEMS), RIKEN, Saitama 351-0198, Japan}
\affiliation{Theoretical Astrophysics, IAAT, University of T\"{u}bingen, 72076 T\"{u}bingen, Germany}

\date{\today}

\begin{abstract}
\noindent In this study, we investigate the impact of dark matter (DM) on neutron stars (NSs) using a two-fluid formalism that treats nuclear matter (NM) and DM as gravitationally coupled components. Employing NM equations of state spanning a wide range of stiffness and a self-interacting asymmetric fermionic DM framework, we explore the emergence of DM core- and halo-dominated structures and their observational implications. Constraints from gravitational waves (GW170817), NICER X-ray measurements (PSR J0030+0451), and pulsar mass limits (PSR J0740+6620) delineate a consistent parameter space for DM properties derived from these multi-messenger observations. DM halo-dominated configurations, while consistent with PSR J0740+6620's mass limits and NICER's radius measurements for PSR J0030+0451, are ruled out by the tidal deformability bounds inferred from the GW170817 event. Consequently, the combined limits inferred from the observational data of GW170817, PSR J0030+0451, and PSR J0740+6620 support the plausibility of DM core-dominated configurations. Constraints on the DM self-interaction strength from galaxy cluster dynamics further refine the DM parameter space permitted by NS observations. This work bridges multi-messenger astrophysics and cosmology, providing insights into DM interactions and their implications for NS structure, evolution, and observational signatures.
\end{abstract}

%

\maketitle

\section{Introduction}
\label{sec:1}
Neutron stars, among the densest known objects in the Universe, serve as natural laboratories to explore fundamental physics under extreme conditions. These compact stellar remnants, with central densities surpassing nuclear saturation density, provide unique insights into the properties of matter at supranuclear densities. At the same time, their immense gravitational potential and extreme environments make them excellent candidates for investigating DM—an enigmatic component constituting approximately 85\% of the Universe's matter content \cite{2020AA...641A...6P, 10.1093/ptep/ptac097}. DM's existence is well-established through its gravitational influence on large-scale structures, galactic dynamics, and the cosmic microwave background \cite{1985ApJ...292..371D, Spergel_2003, BERTONE2005279}. However, its fundamental nature remains one of the most pressing questions in modern physics. Possible DM candidates span a wide range of masses, from very light ($\sim 10^{-20}$ eV) fuzzy bosonic DM~\cite{Hui:2016ltb} to primordial black holes with astrophysical size mass range~\cite{Carr:2020xqk} (see~\cite{cirelli2024darkmatter} for a review). Here we consider an asymmetric fermionic DM model, which links the origin of matter anti-matter asymmetry in the Universe to the observed DM number density~\cite{Kaplan:1991ah, Zurek:2013wia}.

It has been realized in recent times that NS can form a DM admixed configuration where the amount of DM present can vary depending on different formation mechanisms. Although pure gravitational accretion of DM seems unlikely to create a substantial DM core or halo in a NS~\cite{PhysRevD.40.3221}, there are different scenarios via which a DM admixed NS configuration can form. In particular, in the context of asymmetric self-interacting DM, possibly with a subdominant dissipative component, the gravothermal mechanism may lead to the formation of compact dark objects, which can then accrete visible matter to form DM-admixed-NSs~\cite{Kouvaris:2015rea, Ellis:2018bkr}. Motivated by this, we consider an asymmetric fermionic DM model that possesses strong self-interaction due to the mediation of a light vector mediator. We will remain agnostic about the specific formation mechanism of DM admixed NS.
\begin{figure}[htbp]
    \centering
    \vspace{0.10cm}
    \includegraphics[width=\columnwidth]{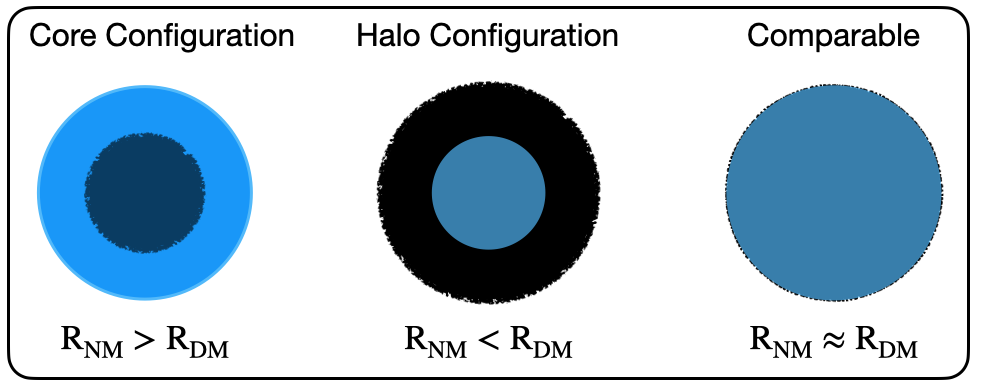} 
    \caption{Configurations of DM admixed NSs with $R_{\rm{NM}}$ and $R_{\rm{DM}}$ representing the surfaces of nuclear matter and DM, respectively.}
    \label{fig:figure1}
\end{figure}

Dynamic properties of the DM-admixed-NS depend crucially on the characteristics of DM. Fermionic DM, for instance, may resist collapse due to degeneracy pressure, while bosonic DM, under certain conditions, can form condensates~\cite{Liang_2023}. Apart from studying the DM capture process or rates, it is vital to explore how the presence of DM captured in an NS reshapes the structural configuration and observational signatures of NSs \cite{PhysRevD.81.123521}. DM accumulation in NSs through gravitational capture leads to distinct structural configurations, influenced by the DM model—its mass, self-interaction strength, and interaction cross-section with baryonic matter~\cite{2022ApJ...936...69M}. These configurations are classified into a core, where DM is confined to the central regions; a halo, where DM extends beyond the baryonic surface with significant impact on the central density; or a comparable distribution, where the DM surface is almost same as that of the baryonic surface (Fig. \ref{fig:figure1}). These varying distributions arise from the balance between gravitational attraction and the internal pressure of DM, dictated by its equation of state (EOS). In the current work, these structural configurations, whether core-dominated, halo-dominated, or comparable, are systematically modeled using the two-fluid formalism, which decouples the dynamics of baryonic and DM, retaining their gravitational coupling. This framework is well-suited for studying the interplay between the nuclear matter EOS and DM properties, providing insights into how DM modifies NS structural configurations.

The presence of DM in these forms can lead to profound changes in the NS’s observable properties, such as deviations in the mass-radius relation, reduced or enhanced tidal deformability, and modifications to the star’s surface characteristics. These alterations, in turn, influence the interpretation of electromagnetic signals, such as X-ray pulse profiles, and gravitational-wave signatures from NS mergers~\cite{ELLIS2018607, PhysRevD.97.043006, PhysRevD.77.023006, Shawqi_2024}. Recent observations, such as those from the Neutron Star Interior Composition Explorer (NICER)~\cite{2016SPIE.9905E..1HG} and gravitational-wave detections by LIGO, Virgo, and KAGRA (LVK collaboration) \cite{Abbott_2023}, have enhanced our ability to constrain NS properties. Understanding these configurations is critical, as they directly affect key observables like the mass-radius relation and tidal deformability, which are probed by data from NICER and LVK. These investigations not only provide insights into DM behavior in extreme astrophysical conditions but also offer constraints on DM theories, with implications for broader cosmological contexts.

DM admixed NSs have recently gained significant attention as a promising avenue for probing DM properties and understanding their impact on extreme astrophysical environments. Various theoretical studies have investigated the influence of different DM models on NSs, including self-interacting fermionic DM \cite{Shawqi_2024, rutherford2024probingfermionicasymmetricdark, Nelson_2019, PhysRevD.102.063028, 2024arXiv241205207B}, asymmetric interacting or non-interacting bosonic DM \cite{PhysRevD.97.123007, 2012arXiv1204.2564F, particles7010011, Konstantinou_2024}, DM interacting with nuclear matter (NM) via Higgs channel \cite{PhysRevD.96.083004, galaxies10010014, PhysRevD.106.043010, PhysRevD.110.063001}, as well as novel approaches like neutron decay to DM within NS interiors \cite{PhysRevLett.120.191801, Husain_2022, 2018IJMPA..3344020M}. This growing interest in DM admixed NSs highlights their significance as natural laboratories for testing theoretical DM models and investigating DM properties, offering a unique interface to bridge the gap between astrophysical observations and the underlying principles of fundamental physics. In this study, we focus on understanding how the presence of DM affects the structure and observable properties of NSs, while simultaneously imposing stringent constraints on the DM self-coupling parameters by leveraging observational data. By solving the coupled Tolman-Oppenheimer-Volkoff (TOV) equations, we investigate how the central density ratios of baryonic and DM fluids, DM self-coupling strengths, and different NM parameter sets influence the mass-radius relation and the tidal deformability of NSs. For baryonic matter, we adopt relativistic mean-field (RMF) EOS, incorporating multiple parameter sets to account for uncertainties in nuclear interactions at supranuclear densities. For the DM component, we adopt a physically motivated Lagrangian that adheres to constraints from the Bullet Cluster (1E 0657-56), ensuring consistency with astrophysical observations across galactic and cluster scales~\cite{Tulin:2017ara, Randall:2008ppe}.

To statistically constrain DM properties, a recent study utilized Bayesian inference within a similar DM model \cite{rutherford2024probingfermionicasymmetricdark}. Incorporating NICER observational data with synthetic mass-radius distributions from prospective missions such as STROBE-X \cite{ray2019strobexxraytimingspectroscopy}, the study inferred posterior distributions for DM parameters such as particle mass, self-repulsion strength, and mass fraction. This Bayesian framework effectively combines prior theoretical assumptions with observational likelihoods to delineate the DM parameter space. In contrast, our study differs significantly in objectives and methodology. While the Bayesian approach relies on probabilistic constraints, this study emphasizes a more direct and physically motivated approach, leveraging existing astrophysical observations—such as NICER's precise mass-radius measurements and gravitational-wave data from GW170817 event—to constrain DM properties. Furthermore, our work incorporates cosmological restrictions obtained from the Bullet Cluster to ensure that the inferred DM configurations remain consistent with large-scale structure observations. This work aims to provide a more realistic and deeper understanding of how DM modifies the structure of NSs within the two-fluid framework, offering insights grounded in current observational capabilities.

The paper is organized as: Section~\ref{sec:2} lays out the formalism used in this study. In Sec.~\ref{sec:2a}, we introduce the NM EOS, detailing the different RMF parameter sets employed to explore the effects of NM stiffness on DM admixed NS configurations. Section~\ref{sec:2b} describes the DM EOS within the framework of asymmetric self-interacting DM, where the DM particles interact through a light mediator and exhibit a number asymmetry, i.e., an asymmetry in the number density of DM over its corresponding anti-DM, similar to baryons in the visible sector.
This formalism allows us to account for the constraints from cluster-scale dynamics while studying DM properties exclusively through gravitational coupling with nuclear matter. Section~\ref{sec:2c} presents the Tolman-Oppenheimer-Volkoff (TOV) equations in the two-fluid formalism, capturing the distinct gravitational contributions of NM and DM to NS structure while accounting for tidal effects essential for connecting theoretical predictions with gravitational wave observations. Section~\ref{sec:3} presents the results and discussion. In Sec.~\ref{sec:3a}, we examine the structural implications of DM on NSs, focusing on the interplay between DM self-coupling and DM to NM central density ratio in shaping NS properties. Section~\ref{sec:3b} examines how multi-messenger astrophysical observations, combined with cosmological constraints from galaxy clusters, are utilized to derive limits on the DM self-coupling constant and mediator mass. Finally, Sec.~\ref{sec:4} concludes the paper, summarizing the key findings and discussing the broader implications of the study for understanding DM properties and their role in compact stars.

\section{Formalism}
\label{sec:2}
The interplay between NM and DM components is central to understanding the behavior of DM admixed NSs. The EOS governs the behavior of matter under the extreme conditions present in NSs, playing a pivotal role in determining their internal structure, stability, and observable properties. In this section, we outline the theoretical framework used in this study, beginning with the EOS for NM and DM, followed by the TOV for two-fluid formalism with gravitational coupling.
\subsection{Nuclear Matter EOS}
\label{sec:2a}
At extreme densities in NS cores (typically 5-8 times the nuclear saturation density), the behavior of matter is dominated by strong nuclear interactions, necessitating a reliable theoretical framework to accurately describe the EOS. In this work, we employ the RMF theory to derive the EOS for baryonic matter, which is a well-established approach recognized for its ability to describe the saturation properties of symmetric NM and extend them to the dense regimes of NS cores. The RMF formalism is grounded in the description of nuclear interactions mediated by scalar and vector meson fields, where baryons are treated as Dirac particles interacting through these mesonic fields. The scalar meson provides attractive interactions, while the vector mesons generate repulsive forces, balancing each other to reproduce the observed properties of finite nuclei and bulk NM. The inclusion of non-linear self-interactions and cross-couplings of mesonic fields further enhance the EOS's adaptability across varying density regimes, making it a versatile tool for exploring NS interiors.

The explicit Lagrangian formulation of the RMF model has been extensively detailed in the literature \cite{MULLER1996508, BOGUTA1977413, KUBIS1997191, PhysRevC.55.540, PhysRevC.62.015802, PhysRevC.90.044305, Kumar2020-zh}. Here we apply this framework to drive the NM EOS, as part of modeling DM admixed NS configurations. To account for the uncertainties in nuclear interactions at high densities, we adopt three RMF parameter sets: QMC-RMF4 \cite{PhysRevC.106.055804}, BigApple \cite{PhysRevC.102.065805}, and NL3 \cite{PhysRevC.55.540}. These parameter sets span a wide range of NM stiffness, allowing for a comprehensive exploration of baryonic matter effects on NS configurations. QMC-RMF4 represents a soft EOS, predicting compact stars with smaller radii and lower maximum mass; BigApple provides a moderately stiff EOS, balancing compactness with higher maximum mass and canonical star ($1.4 M_{\odot}$) tidal deformability. NL3 offers a highly stiff EOS, capable of supporting NSs with significantly larger radii and maximum mass. Each parameter set reflects different assumptions about the high-density nuclear interactions and serves to explore how uncertainties in the NM EOS influence NS properties. To characterize these parameter sets, we calculated key NS properties (maximum gravitational mass, $M_{\rm{max}}$; the corresponding radius, $R_{\rm{max}}$; and the dimensionless tidal deformability of canonical stars, $\Lambda_{1.4 M_{\odot}}$) for pure baryonic NS configurations. These calculations used single-fluid TOV equations \cite{PhysRev.55.364, PhysRev.55.374}, incorporating the SLy4a as crust EOS \cite{refId0} to ensure realistic behavior at low densities:
\begin{itemize}
    \item QMC-RMF4: $M_{\rm{max}} = 2.206 \,M_{\odot}$, $R_{\rm{max}} = 10.933\, \rm{km}$, $\Lambda_{1.4 M_{\odot}} = 453.978$.
    \item BigApple: $M_{\rm{max}} = 2.599 \,M_{\odot}$, $R_{\rm{max}} = 12.407\, \rm{km}$, $\Lambda_{1.4 M_{\odot}} = 716.427$.
    \item NL3: $M_{\rm{max}} = 2.773 \,M_{\odot}$, $R_{\rm{max}} = 13.316\, \rm{km}$, $\Lambda_{1.4 M_{\odot}} = 1277.202$.
\end{itemize}
Among these RMF models, QMC-RMF4 parameter set satisfies the observational constraints on the maximum mass NS~\cite{2020NatAs...4...72C} and the tidal deformability derived from GW170817~\cite{PhysRevLett.121.161101}. BigApple and NL3, while exceeding these constraints, provide insights into how the stiffness of the nuclear EOS influences NS properties. By covering a wide range of EOS behaviors, we aim to encompass the uncertainties in the NM EOS at supranuclear densities, thereby enabling a comprehensive investigation of the interplay between baryonic and DM components in admixed configurations. 
\subsection{Dark Matter EOS}
\label{sec:2b}
The paradigm of asymmetric DM is motivated by the fact that the observed DM and baryon abundances are close to each other, namely $\rho_{\rm DM}/\rho_b \simeq 5$. In this framework, the DM possesses a number asymmetry, i.e., an asymmetry in the number density of DM over anti-DM, similar to the visible sector (see~\cite{Zurek:2013wia} for a review). The imbalance between the matter (baryon) over anti-matter (anti-baryons) observed in the present Universe is parametrized by the ratio of the asymmetry in the baryon number density to the entropy density today $Y_{\rm B} = (n_b-n_{\bar{b}})/s = (0.82$-$0.92) \times 10^{-10}$, as obtained by the Cosmic Microwave Background data~\cite{2020AA...641A...6P}, which is a result of a primordial matter-antimatter asymmetry in the visible sector. The proximity of the observed abundances for the DM and baryons can be explained naturally if the abundance of DM in the present Universe is determined by the asymmetry between the number density of DM over anti-DM and its origin is related to the source of the baryon asymmetry. 

An asymmetry is created in the dark and/or visible sector(s) in the early Universe and is communicated between the sectors through some portal operator. Once the communicating portal decouples below a certain temperature, the asymmetry is separately frozen in the two sectors, therefore, one expects 
\begin{equation}
n_\chi -n_{\bar \chi} \sim n_b-n_{\bar b} \ ,
\label{Eq:Asymm}
\end{equation}
where $n_\chi$ ($n_{\bar \chi}$) denotes the number density of DM (anti-DM), and $n_b$ ($n_{\bar b}$) represents the number density of baryons (anti-baryons). This suggests that the DM mass should be $m_\chi \simeq 5$ GeV within the factor of a few that depends on the specific model. By the time of structure formation, the symmetric component of DM annihilates away to lighter degrees of freedom, and only the asymmetric part remains. Thus we will take the DM mass to be $5$ GeV for concreteness in this study.

Although cold collisionless dark matter (CDM) is remarkably successful in describing the dynamics at large scales ($\sim 10$ Mpc), it faces some challenges in explaining the small-scale ($\sim 1$-$10$ kpc) structures (see~\cite{Tulin:2017ara} for a review). In particular, the observed diversity in the rotation curves of the spiral galaxies can be explained if the DM has a substantial self-interaction cross-section, characterized by the cross-section per unit DM mass being ${\sigma}/{m_{\chi}} \approx  {\cal O}(1)$ cm$^2$/g. This is even more pronounced in low surface brightness galaxies, where the DM dominates the dynamics, and the baryon feedback effects are insufficient in the CDM case~\footnote{Note that recently a larger cross-section in self-interaction, e.g., $\sigma/m_{\chi} \approx {\cal O}(20-40)$ cm$^2$/g, at the dwarf galactic halo scale has been advocated to explain the diversity problem~\cite{Roberts:2024uyw, Correa:2022dey}.}. The required magnitude of the cross-section is understood as follows: to have at least one DM-DM scattering per particle over the 10 Gyr halo lifetime scale, with the given scattering rate 
\begin{align}
    \nonumber
    R_{\rm scat} &= \left( \frac{\sigma}{m_{\chi}} \right) \rho_{\rm DM}^{\rm gal} v_{\rm rel} \\
    &= 0.1 {\rm Gyr}^{-1} \left( \frac{\sigma/m_{\chi}}{1 \  {\rm cm}^2/{\rm g}} \right) \left( \frac{v_{\rm rel}}{50 \  {\rm km}/{\rm s}} \right) \left( \frac{\rho_{\rm DM}^{\rm gal}}{0.1 \ M_\odot/{\rm pc}^3}\right),
\end{align}
where $v_{\rm rel}$ is the relative velocity of the DM particles, and $\rho_{\rm DM}^{\rm gal}$ is the DM density in a typical dwarf galaxy, while the self-interaction strength should be at least of the required quoted value. However, to fit the observational data at the galaxy cluster scale and retain the success of CDM in that regime, the self-interaction cross-section should decrease as the relative velocity between the DM particles increases. This desired property is common in models where the DM-DM scattering proceeds via a light mediator.

Let us consider the dark sector consisting of a spin $1/2$ Dirac fermion DM, $\chi$, and a light vector mediator, $V_\mu$. $\chi$ is charged under a dark $U(1)_{\rm V}$ gauge symmetry, where the charge can be taken to unity without the loss of generality. The gauge coupling is denoted as $g_\chi$. $V_\mu$ can obtain a mass through a dark Higgs mechanism which breaks $U(1)_{\rm V}$ spontaneously. We have considered a single Dirac $\chi$ minimally to avoid gauge anomalies. The  Lagrangian for dark sector is
\begin{align}
    {\cal L}_{\rm DS} &= \bar\chi(i \gamma^\mu D_\mu-m_\chi)\chi-\frac{1}{2} m_{\rm v}^2 V^\mu V_\mu - \frac{1}{4} Z^{\mu \nu} Z_{\mu \nu} \ ,
    \label{Eq:LDS}
\end{align}
where the gauge covariant derivative, $D_\mu = \partial_\mu + i g_\chi V_\mu$, determines the DM-vector mediator coupling; $m_{\rm v}$ denotes the mass of vector mediator; and $Z^{\mu \nu} = \partial^\mu V^\nu-\partial^\nu V^\mu$. We consider $V_{\mu}$ to be much lighter than $m_\chi$ so that the symmetric component of the DM in the early Universe can be annihilated via $\bar \chi \chi \to V V$. As the energy density of the symmetric component ends up in the massive vector mediator, it should get transferred to the visible sector before Big Bang Nucleosynthesis so as not to spoil the success of standard cosmology. This entails the introduction of portal operators, like kinetic mixing with the photon, which also gives rise to a complementary search through direct detection experiments~\cite{Kaplinghat:2013yxa, Fujikura:2024jto}. In our current study, we restrict ourselves to the case where the DM and NM are connected only gravitationally, however, including the effect of such portal operators in a framework of two-fluid interaction formalism will be interesting for future pursuit~\cite{PhysRevD.99.083008}. To ensure that the annihilation of the symmetric DM component in the early Universe is effective, the annihilation cross-section (in natural units) should be
\begin{equation}
    \langle \sigma_{\rm ann} v_{\rm rel} \rangle_{\bar \chi \chi \to VV} \simeq \frac{2 \pi \alpha_\chi^2}{m_\chi^2} \gtrsim 0.6 \times 10^{-25}  \ {\rm cm}^3/{\rm s} \ ,
    \label{Eq:annCon}
\end{equation}
where $\alpha_\chi \equiv g_\chi^2/(4 \pi)$, and we have assumed $m_{\rm v} \ll m_\chi$. The condition in Eq.~\eqref{Eq:annCon} is satisfied when
\begin{equation}
    g_\chi \gtrsim 0.05 \left( \frac{m_\chi}{5 \ {\rm GeV}}\right)^{1/2} \ .
    \label{Eq:gchiCon}
\end{equation}

The self-interaction of the DM essentially causes a thermalization of the interior of the DM halo, and to quantify this effect, effective transfer, $\sigma_{\rm T}$, and viscosity cross-sections, $\sigma_{\rm V}$, are defined by giving greater weightage to large-angle DM scattering:
\begin{align}
    \frac{d \sigma_{\rm T}}{d \Omega} &= (1-\cos{\theta}) \left( \frac{d \sigma}{d \Omega} \right)_{\rm CM}\ , \\
    \frac{d \sigma_{\rm V}}{d \Omega} &= (1-\cos^2{\theta}) \left( \frac{d \sigma}{d \Omega} \right)_{\rm CM} \ ,
\end{align}
where CM denotes the cross-section in the center-of-mass frame, and $\theta$ symbolizes the scattering angle. Analytical formulas are available for the transfer and viscosity cross-section including the identical particle exchange effects~\cite{Girmohanta:2022dog, Girmohanta:2022izb} in the Born regime, and classical regime~\cite{Khrapak:2003kjw}, while numerical results are presented in the intermediate regime~\cite{Zurek:2013wia}, including an approximate solution utilizing the Hulthén potential. Here we quote the analytical results for the transfer cross-sections available in different regimes of the parameter space. In the Born regime ($\alpha_\chi m_\chi/m_{\rm v} \ll 1$)~\cite{Girmohanta:2022dog}, 
\begin{align}
   \sigma_{\rm T}^{\rm Born} &= \frac{4 \pi \alpha_\chi^2}{m_{\rm v}^2 v_{\rm rel}^2} \left[ \frac{{\cal R}^2}{1+ {\cal R}^2} -\frac{\ln(1+{\cal R}^2)}{ (2+{\cal R}^2)}  \right]  \ , 
\end{align}
where ${\cal R} \equiv m_\chi v_{\rm rel}/m_{\rm v}$. In the classical regime ($\alpha_\chi m_\chi/m_{\rm v} \gtrsim 1$, and ${\cal R} \gg 1$), for the repulsive potential we currently consider~\cite{Khrapak:2003kjw}
\begin{equation}
    \sigma_{\rm T}^{\rm clas} =  \begin{cases}
                                 2 \pi \beta_\chi^2 \ln{\left( 1+ \beta_\chi^{-2}\right)}/m_{\rm v}^2 & \beta_{\chi} \lesssim 1 \\
                                 \pi \left( \ln 2 \beta_\chi- \ln \ln 2 \beta_\chi \right)^2/m_{\rm v}^2 & \beta_\chi \gtrsim 1
                                \end{cases} \ ,
\label{Eq:sigmaTClassical}
\end{equation}
where $\beta_\chi \equiv 2 \alpha_\chi m_{\rm v}/(m_\chi v_{\rm rel}^2)$. Further, when ${\cal R} \lesssim 1$, i.e., when the cross-section is primarily determined by the s-wave scattering, an approximate analytical formula is available for Hulthén potential~\cite{Tulin:2013teo} (see the references mentioned for viscosity cross-section formulae).

We utilize these to infer constraints on the parameter space of the model from dynamics at the cluster and galactic scales for comparison, which demands $\sigma/m_{\chi} \lesssim 0.1$-$1$ cm$^2$/g at $v_{\rm rel} \simeq 10^{3}$ km/s. As the inferred constraints on the self-interaction cross-section vary from considerations of halo shapes/ellipticity constraints from cluster lensing surveys, and merging cluster constraints~\cite{Tulin:2017ara}, we impose a representative constraint $\sigma/m_{\chi} < 0.1$ cm$^2$/g at $v_{\rm rel} \sim 4 \times 10^{3}$ km/s. For DM mass $m_\chi = 5$ GeV, and mediator mass ranging from $1$-$30$ MeV, as is suitable for the cross-section to decrease sufficiently from the dwarf scale to the cluster scale as alluded to earlier, with $v_{\rm rel} \sim 4 \times 10^{3}$ km/s, which leads the parameter ${\cal R}$ ranging from $2$ to $67$. Now, to show the constraint $\sigma/m_\chi < 0.1$ cm$^2$/g on the parameter space of $g_\chi-m_{\rm v}$, we checked that the contour of $\sigma/m_\chi = 0.1$ cm$^2$/g lies entirely in the validity range of the classical approximation. Therefore, we utilize Eq.~\eqref{Eq:sigmaTClassical} to obtain the region corresponding to $\sigma/m_\chi < 0.1$ cm$^2$/g, which is the region allowed by the Bullet cluster constraint. This will be shown in Figs.~\ref{fig:figure11} and  \ref{fig:figure12} as a black hatched region and is utilized to compare with the other astrophysical allowed region in the parameter space as described later in the text.

The above framework for a self-interacting fermionic DM is utilized to evaluate EOS and is characterized by parameters that directly affect the spatial distribution and stability of DM within NSs. Under the mean-field approximation \cite{rutherford2024probingfermionicasymmetricdark}, the energy density and pressure for the DM component is expressed as
\begin{align}
    {\cal E}_{\rm{DM}} &= \frac{2}{\left(2\pi\right)^{3}} \int^{k_{\chi}^{F}}_{0} \sqrt{k^{2}+m_{\chi}^{2}}\, d^{3}k \,+\,  \frac{1}{2} \left(\frac{g_{\chi}}{m_{\rm{v}}}\right)^{2} n_{\chi}^{2} \nonumber \\
    P_{\rm{DM}} &= \frac{2}{3\left(2\pi\right)^{3}} \int^{k_{\chi}^{F}}_{0} \frac{k^{2}}{\sqrt{k^{2}+m_{\chi}^{2}}}\, d^{3}k \,+\,  \frac{1}{2} \left(\frac{g_{\chi}}{m_{\rm{v}}}\right)^{2} n_{\chi}^{2}
    \label{eq:eq2}
\end{align}
where $n_{\chi} \left(= \frac{2}{\left(2\pi\right)^{3}}\int^{k_{\chi}^{F}}_{0} d^{3}k \right)$ is  DM number density. The first term in ${\cal E}_{\rm{DM}}$ accounts for the kinetic and rest-mass energy of DM particles, while the second term describes the energy contribution from the self-interactions mediated by $V_{\mu}$. The constraints from cluster scale dynamics imposed play a crucial role in defining the DM self-coupling parameter space, ensuring consistency with large-scale astrophysical observations while allowing for realistic modeling of DM in NS interiors.

The DM EOS interacts with the baryonic EOS exclusively through gravity, consistent with the two-fluid formalism employed in this study. This gravity-only coupling ensures that DM properties can be studied independently, reflecting the absence of interaction terms in the DM Lagrangian with normal matter. We note that as the anti-DM particle is depleted in the present Universe, while there is a conserved global number symmetry for the DM, broken only by effective operators in the early Universe that share the asymmetry with the visible sector, the framework of two-fluid formalism employed in current study is justified. By varying $g_{\chi}/m_{\rm{v}}$ we explore a range of DM EOS behaviors, spanning from weakly interacting scenarios to strongly self-repulsive configurations. The resulting EOS is integrated within the two-fluid formalism, enabling us to investigate how DM properties influence the structural and observable characteristics of DM-admixed NSs.
\subsection{Two-Fluid TOV and Tidal Effects}
\label{sec:2c}
The structural properties of a NS, in the case of a perfect single-fluid composition, are governed by the TOV equations \cite{PhysRev.55.364, PhysRev.55.374}. However, when accounting for a multi-component system like DM admixed NSs, these equations are generalized into a two-fluid formalism, where baryonic matter and DM are treated as independent components interacting only gravitationally \cite{CIARCELLUTI201119, 2009APh....32..278S}. In this framework, the pressures and energy densities of the two fluids evolve under the influence of their combined gravitational field, ensuring mutual coupling of their respective structures. The two-fluid TOV equations are expressed as \cite{GOLDMAN2013200, Sagun_2023}
\begin{align}
    \frac{dP_{\rm{NM}}}{dr} =& - \left(P_{\rm{NM}}+\rm{{{\cal E}_{\rm{NM}}}}\right)\frac{\left(m+4\pi r^3 \left\{P_{\rm{NM}}+P_{\rm{DM}}\right\}\right)}{r\,\left(r-2m\right)}, \nonumber \\
    \frac{dP_{\rm{DM}}}{dr} =& - \left(P_{\rm{DM}}+\rm{{{\cal E}_{\rm{DM}}}}\right)\frac{\left(m+4\pi r^3 \left\{P_{\rm{NM}}+P_{\rm{DM}}\right\}\right)}{r\,\left(r-2m\right)},
    \nonumber \\
    \frac{dm_{\rm{NM}}}{dr} =& \,\, 4\pi r^2 {\cal E}_{\rm{NM}}\,\,\,\,\,\,\,\,\,\,\,\,\,\,\, \& \,\,\,\,\,\,\,\,\,\,\,\,\,\,\, \frac{dm_{\rm{DM}}}{dr} = 4\pi r^2 {\cal E}_{\rm{DM}},
\end{align}
where $P_{\rm{NM}}$ and $P_{\rm{DM}}$ are the pressures, ${\cal E}_{\rm{NM}}$ and ${\cal E}_{\rm{DM}}$ are the energy densities of baryonic and DM components, respectively. The total gravitational mass enclosed within a radius $r$ is given by $m(r) = m_{\rm{NM}}(r) + m_{\rm{DM}}(r)$, ensuring that the gravitational field reflects the combined contributions of both fluids. To numerically solve these coupled equations, appropriate boundary conditions are required—\\
At the center ($r = 0$):
    \begin{itemize}
        \item The central energy densities for NM $\left({\cal E}_{\rm c, 0}^{\rm{NM}} \equiv {\cal E}_{\rm c}^{\rm{NM}} (r=0)\right)$ and DM $\left({\cal E}_{\rm c, 0}^{\rm{DM}} \equiv {\cal E}_{\rm c}^{\rm{DM}} (r=0)\right)$ must be specified as free parameters. These central densities govern the boundary conditions for the respective EOSs and allow for the exploration of different DM admixed NS configurations.
        \item The central pressures are determined by the respective EOSs: $P_{\rm{NM}, 0} = P_{\rm{NM}} ({\cal E}_{\rm c, 0}^{\rm{NM}})$ and $P_{\rm{DM}, 0} = P_{\rm{DM}} ({\cal E}_{\rm c, 0}^{\rm{DM}})$.
        \item The masses, $m_{\rm{NM}}$ and $m_{\rm{DM}}$, should be set to zero, reflecting the absence of mass at $r = 0$: $m_{\rm{NM}}(0) = m_{\rm{DM}}(0) = 0$.
    \end{itemize}
At the surface:
    \begin{itemize}
        \item The surface of each component is defined by the vanishing of its pressure: $P_{\rm{NM}} (R_{\rm{NM}}) = 0$ for the baryonic matter and $P_{\rm{DM}} (R_{\rm{DM}}) = 0$ for the DM component.
        \item The total radius $R$ is determined by the outermost boundary of the two components through $R = {\rm{max}} (R_{\rm{NM}},\, R_{\rm{DM}})$.
    \end{itemize}
The coupled equations are integrated outward from $r = 0$ to the surface of the star ($r=R$). The spatial configuration of DM—whether core-dominated, halo-dominated, or comparable—is governed by the interplay between the baryonic and DM EOSs, along with their respective central densities.

The coupled TOV equations not only provide insight into the structural properties of DM admixed NSs but also serve as the foundation for calculating their tidal deformability. In a binary NS system, each star experiences a quadrupole deformation due to the gravitational field of its companion. The degree of this deformation is quantified by tidal deformability, a parameter that directly links the internal structure of the star to its response under tidal forces. The dimensionless tidal deformability parameter is defined as
\begin{align}
    \Lambda = \frac{2}{3} k_{2} \left(\frac{R}{M}\right)^{5},
\end{align}
where $R$ is the radius of the star, $M$ is its mass, and $k_{2}$ is the second Love number, linking tidal effects to the star's internal composition. The derivation of $k_{2}$ for single-fluid NSs, involves solving the perturbed Einstein field equations under the assumption of a static, spherically symmetric star and has been comprehensively detailed in Refs. \cite{Hinderer_2008, PhysRevD.107.115028, PhysRevD.81.123016}. 
 
In the context of DM admixed NSs, the derivation of tidal deformability is extended to account for the two-fluid formalism. In this framework, baryonic matter and DM are treated as independent components, interacting only through their shared gravitational field. Each fluid contributes to the star's overall tidal response, necessitating the inclusion of both components in the perturbed equations. The explicit derivation of these equations for two-fluid systems is presented in Ref.~\cite{PhysRevD.105.123010, PhysRevD.105.123034}, and it accounts for the gravitational coupling between the two components. For completeness, we provide the detailed expressions for the Love number, $k_{2}$, and the associated differential equations in Appendix \ref{sec:appendx2}. These equations are solved numerically alongside the two-fluid TOV equations, ensuring a consistent background structure for the star. To compute the total tidal deformability $\Lambda$, we use the total gravitational mass ($M$) and radius ($R= {\rm{max}} (R_{\rm{NM}}, R_{\rm{DM}})$) of the star.
\section{Results and Discussion}
\label{sec:3}
\subsection{Structural implications of DM in NSs}
\label{sec:3a}
The mass-radius ($M-R$) profiles of DM admixed NSs are systematically constructed using the two-fluid TOV formalism as outlined in Sec. \ref{sec:2c}, which decouples the dynamics of NM and DM while retaining their gravitational coupling. 
\begin{figure}[tbp]
    \centering
    \includegraphics[width=\columnwidth]{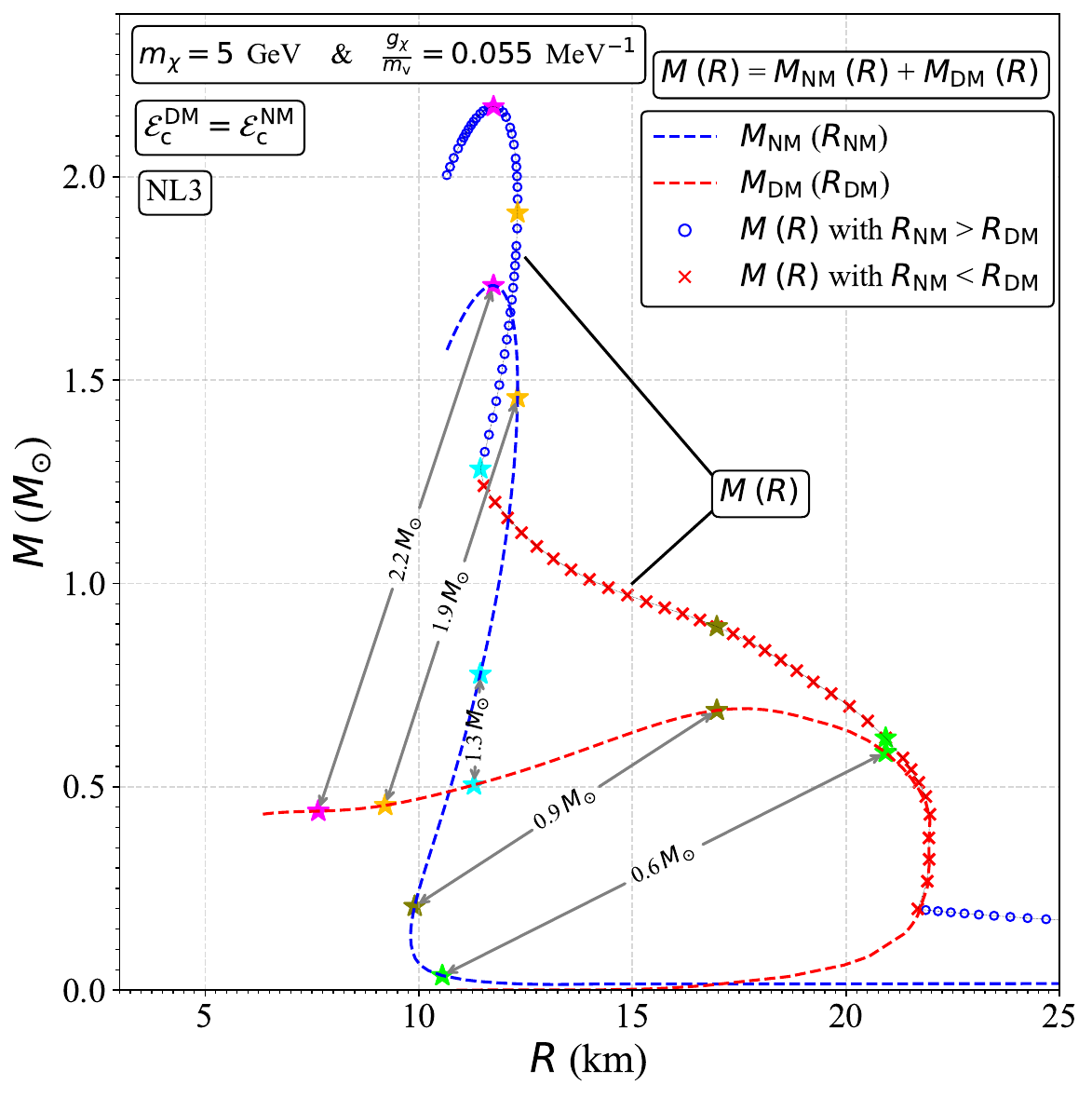} 
    \caption{Mass-radius ($M-R$) profile for a DM admixed NS with $m_{\chi} = 5$ GeV and $g_{\chi}/m_{\rm{v}} = 0.055$ MeV$^{-1}$, employing the NL3 EOS for baryonic matter, with the same central energy density for baryonic matter and DM. The blue dashed line represents the baryonic mass $M_{\rm{NM}}$ as a function of its radius $R_{\rm{NM}}$, while the red dashed line shows the DM mass $M_{\rm{DM}}$ as a function of its radius $R_{\rm{DM}}$. The total mass-radius relationship $M(R)$ is shown by markers, differentiated based on whether the baryonic radius is larger ($R_{\rm{NM}}$ $>$ $R_{\rm{DM}}$, blue open circle) or smaller ($R_{\rm{NM}}$ $<$ $R_{\rm{DM}}$, red crosses) than the DM radius. Five specific configurations are highlighted at $M=0.6 M_{\odot}$, $0.9 M_{\odot}$, $1.3 M_{\odot}$, $1.9 M_{\odot}$ and $2.2 M_{\odot}$. The total mass $M$ for these configurations are shown as star markers on the $M (R)$ curve, connected by arrows to their corresponding points on the $M_{\rm{NM}}$ ($R_{\rm{NM}}$) and $M_{\rm{DM}}$ ($R_{\rm{DM}}$) curves.} 
    \label{fig:figure2}
\end{figure}
By varying the central density ratios of NM and DM, as well as the DM self-coupling parameter ($g_{\chi}/m_{\rm v}$), we explore their influence on the $M-R$ relationship and the structural properties of these stars. To illustrate the construction of these two fluid configurations, we begin with a representative case assuming that NM and DM share the same central energy density
(${\cal E}_{\rm c}^{\rm DM} = {\cal E}_{\rm c}^{\rm NM}$). For this case, the NM EOS is described using the NL3 parameter set, while the DM EOS is characterized by a DM particle mass $m_{\chi} = 5$ GeV and the self-coupling parameter $g_{\chi}/m_{\rm v} = 0.055$ MeV$^{-1}$. 
\begin{figure*}[tbp]
    \centering
    \includegraphics[width=\textwidth]{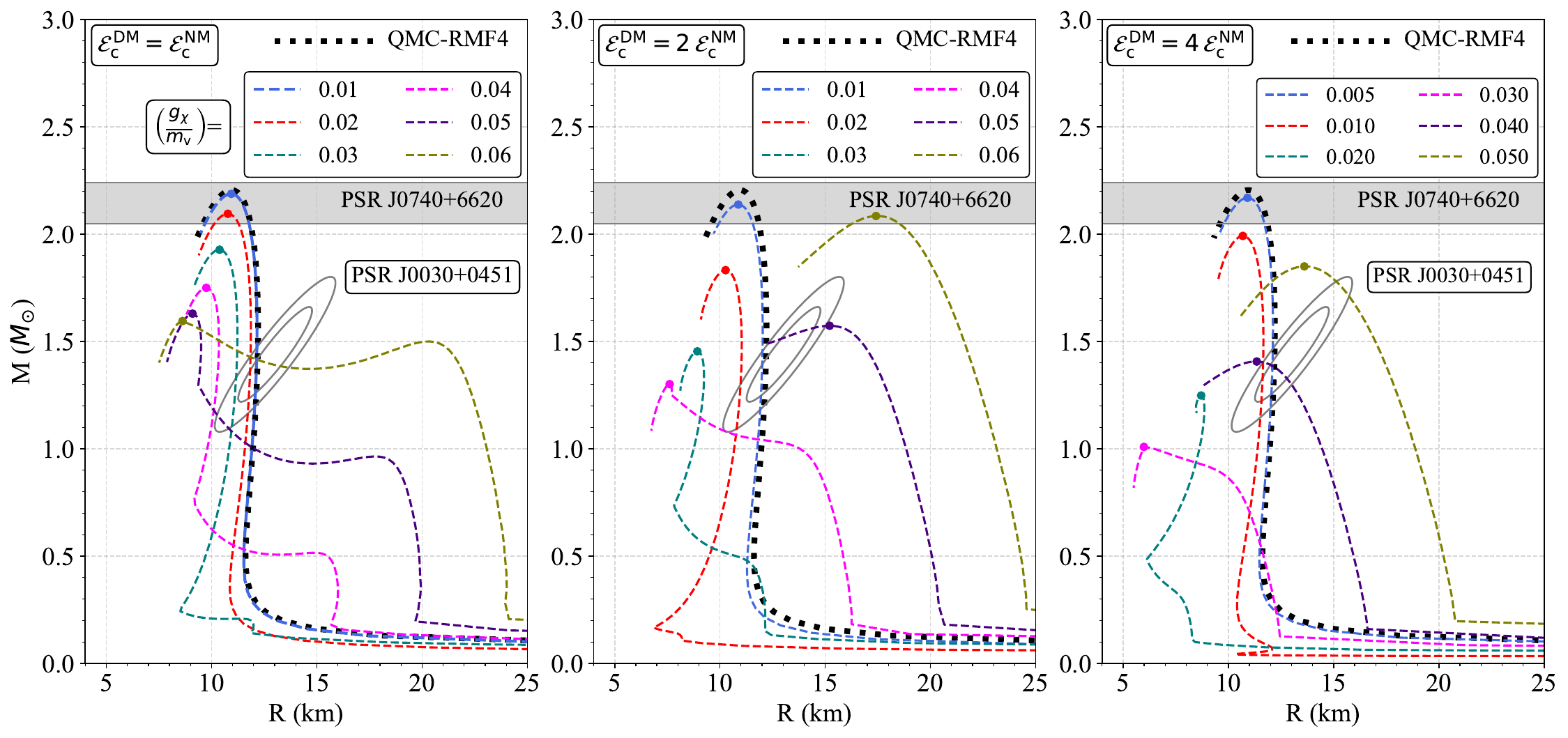} 
    \caption{Total mass-radius ($M-R$) profiles for DM admixed NSs using QMC-RMF4 as the NM EOS, plotted for different values of the DM self-coupling parameter ($g_{\chi}/m_{\rm{v}}$), which are in the unit of MeV$^{-1}$. The black dotted line in each panel represents the $M$-$R$ profile of a pure baryonic NS, calculated using the single-fluid TOV with QMC-RMF4 EOS. Each grid corresponds to a different ratio of DM to NM central energy density: ${\cal E}_{\rm c}^{\rm DM} = {\cal E}_{\rm c}^{\rm NM}$ (left), ${\cal E}_{\rm c}^{\rm DM} = 2 {\cal E}_{\rm c}^{\rm NM}$ (center), and ${\cal E}_{\rm c}^{\rm DM} = 4 {\cal E}_{\rm c}^{\rm NM}$ (right). The color-coded curves represent various values of $g_{\chi}/m_{\rm{v}}$ , ranging from 0.01 to 0.06 MeV$^{-1}$. Observational constraints from PSR J0740+6620 and PSR J0030+0451 are depicted as shaded bands and contours.}
    \label{fig:figure3}
\end{figure*}
Figure~\ref{fig:figure2} presents the individual contributions from NM and DM to the total gravitational mass and radius, as well as their combined mass-radius profile. The blue dashed curve in Fig. \ref{fig:figure2} represents the NM mass-radius profile ($M_{\rm{NM}}$ and $R_{\rm{NM}}$), while the red dashed curve corresponds to the DM mass-radius relationship ($M_{\rm{DM}}$ and $R_{\rm{DM}}$). The total mass-radius curve ($M = M_{\rm{NM}} + M_{\rm{DM}}$ and $R= {\rm{max}} (R_{\rm{NM}}, R_{\rm{DM}})$) is shown with markers, which distinguish between two structural regimes: blue open circles indicate core-dominated configurations where the NM radius ($R_{\rm NM}$) exceeds the DM radius ($R_{\rm DM}$) (i.e. $R_{\rm{NM}} > R_{\rm{DM}}$), while red crosses represent halo-dominated configurations where $R_{\rm DM} > R_{\rm NM}$. These markers highlight the relative contributions of each fluid across the mass spectrum and provide insight into the structural sequence of the star. 

Specific configurations with total gravitational masses of $0.60 M_{\odot}$, $0.90 M_{\odot}$, $1.30 M_{\odot}$, $1.90 M_{\odot}$ and $2.20 M_{\odot}$ are marked (star markers) on the total $M-R$ curve. These configurations are solutions of the TOV equations for specific values of the shared central energy density. These markers on the $M(R)$ curve indicate the total gravitational mass of the star, while the arrows connect these markers to their corresponding points on the $M_{\rm NM}$ and $M_{\rm DM}$ curves, illustrating the respective contributions of NM and DM to the total mass. 

Another feature of this representative case is the transition from halo-dominated to core-dominated configurations as the total mass increases. For lower masses ($M \lesssim 1.30 M_{\odot}$), the DM component forms an extended halo that dominates the gravitational field, leading to $R = R_{\rm DM}$. In these configurations, the softer DM EOS results in a larger radius for the DM component relative to the NM core. However, as the total mass increases ($M \gtrsim 1.30 M_{\odot}$), the stronger gravitational forces compress the DM halo, causing the NM core to dominate the structure and resulting in $R = R_{\rm NM}$. This transition reflects the interplay between DM self-interactions and the gravitational coupling between NM and DM. This structural evolution highlights the nuanced role of DM in shaping the properties of DM admixed NSs, particularly the gravitational mass, radii, and the transition between halo-dominated and core-dominated configurations (see also Fig.~\ref{fig:figure13} in appendix~\ref{sec:appendx1}).
\begin{figure*}[tbp]
    \centering
    \includegraphics[width=\textwidth]{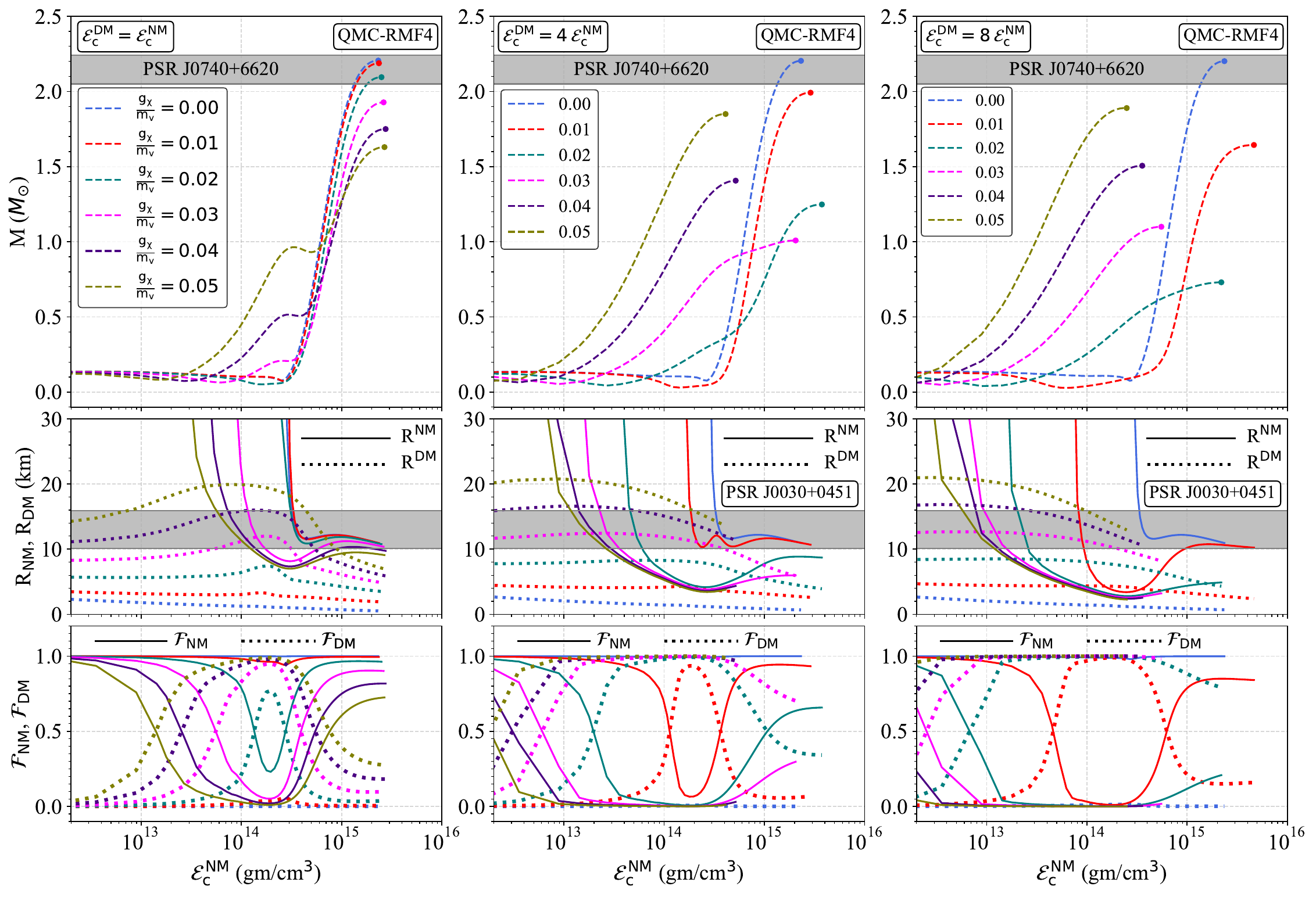} 
    \caption{Total mass $M$, radii ($R_{\rm{NM}}$ and $R_{\rm{DM}}$), and mass fractions for admixed NS configurations as functions of NM central energy density (${\cal E}_{\rm c}^{\rm NM}$), constructed using QMC-RMF4 EOS for baryonic matter. The three columns correspond to different central energy density ratios: ${\cal E}_{\rm c}^{\rm DM} = {\cal E}_{\rm c}^{\rm NM}$ (left), ${\cal E}_{\rm c}^{\rm DM} = 4 {\cal E}_{\rm c}^{\rm NM}$ (center), and ${\cal E}_{\rm c}^{\rm DM} = 8{\cal E}_{\rm c}^{\rm NM}$ (right). The color coding in each panel represents varying DM self-coupling strengths ($g_{\chi}/m_{\rm{v}}$) ranging from 0.00 to 0.05 MeV$^{-1}$. The total mass ($M$) is presented in the first row, with observational constraints from PSR J0740+6620 depicted as shaded regions and the maximum mass for each case marked with solid dots. The middle row compares the radii of NM (solid lines) and DM (dotted lines), with observational constraints from PSR J0030+0451 highlighted as shaded regions. The lower row illustrates the mass fractions of NM $\left({\cal F}_{\rm{NM}}=M_{\rm{NM}}/M, \,{\rm{solid\,\,lines}} \right)$ and DM $\left({\cal F}_{\rm{DM}}=M_{\rm{DM}}/M, \,{\rm{dotted\,\,lines}} \right)$.}
    \label{fig:figure4}
\end{figure*}

Having established the methodology for constructing of $M-R$ profiles for DM admixed NSs using the two-fluid formalism, we extend the analysis to explore the effects of varying DM to NM central energy density ratios (${\cal E}_{\rm c}^{\rm DM}/{\cal E}_{\rm c}^{\rm NM}$) and the DM self-coupling parameter ($g_{\chi}/m_{\rm v}$) on the gravitational mass ($M$) and structural regime of these stars, as shown in Fig.~\ref{fig:figure3}. This figure illustrates $M-R$ profiles for three distinct central energy density ratios: ${\cal E}_{\rm c}^{\rm DM} = {\cal E}_{\rm c}^{\rm NM}$ (left panel), ${\cal E}_{\rm c}^{\rm DM} = 2\, {\cal E}_{\rm c}^{\rm NM}$ (middle panel), ${\cal E}_{\rm c}^{\rm DM} = 4\, {\cal E}_{\rm c}^{\rm NM}$ (right panel), with QMC-RMF4 as the NM EOS and DM characterized by varying DM self-coupling parameter ($g_{\chi}/m_{\rm v}$) values spanning from 0.01 to 0.06 MeV$^{-1}$ (from 0.005 to 0.05 MeV$^{-1}$ for ${\cal E}_{\rm c}^{\rm DM} = 4\, {\cal E}_{\rm c}^{\rm NM}$). For reference, we also show in each panel the $M-R$ relations constructed from the pure baryonic EOS, without any DM contribution, using the dotted line.
Observational constraints from PSR J0740+6620 (gray-shaded band) and PSR J0030+0451 (elliptical contours) are overlaid for reference, providing key benchmarks. Configurations falling outside these constraints are excluded as physically implausible in light of currently available observational data. In particular, the compatibility of these $M-R$ profiles with the PSR J0030+0451 contours is critical in validating the role of DM in admixed NSs. The solid dots on each curve represent the maximum gravitational mass ($M_{\rm max}$) for the corresponding $M-R$ profile.

For configurations with smaller $g_{\chi}/m_{\rm v}$ values (e.g. 0.01 MeV$^{-1}$ for ${\cal E}_{\rm c}^{\rm DM} = {\cal E}_{\rm c}^{\rm NM}$ and 0.005 MeV$^{-1}$ for ${\cal E}_{\rm c}^{\rm DM} = 4\, {\cal E}_{\rm c}^{\rm NM}$), the $M-R$ profiles closely resemble those of pure baryonic stars. In this regime, the DM EOS remains relatively soft, resulting in minimal contributions to the overall structure. The maximum mass and radius remain dominated by NM, and the structural impact of DM is negligible. However, as $g_{\chi}/m_{\rm v}$ increases for each ${\cal E}_{\rm c}^{\rm DM}/{\cal E}_{\rm c}^{\rm NM}$ ratio (i.e. 0.04 MeV$^{-1}$ for ${\cal E}_{\rm c}^{\rm DM} = {\cal E}_{\rm c}^{\rm NM}$, 0.03 MeV$^{-1}$ for ${\cal E}_{\rm c}^{\rm DM} = 2\, {\cal E}_{\rm c}^{\rm NM}$, and 0.02 MeV$^{-1}$ for ${\cal E}_{\rm c}^{\rm DM} = 4\, {\cal E}_{\rm c}^{\rm NM}$), the DM EOS stiffens, leading to exert a greater influence on the $M-R$ profile, particularly in the low-mass region. This results in an extended halo around the NM core, resulting in halo-dominated configurations characterized by larger radii for low-mass stars. For such configurations, a reduction in the maximum mass ($M_{\rm{max}}$) of the star is also observed compared to the pure baryonic star. This reduction occurs because, for these values of $g_{\chi}/m_{\rm v}$, the DM EOS remains softer than the NM EOS. Although the DM component contributes significantly to the overall structure, its softer EOS (as compared to NM EOS) supports gravitational equilibrium less effectively than the NM EOS, leading to reduced $M_{\rm{max}}$. The DM component for these $g_{\chi}/m_{\rm v}$ values, while exerting a greater influence on the low-mass star's structure, is less efficient in providing the stiffness required for higher maximum masses.

For higher values of $g_{\chi}/m_{\rm v}$ (e.g., 0.05 MeV$^{-1}$ for ${\cal E}_{\rm c}^{\rm DM} = 2\, {\cal E}_{\rm c}^{\rm NM}$ and ${\cal E}_{\rm c}^{\rm DM} = 4\, {\cal E}_{\rm c}^{\rm NM}$), the DM EOS becomes significantly stiffer due to stronger self-repulsion among DM particles. This stiffening stabilizes the DM component, leading to maximum mass configurations dominated by the DM halo as the DM EOS becomes comparable to or stiffer than the NM EOS. However, because of the characteristic large radii predicted by the DM EOS, DM halo-dominated maximum mass stars tend to have extended radii compared to their core-dominated counterparts, even for similar masses. This behavior reflects the ability of DM with increased stiffness to contribute effectively to the star's gravitational equilibrium. A more detailed step-by-step analysis of the influence of DM self-coupling on NS structure is presented in Appendix \ref{sec:appendx1} (Figure \ref{fig:figure13}), highlighting the transition between core- and halo-dominated configurations with varying $g_\chi/m_{\rm v}$ values.

A particularly intriguing feature in the left panel (${\cal E}_{\rm c}^{\rm DM} = {\cal E}_{\rm c}^{\rm NM}$) of Fig.~\ref{fig:figure3} is the appearance of two distinct maxima in some $M-R$ profiles, such as for $g_{\chi}/m_{\rm v} = 0.06$ MeV$^{-1}$. This unusual behavior arises due to a transition in structural dominance between the NM core and the DM halo. For such high values of $g_{\chi}/m_{\rm v}$ at this particular central density ratio, the increased stiffness of the DM EOS creates localized regions of stability, where both the NM core and the DM halo contribute comparably to the overall gravitational equilibrium. This behavior reflects the intricate interplay between the two fluids, with localized regions of stability emerging from the gravitational coupling of the NM and DM components.

Figure~\ref{fig:figure4} offers a detailed breakdown of the interplay between NM and DM components in determining the structural and compositional properties of DM admixed NSs, using QMC-RMF4 EOS for baryonic matter. Complementing the mass-radius profiles from Fig.~\ref{fig:figure3}, this figure expands the discussion by presenting total gravitational mass ($M$), radii ($R_{\rm NM}$ and $R_{\rm DM}$), and mass fractions (${\cal F}_{\rm NM}\equiv M_{\rm NM}/M$ and ${\cal F}_{\rm DM}\equiv M_{\rm DM}/M$) as functions of the NM central energy density (${\cal E}_{\rm c}^{\rm NM}$) for different DM to NM central energy density ratios (${\cal E}_{\rm c}^{\rm DM}/{\cal E}_{\rm c}^{\rm NM}$): ${\cal E}_{\rm c}^{\rm DM} = {\cal E}_{\rm c}^{\rm NM}$ (left), ${\cal E}_{\rm c}^{\rm DM} = 4\, {\cal E}_{\rm c}^{\rm NM}$ (center), and ${\cal E}_{\rm c}^{\rm DM} = 8\, {\cal E}_{\rm c}^{\rm NM}$ (right). The rows, from top to bottom, present the total gravitational mass ($M$), radii of NM and DM components ($R_{\rm NM}$ and $R_{\rm DM}$), and the mass fractions of NM and DM (${\cal F}_{\rm NM}$ and ${\cal F}_{\rm DM} $). DM self-coupling parameter ($g_{\chi}/m_{\rm v}$) is varied from 0.00 to 0.05 MeV$^{-1}$, with color-coded lines denoting specific values.

The top row highlights the total gravitational mass $M$, demonstrating a clear dependence on both ${\cal E}_{\rm c}^{\rm DM}/{\cal E}_{\rm c}^{\rm NM}$ and $g_{\chi}/m_{\rm v}$. For $g_{\chi}/m_{\rm v} \approx 0$, the $M$ profiles across all panels exhibit a monotonic increase with ${\cal E}_{\rm c}^{\rm NM}$ once the stellar model enters into a stable sequence, culminating in a maximum gravitational mass ($M_{\rm max}$), which is marked with solid dots. In this regime, the contribution of DM to the total gravitational structure is minimal, as evidenced by the dominance of NM in both radii ($R_{\rm NM} \gg R_{\rm DM}$) and mass fractions (${F}_{\rm NM} > {F}_{\rm DM}$). However, as $g_{\chi}/m_{\rm v}$ increases, the subtle deviations in the intermediate and low-mass portion of $M$ profiles emerges showing significant DM contributions (${\cal F}_{\rm DM} > {\cal F}_{\rm NM}$ for intermediate values of ${\cal E}_{\rm c}^{\rm NM}$, bottom row of Fig. \ref{fig:figure4}). These deviations reflect the gradual stiffening of the DM EOS, which begins to exert a noticeable effect on the star's structure. 

For ${\cal E}_{\rm c}^{\rm DM}/{\cal E}_{\rm c}^{\rm NM} = 4$ and 8, $M_{\rm{max}}$ is significantly reduced compared to the ${\cal E}_{\rm c}^{\rm DM} = {\cal E}_{\rm c}^{\rm NM}$ case for smaller $g_{\chi}/m_{\rm v}$ values, with $M_{\rm{max}}$ configurations transitioning from core-dominated to halo-dominated structures as $g_{\chi}/m_{\rm v}$ increases. Notably, we observe comparable $M_{\rm{max}}$ configurations ($R_{\rm NM} \approx R_{\rm DM}$, middle row) for ${\cal E}_{\rm c}^{\rm DM} = 4\, {\cal E}_{\rm c}^{\rm NM}$ with $g_{\chi}/m_{\rm v}=0.03$ MeV$^{-1}$ and for ${\cal E}_{\rm c}^{\rm DM} = 8\, {\cal E}_{\rm c}^{\rm NM}$ with $g_{\chi}/m_{\rm v}\approx$ 0.02 MeV$^{-1}$. However, when focusing on ${\cal E}_{\rm c}^{\rm NM}$, which corresponds to the maximum mass configuration for comparable $M_{\rm{max}}$ structures, we observe that this value for DM-dominated $M_{\rm{max}}$ configurations becomes smaller compared to their NM-dominated counterparts. Additionally, DM-dominated $M_{\rm{max}}$ structures exhibit larger radii, consistent with the characteristics of halo-dominated configurations.

\begin{figure}[tbp]
    \centering
    \includegraphics[width=\columnwidth]{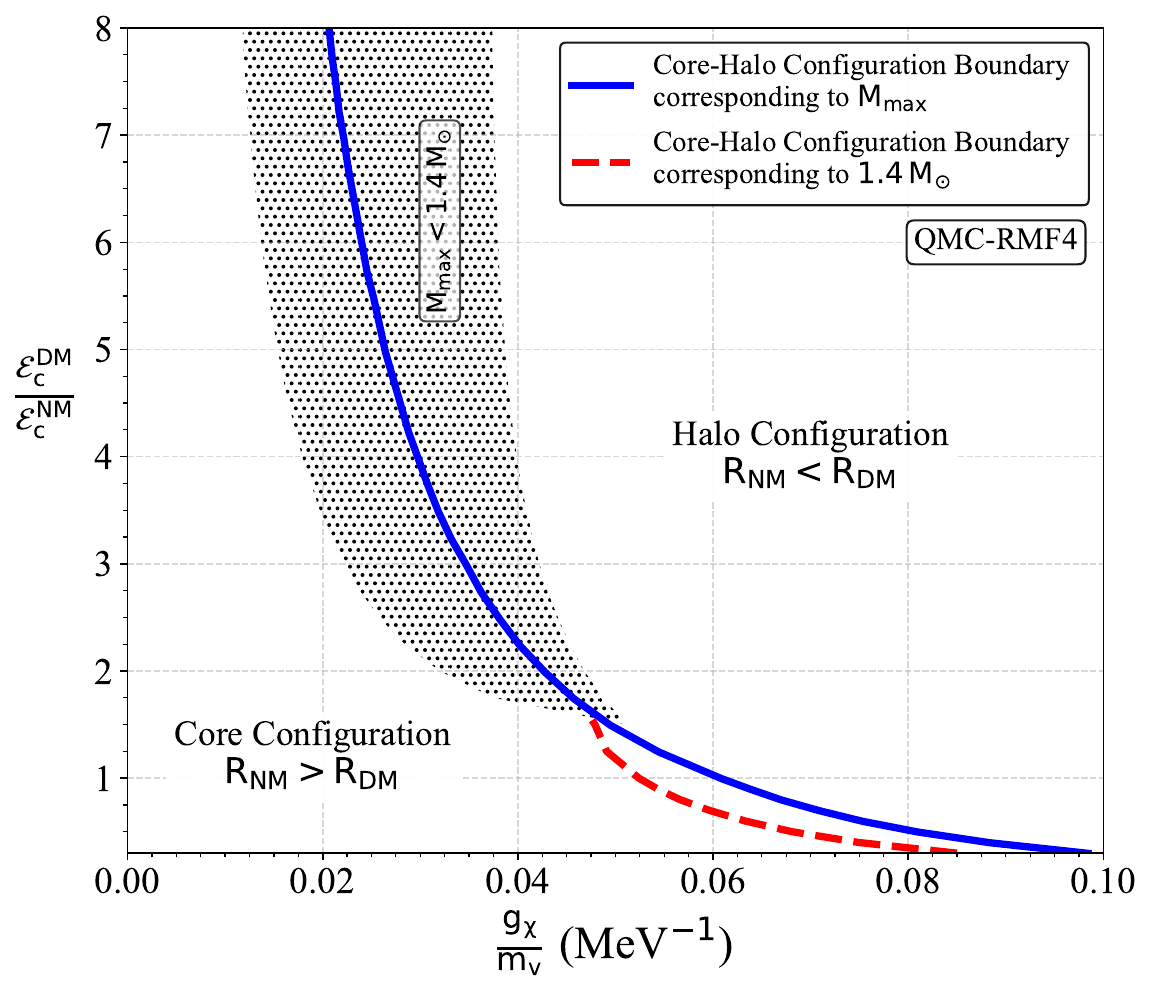} 
    \caption{Core-halo configuration boundaries for DM admixed NSs, constructed using QMC-RMF4 EOS for NM and varying DM self-coupling strengths ($g_{\chi}/m_{\rm v}$). The plot shows the transition between core-dominated ($R_{\rm NM} > R_{\rm DM}$) and halo-dominated ($R_{\rm NM} < R_{\rm DM}$) configurations as a function of the central energy density ratio ($\mathcal{E}_{\rm c}^{\rm DM}/\mathcal{E}_{\rm c}^{\rm NM}$) and $g_{\chi}/m_{\rm v}$. The blue solid line represents the core-halo boundary corresponding to configurations with the maximum mass, while the green solid line indicates the boundary for $1.4 \,M_{\odot}$ configurations. The hatched region with black dots indicates configurations where the maximum mass is less than $1.4 \,M_{\odot}$.}
    \label{fig:figure5}
\end{figure}
The middle row illustrates $R_{\rm NM}$ (solid lines) and $R_{\rm DM}$ (dotted lines) as functions of ${\cal E}_{\rm c}^{\rm NM}$, emphasizing a clear transition from core-dominated to halo-dominated configurations. Interestingly, $R_{\rm NM}$ decreases as ${\cal E}_{\rm c}^{\rm DM}/{\cal E}_{\rm c}^{\rm NM}$ increases for almost all $g_{\chi}/m_{\rm v}$ values, indicating the contraction of NM surface due to increased gravitational attraction by DM component. The mass fraction curves (${\cal F}_{\rm NM}$ and ${\cal F}_{\rm DM}$) further highlight the compositional shifts in these two fluid stars. The DM mass fraction ${\cal F}_{\rm DM}$ remains significant at intermediate ${\cal E}_{\rm c}^{\rm NM}$ across all three panels and $g_{\chi}/m_{\rm v}$ values, underscoring the role of DM EOS stiffness in determining compositional transitions. The DM fraction approaches unity for the case of ${\cal E}_{\rm c}^{\rm DM} = 4\, {\cal E}_{\rm c}^{\rm NM}$, particularly for configurations with $g_{\chi}/m_{\rm v} \geq 0.03$ MeV$^{-1}$, while DM accounts for the majority of the total mass across most of the ${\cal E}_{\rm c}^{\rm NM}$ spectrum for ${\cal E}_{\rm c}^{\rm DM} = 8\, {\cal E}_{\rm c}^{\rm NM}$ and $g_{\chi}/m_{\rm v} \geq 0.02$ MeV$^{-1}$, highlighting the structural dominance of DM in these scenarios. Also, the transition from core-dominated to halo-dominated configurations occurs earlier for higher ${\cal E}_{\rm c}^{\rm DM}/{\cal E}_{\rm c}^{\rm NM}$ ratios i.e. with smaller $g_{\chi}/m_{\rm v}$ values. This shift in dominance is also consistent with the trends observed in gravitational mass and radii (Fig.~\ref{fig:figure3}), where DM halo-dominated configurations exhibit extended radii with reduced $M_{\rm{max}}$ for smaller $g_{\chi}/m_{v}$ and higher $M_{\rm{max}}$ as $g_{\chi}/m_{v}$ further increases.

Figure \ref{fig:figure5} provides a comprehensive overview of the structural transition boundaries between core and halo-dominated configurations for DM admixed NSs in the parameter space of $g_{\chi}/m_{\rm v}$ and ${\cal E}_{\rm c}^{\rm DM}/{\cal E}_{\rm c}^{\rm NM}$, constructed using the QMC-RMF4 EOS for NM. The blue solid line represents the transition boundary between core- and halo-dominated configurations for maximum mass ($M_{\rm max}$) stars, while the red dashed line delineates the same transition boundary for configurations with a gravitational mass of $1.4 M_{\odot}$. Configurations below these boundaries correspond to core-dominated structures, where $R_{\rm NM} > R_{\rm DM}$, conversely, configurations above these boundaries are halo-dominated, with $R_{\rm DM} > R_{\rm NM}$, where the DM halo envelops the NM. 

The hatched region with black dots highlights unphysical configurations where $M_{\rm max}$ falls below $1.4 M_{\odot}$, indicating the inability of these parameter combinations to support canonical NS masses. The boundary of this hatched region separates core- and halo-dominated structures specifically for $1.4 M_{\odot}$ stars. To the left of this boundary, $1.4 M_{\odot}$ stars are core-dominated, while to the right, they transition to halo-dominated structures. The core-halo boundary shifts toward smaller values of $g_{\chi}/m_{\rm v}$ as ${\cal E}_{\rm c}^{\rm DM}/{\cal E}_{\rm c}^{\rm NM}$ increases, reflecting that higher central energy density ratios favor halo-dominated structures even for relatively small DM self-coupling strengths.

\subsection{Constraining DM parameters from astrophysical observations}
\label{sec:3b}
To constrain the DM self-coupling parameter ($g_{\chi}/m_{\rm v}$) and its impact on NS structure, we leverage observational data, including the maximum mass constraint from PSR J0740+6620~\cite{2020NatAs...4...72C, Fonseca_2021}, NICER's analysis of PSR J0030+0451~\cite{Miller_2019, Riley_2019}, and tidal deformability limits for canonical stars from GW170817~\cite{PhysRevLett.121.161101}. These data, combined with theoretical models, allow us to systematically explore the permissible parameter space for DM in NSs.
\begin{figure}[tbp]
    \centering
    \includegraphics[width=\columnwidth]{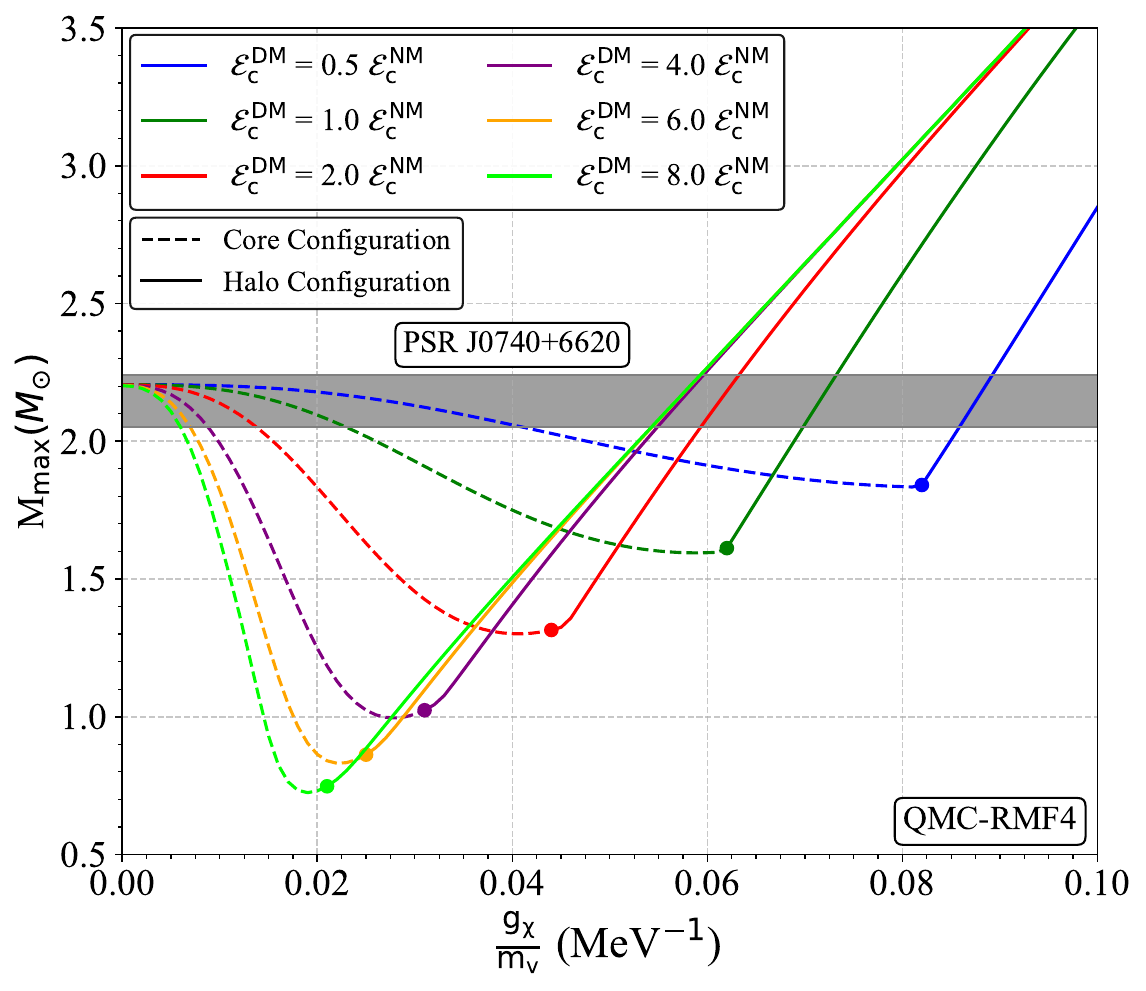} 
    \caption{Maximum gravitational mass ($M_{\rm max}$) of DM admixed NSs as a function of DM self-coupling parameter ($g_{\chi}/m_{\rm v}$), with the QMC-RMF4 EOS for NM. Each line corresponds to a different central energy density ratio, ranging from ${\cal E}_{\rm c}^{\rm DM} = 0.5\, {\cal E}_{\rm c}^{\rm NM}$ to ${\cal E}_{\rm c}^{\rm DM} = 8.0\, {\cal E}_{\rm c}^{\rm NM}$. Dashed lines indicate configurations where the maximum mass corresponds to a core-dominated structure, while solid lines represent configurations dominated by a DM halo. The shaded region represents the observational constraint on the maximum NS mass from PSR J0740+6620. The transitions between core-dominated and halo-dominated configurations are marked by circles on the curves.}
    \label{fig:figure6}
\end{figure}

Figure \ref{fig:figure6} illustrates the maximum gravitational mass ($M_{\rm max}$) of DM admixed NSs with various values of ${\cal E}_{\rm c}^{\rm DM}/{\cal E}_{\rm c}^{\rm NM}$ as a function of the DM self-coupling parameter ($g_{\chi}/m_{\rm v}$). The analysis spans a range of ${\cal E}_{\rm c}^{\rm DM}/{\cal E}_{\rm c}^{\rm NM}$ from 0.5 to 8.0, with each curve corresponding to a specific ratio. Dashed curves indicate maximum mass structures classified as core configurations, where $R_{\rm NM} > R_{\rm DM}$, while solid curves represent halo configurations, where $R_{\rm DM} > R_{\rm NM}$. Transition points between core and halo configurations are marked by circles on each curve. The shaded region represents the observational constraint on $M_{\rm max}$ for NSs, derived from PSR J0740+6620, which is considered the most massive pulsar observed to date with $M = 2.14^{+0.10}_{-0.09} \,M_{\odot}$ at a 68.3\% credibility interval \cite{2020NatAs...4...72C}. 

The maximum gravitational mass demonstrates a clear dependence on both ${\cal E}_{\rm c}^{\rm DM}/{\cal E}_{\rm c}^{\rm NM}$ and $g_{\chi}/m_{\rm v}$. The transitions between core and halo configurations, marked by circles, reveal the sensitivity of structural regimes to both $g_{\chi}/m_{\rm v}$ and ${\cal E}_{\rm c}^{\rm DM}/{\cal E}_{\rm c}^{\rm NM}$. At low ${\cal E}_{\rm c}^{\rm DM}/{\cal E}_{\rm c}^{\rm NM}$ ratios, the transition to halo configurations occurs at relatively high $g_{\chi}/m_{\rm v}$ values, as NM remains the dominant contributor to the star's stiffness. Conversely, for high ${\cal E}_{\rm c}^{\rm DM}/{\cal E}_{\rm c}^{\rm NM}$ ratios, the transition occurs at lower $g_{\chi}/m_{\rm v}$ values, reflecting the increased dominance of DM in maintaining gravitational equilibrium.

\begin{figure}[tbp]
    \centering
    \includegraphics[width=\columnwidth]{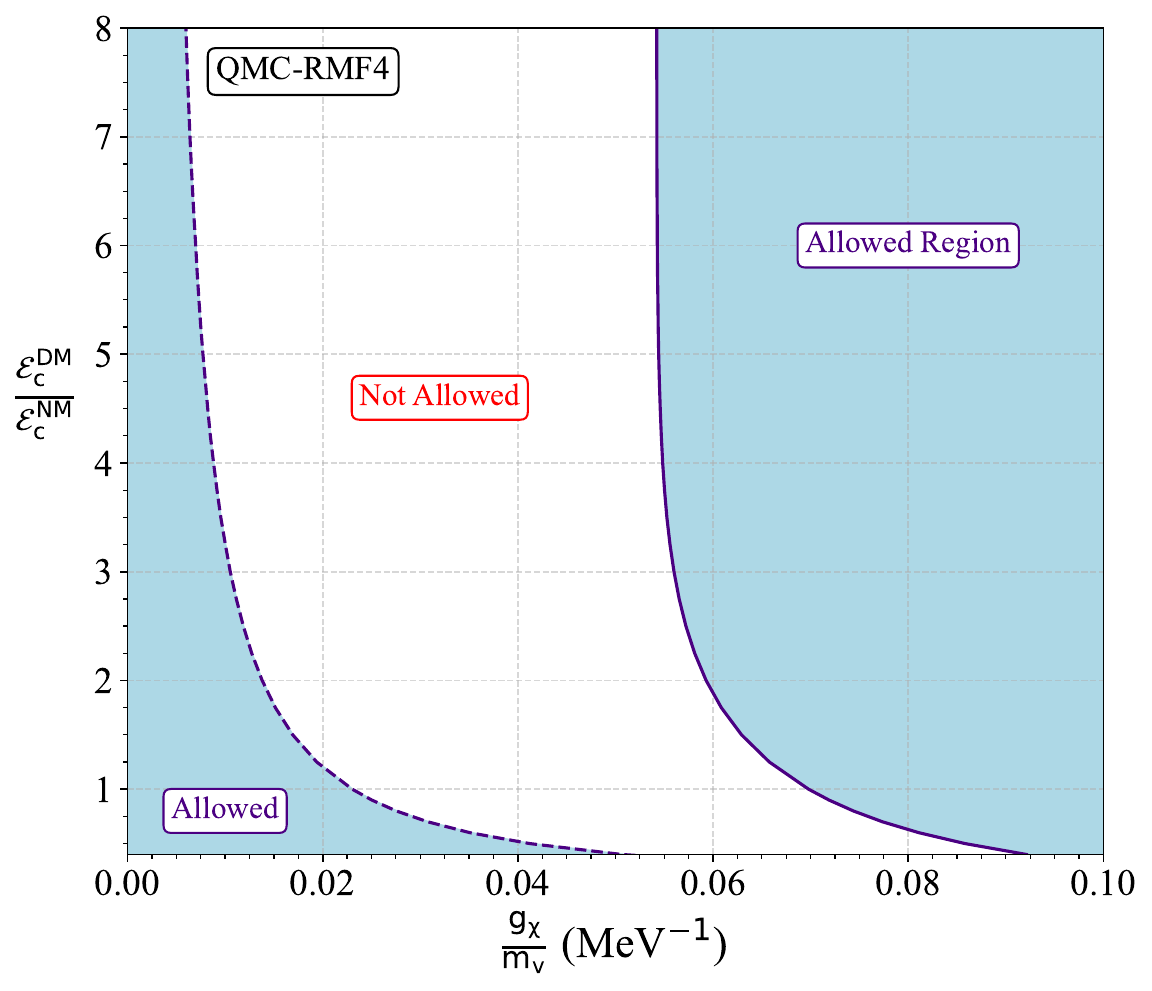} 
    \caption{Allowed parameter space for $g_{\chi}/m_{\rm v}$ as a function of the central energy density ratio ${\cal E}_{\rm{c}}^{\rm DM}/{\cal E}_{\rm c}^{\rm NM}$ with the QMC-RMF4 EOS for NM, constrained by the observational lower limit of the maximum mass from PSR J0740+6620. The shaded blue region represents the configurations where this observational constraint is satisfied. The left boundary of the allowed region corresponds to core-dominated configurations, while the right boundary corresponds to halo-dominated configurations.}
    \label{fig:figure7}
\end{figure}
Linking this analysis to observations, the $M_{\rm max}$ constraint from PSR J0740+6620 provides a critical upper bound on DM contributions to NS structure. For core configurations, this constraint is satisfied for a wide range of ${\cal E}_{\rm c}^{\rm DM}/{\cal E}_{\rm c}^{\rm NM}$ and $g_{\chi}/m_{\rm v}$ values, due to the inherent stiffness of NM, which ensures gravitational equilibrium. In contrast, halo configurations can satisfy the $M_{\rm max}$ constraint only for sufficiently stiff DM EOSs, typically achieved at higher $g_{\chi}/m_{\rm v}$ values. This underscores the necessity of a moderately stiff DM EOS to maintain gravitational stability while accommodating the extended radii characteristic of halo configurations.

Observationally, the existence of PSR J0740+6620 confirms that $M_{\rm max}$ must exceed 2.05 $M_{\odot}$, ruling out parameter combinations that fail to support this mass, which reinforces the need for consistent gravitational coupling between NM and DM. Figure \ref{fig:figure7} builds on this constraint by mapping the allowed parameter space for $g_{\chi}/m_{\rm v}$ as a function of ${\cal E}_{\rm{c}}^{\rm DM}/{\cal E}_{\rm c}^{\rm NM}$. The shaded blue region represents the combinations of $g_{\chi}/m_{\rm v}$ and ${\cal E}_{\rm{c}}^{\rm DM}/{\cal E}_{\rm c}^{\rm NM}$ that satisfy the observational lower limit on $M_{\rm max}$, namely $M_{\rm max} \geq 2.05 M_{\odot}$, as observed for PSR J0740+6620.

\begin{figure*}[tbp]
    \centering
    \includegraphics[width=\textwidth]{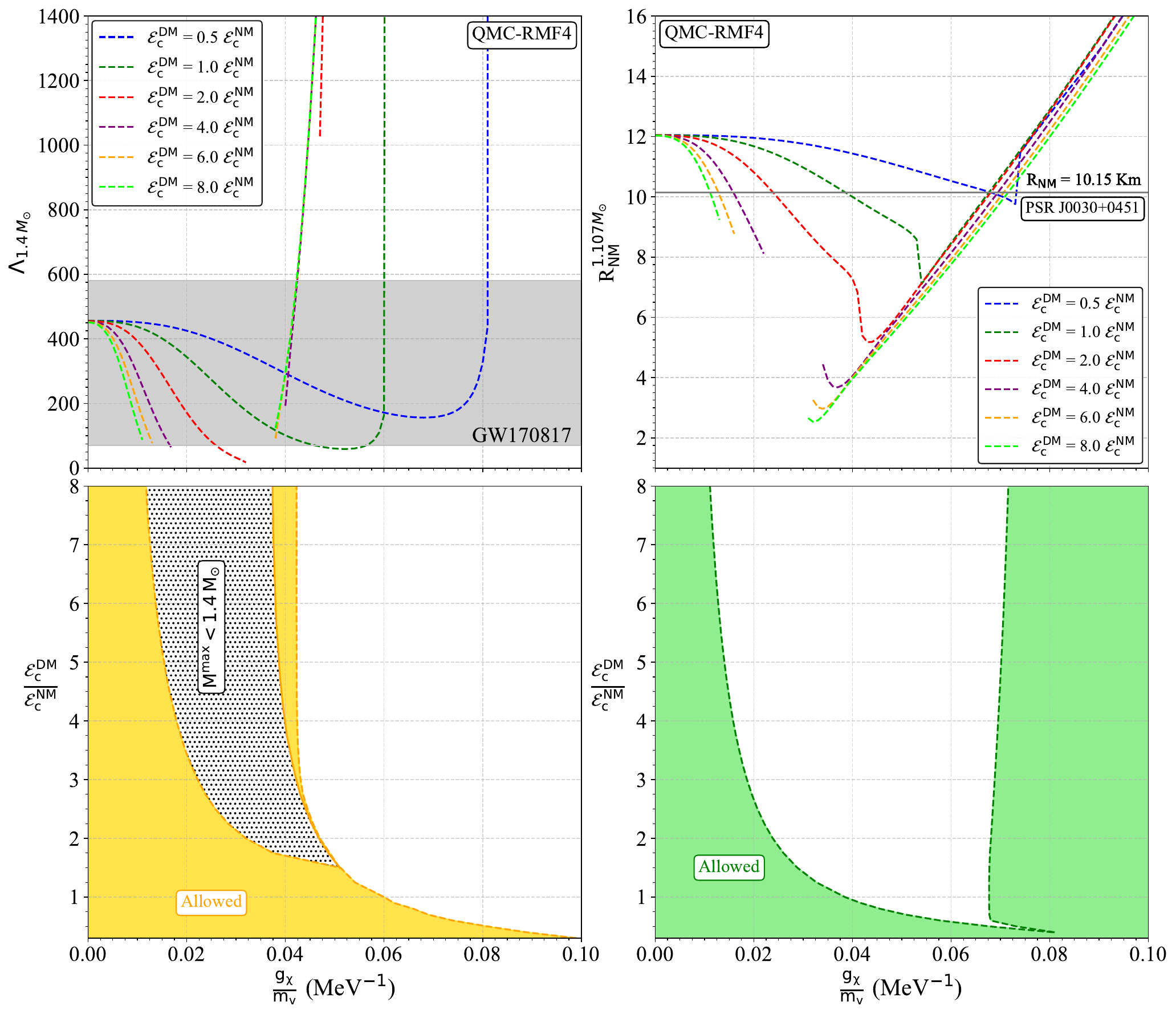} 
    \caption{This figure presents the constraints on DM admixed NS configurations with QMC-RMF4 EOS for NM. \textbf{Upper left panel :} Dimensionless tidal deformability $\Lambda_{1.4M_{\odot}}$ for canonical stars ($1.4M_\odot$) is plotted against $g_{\chi}/m_{\rm v}$ for different central energy density ratios ($\mathcal{E}_{\rm c}^{\rm DM}/\mathcal{E}_{\rm c}^{\rm NM}$). The shaded region indicates the allowed range of $\Lambda_{1.4M{\odot}}$ extracted from GW170817 data, with an upper limit of $\Lambda_{1.4M_{\odot}} = 580$. Gaps or discontinuities in the curves correspond to configurations where the maximum mass is less than $1.4 M_{\odot}$. \textbf{Upper Right :} Nuclear matter radius $R_{\rm NM}$ for a total mass $M = 1.107 M_{\odot}$ (corresponding to the leftmost point in the contour for PSR J0030+0451, see Fig.~\ref{fig:figure3}) is shown as a function of $g_{\chi}/m_{\rm v}$. Different color-coded lines represent results for varying central energy density ratios, $\mathcal{E}_{\rm c}^{\rm DM}/\mathcal{E}_{\rm c}^{\rm NM}$. The horizontal line at $R_{\rm NM} = 10.15$ km represents the NM radius for the $M = 1.107 M_{\odot}$ stars, consistent with NICER constraints for PSR J0030+0451. \textbf{Lower left panel :} The allowed region (yellow shaded) in the parameter space of $g_{\chi}/m_{\rm v}$ and $\mathcal{E}_{\rm c}^{\rm DM}/\mathcal{E}_{\rm c}^{\rm NM}$ based on the canonical tidal deformability constraint from GW170817 event. The upper boundary of the shaded region corresponds to the combinations of $g_{\chi}/m_{\rm v}$ and $\mathcal{E}_{\rm c}^{\rm DM}/\mathcal{E}_{\rm c}^{\rm NM}$ that produce $\Lambda_{1.4M{\odot}} = 580$. Configurations outside this region, yield tidal deformability values for canonical stars exceeding the observational limit and are therefore ruled out. The hatched region filled with black dots represents configurations where the maximum mass of the DM admixed NS is less than $1.4M_{\odot}$, excluding them due to the absence of a viable canonical star. \textbf{Lower right panel :} Allowed region (green shaded area) in the parameter space of $g_{\chi}/m_{\rm v}$ and $\mathcal{E}_{\rm c}^{\rm DM}/\mathcal{E}_{\rm c}^{\rm NM}$ is determined by the NICER constraint of NM radius $R_{\rm NM}$ for a $1.107\, M_{\odot}$ star. The shaded region represents configurations where the NM radius for the $M = 1.108\, M_{\odot}$ star is greater than $10.15$ km, consistent with the NICER analysis of PSR J0030+0451's observational data.}
    \label{fig:figure8}
\end{figure*}

The left boundary of the allowed region corresponds to the core configurations, where $R_{\rm NM} > R_{\rm DM}$ and NM remains the dominant structural component. On the other hand, the right boundary represents halo configurations, where $R_{\rm DM} > R_{\rm NM}$ and DM dominates the structure. The right boundary is largely independent of $g_{\chi}/m_{\rm v}$ for ${\cal E}_{\rm{c}}^{\rm DM}/{\cal E}_{\rm c}^{\rm NM} \geq 3.5$ (as evident from Fig. \ref{fig:figure6} as well), indicating that for such halo configurations, the DM EOS achieves sufficient stiffness to support gravitational equilibrium. This insensitivity to $g_{\chi}/m_{\rm v}$ reflects how the halo boundary depends primarily on the intrinsic properties of the DM EOS, with NM existing as a minor (or negligible) fraction in these scenarios.

Figure \ref{fig:figure8} provides a comprehensive examination of constraints on DM admixed NS configurations by combining insights from two key astrophysical observations: tidal deformability limits from GW170817 and NM radius constraints derived from NICER observations of PSR J0030+0451. The four panels are divided into two categories: the left side focuses on tidal deformability constraints for canonical stars ($1.4M_\odot$), while the right side examines NM radius constraints for $1.107 M_\odot$ stars ($R_{\rm{NM}}^{1.107 M_\odot}$). The upper panels illustrate parameter dependencies, and the lower panels map the corresponding allowed parameter space. 

The upper left panel illustrates the dimensionless tidal deformability of canonical star ($\Lambda_{1.4M_\odot}$) as a function of $g_\chi/m_{\rm v}$ for different ${\cal E}_{\rm c}^{\rm DM}/{\cal E}_{\rm c}^{\rm NM}$ ratios. The observational limit $70 \leq \Lambda_{1.4 M_\odot} \leq 580$, derived from GW170817 data~\cite{PhysRevLett.121.161101}, is represented as a shaded region. Gaps or discontinuities in the curves with ${\cal E}_{\rm c}^{\rm DM}/{\cal E}_{\rm c}^{\rm NM} = 2, 4, 6,$ and 8 correspond to configurations where $M_{\rm max} < 1.4 M_\odot$, highlighting parameter combinations unable to sustain canonical NS masses and making such configurations physically inaccessible for this analysis. Such excluded combinations due to the absence of a viable canonical star mass are shown as hatched region with black dots in the bottom left panel which maps the allowed parameter space in $g_\chi/m_{\rm v}$ and ${\cal E}_{\rm c}^{\rm DM}/{\cal E}_{\rm c}^{\rm NM}$ based on the $\Lambda_{1.4M_\odot} \leq 580$ constraint. Configurations satisfying this tidal deformability limit are shaded in yellow, while those outside this region are forbidden due to exceeding the observational limit. Future gravitational wave observations, especially with improved precision in $\Lambda$, could further refine these constraints.

The calculation of tidal deformability $\Lambda$ reflects the integrated response of the star to external tidal fields, accounting for the combined contributions of both NM and DM components to the overall gravitational structure. Importantly, $\Lambda$ is computed using the total gravitational mass ($M$) and radius ($R$) of these admixed stars, as these parameters encapsulate the cumulative gravitational effects of both NM and DM. In halo configurations, the extended DM component contributes comparatively less to the star's gravitational compactness relative to the significant increase in the total radius. This pronounced shift in $R$, combined with the steep scaling of $\Lambda$ with $R$ ($\Lambda \propto R^5$), explains the disproportionately higher tidal deformabilities observed in halo-dominated regimes. 

The upper right panel shows the NM radius, $R_{\rm NM}$, for a DM adimixed NS with $M = 1.107 M_\odot$ as a function of $g_\chi/m_{\rm v}$ for varying ${\cal E}_{\rm c}^{\rm DM}/{\cal E}_{\rm c}^{\rm NM}$. This specific stellar mass corresponds to the leftmost point on the elliptical contour for PSR J0030+0451 constrained from NICER observations (see Fig.~\ref{fig:figure3}). Observational data analysis suggests that $R_{\rm NM}$ for this star should be greater than $10.15$ km. This threshold is represented by a black horizontal line on the plot.

Notably, our theoretical analysis relies on the NM surface radius, not the total radius, because NICER's pulse-profile modeling technique primarily probes the surface where thermal X-ray emission originates. The NM surface directly interfaces with NICER's observational window, as its pulse-profile analysis is sensitive to the thermal X-ray emissions originating from the NM surface. In contrast, the DM halo, being non-luminous and composed of weakly interacting (or non-interacting) particles, lacks thermal emissions and does not directly contribute to the observed profile. Furthermore, the determination of NM surface properties involves relativistic ray-tracing, which maps the thermal emission to observed pulse profiles while accounting for general relativistic effects such as light bending and Doppler shifts, making the NM surface the observable boundary for radius constraints.

\begin{figure}[tbp]
    \centering
    \includegraphics[width=\columnwidth]{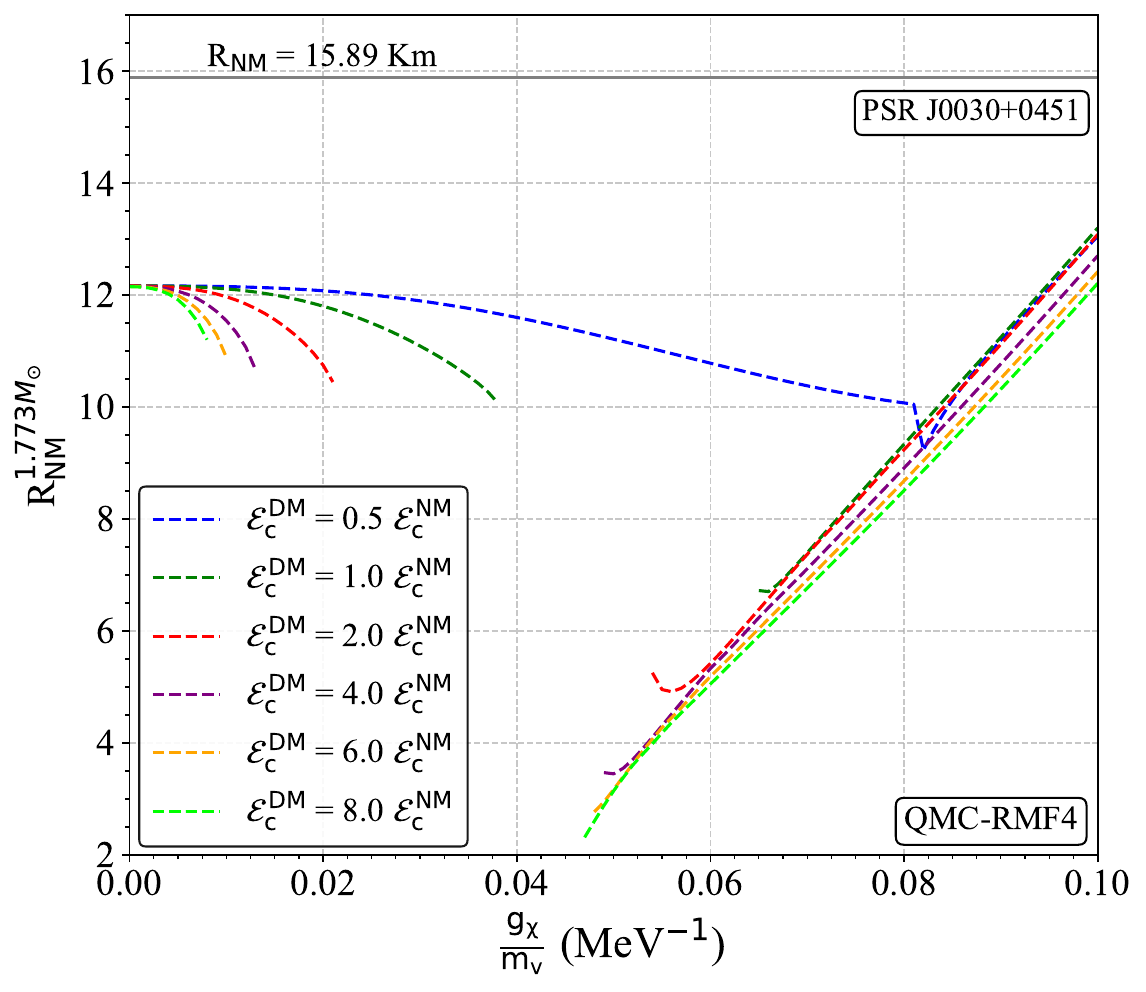} 
    \caption{NM radius $R_{\rm NM}$ for a DM admixed NS with total mass $M = 1.773 M_{\odot}$ is shown as a function of the DM self-coupling parameter $g_{\chi}/m_{\rm v}$ for various central energy density ratios $\mathcal{E}_{\rm c}^{\rm DM}/\mathcal{E}_{\rm c}^{\rm NM}$, constructed using the QMC-RMF4 EOS for NM. The horizontal line at $R_{\rm NM} = 15.89$ km marks the NICER constraint on the NM radius for a $1.773\, M_{\odot}$ star, derived from the analysis of PSR J0030+0451 observational data. Gaps or discontinuities in some curves for specific values of $g_{\chi}/m_{\rm v}$ indicate configurations where the maximum mass of the DM admixed NS is less than $1.773\, M_{\odot}$.}
    \label{fig:figure9}
\end{figure}

The lower right panel maps the allowed parameter space in $g_\chi/m_{\rm v}$ and ${\cal E}_{\rm c}^{\rm DM}/{\cal E}_{\rm c}^{\rm NM}$ based on the NICER radius constraint of $R_{\rm NM} \geq 10.15$ km for $1.107 M_\odot$ stars. The green shaded region represents viable parameter combinations that satisfy this radius constraint, while configurations falling below this threshold are excluded. Interestingly, gaps or discontinuities in the $R_{\rm NM}$ curves for ${\cal E}_{\rm c}^{\rm DM}/{\cal E}_{\rm c}^{\rm NM} \geq 4$ (upper right panel) correspond to configurations where the maximum mass is less than $1.107 M_\odot$. In these cases, the NM radii in the vicinity of the corresponding $g_{\chi}/m_{\rm v}$ region fall below the NICER constraint for $1.107 M_\odot$, represented by the horizontal line at $R_{\rm NM} = 10.15$ km. As a result, such configurations are excluded, ensuring that only physically viable parameter combinations remain within the allowed region.

Figure \ref{fig:figure9} examines the behavior of NM radius for DM admixed NSs with a total mass of $1.773 M_\odot$ ($R_{\rm{NM}}^{1.773 M_\odot}$), modeled using the QMC-RMF4 EOS. The horizontal line at $R_{\rm NM} = 15.89$ km represents the NICER-derived upper limit for this mass, corresponding to the rightmost point of the elliptical contours for PSR J0030+0451 (see Fig.~\ref{fig:figure3}). Consequently, stellar models with $R_{\rm{NM}}^{1.773 M_\odot}$ exceeding 15.89 km are inconsistent with this constraint and should be excluded. Across the range of DM self-coupling parameters ($g_\chi/m_{\rm v}$) and central energy density ratios (${ \mathcal{E}}_{\rm c}^{\rm DM}/{ \mathcal{E}}_{\rm c}^{\rm NM}$), all configurations satisfy this constraint, with $R_{\rm NM}^{1.773 M_\odot}$ remaining below the observational threshold.

Discontinuities in the $R_{\rm NM}^{1.773 M_\odot}$ curves, for ${\cal E}_{\rm c}^{\rm DM}/{\cal E}_{\rm c}^{\rm NM} \geq 0.5$, correspond to ${\cal E}_{\rm c}^{\rm DM}/{\cal E}_{\rm c}^{\rm NM}$ and $g_{\chi}/m_{\rm v}$ combinations where the maximum mass falls below $1.773 M_\odot$. Since all configurations satisfy the NICER constraint, this analysis does not impose stringent limitations on DM properties. However, it reaffirms the compatibility of DM admixed NS models with observational data and highlights the NM surface as the key observational interface for radius measurements.
\begin{figure}[tbp]
    \centering
    \includegraphics[width=\columnwidth]{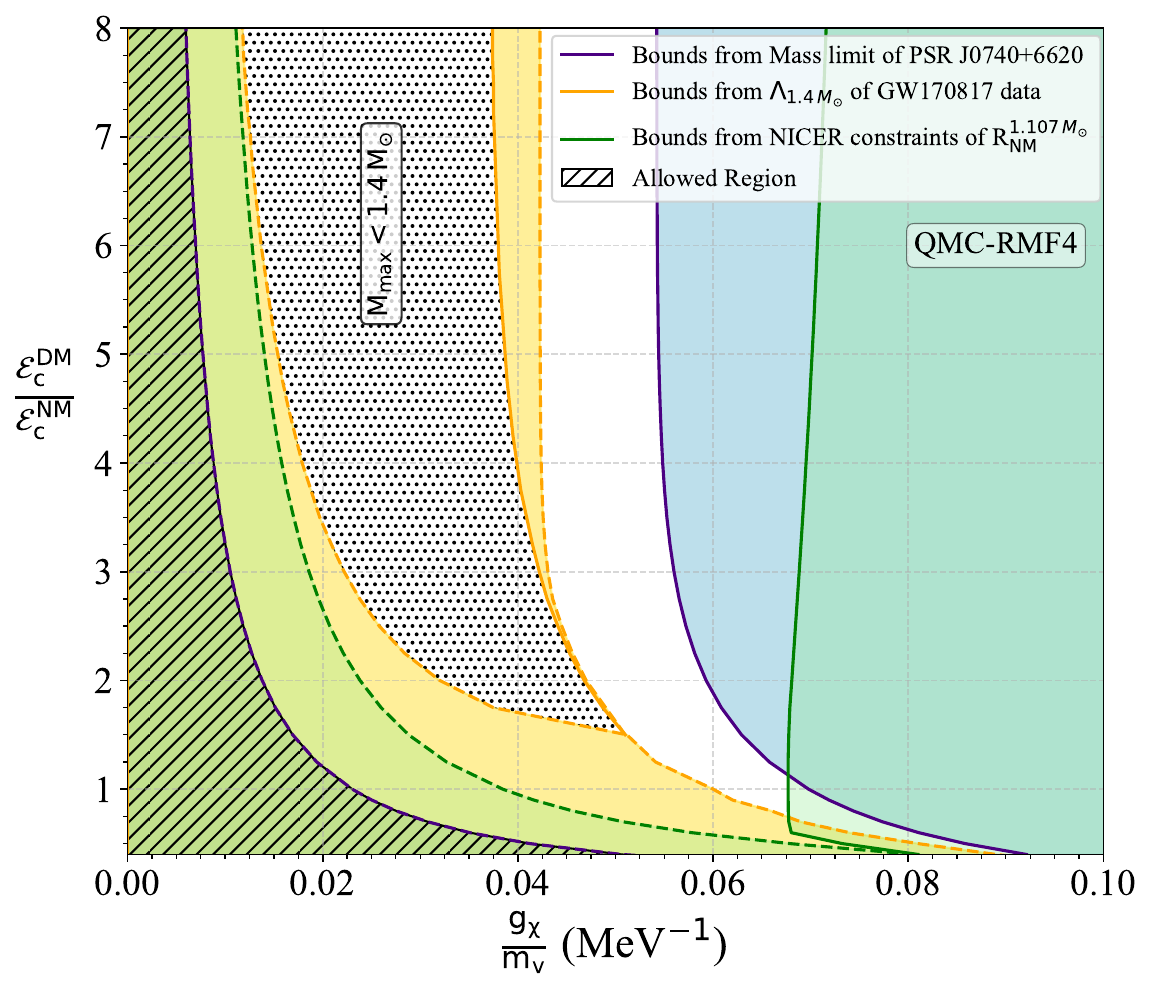}
    \caption{Allowed parameter space for DM admixed NS configurations constructed using the QMC-RMF4 EOS for NM, showing the permissible values of $g_{\chi}/m_{\rm{v}}$ as a function of $\mathcal{E}_{\rm{c}}^{\rm{DM}}/\mathcal{E}_{\rm{c}}^{\rm{NM}}$, constrained by multiple observational limits. The blue-shaded region satisfies the maximum mass limit from PSR J0740+6620, requiring $M_{\rm max} \geq 2.05 M_{\odot}$.  The yellow region satisfies the canonical tidal deformability constraint $\Lambda_{1.4M_{\odot}} \leq 580$ from GW170817 data. The green region corresponds to NICER’s radius constraint for a $1.107 M_{\odot}$ star (leftmost point in the contour for PSR J0030+0451), ensuring $R_{\rm NM}^{1.107M_{\odot}} \geq 10.15$ km. The final allowed parameter space, consistent with all observational constraints, is highlighted by the hatched region with slanted lines.}
    \label{fig:figure10}
\end{figure}

Figure \ref{fig:figure10} presents a synthesized view of the parameter space allowed for DM admixed NS configurations, constrained by three major astrophysical observations: the maximum mass from PSR J0740+6620, the canonical tidal deformability from GW170817, and NM radius constraints from NICER for PSR J0030+0451, already shown in Fig.~\ref{fig:figure7} and the bottom panels in Fig.~\ref{fig:figure8}. The parameter space is delineated in terms of the DM self-coupling parameter $g_\chi/m_{\rm v}$ and the central energy density ratio ${\cal E}_{\rm c}^{\rm DM}/{\cal E}_{\rm c}^{\rm NM}$, two critical parameters governing the interplay between NM and DM in determining NS structure and stability.

The blue-shaded region represents configurations satisfying the maximum mass limit $M_{\rm max} \geq 2.05 M_\odot$ derived from PSR J0740+6620 data. This ensures that the combined NM and DM EOS is capable of supporting the most massive NS observed, effectively placing an upper limit on DM contributions. The yellow-shaded region corresponds to the canonical tidal deformability constraint $\Lambda_{1.4M_\odot} \leq 580$ from GW170817 event. This limit imposes restrictions on the overall stiffness of the EOS, particularly penalizing configurations with extended radii that lead to excessive tidal deformabilities. The green shaded region satisfies NICER’s NM radius constraint of $R_{\rm NM}^{1.107M_\odot} \geq 10.15$ km for PSR J0030+0451, ensuring the NM surface radius aligns with NICER’s pulse-profile modeling and relativistic ray-tracing analysis. The intersection of these constraints, represented by the hatched region with slanted lines (////), highlights the parameter space consistent with all three observations, identifying physically viable configurations of DM admixed NSs based on current astrophysical data.

Notably, the blue and green regions demonstrate that halo configurations (right side of Fig.~\ref{fig:figure10}) can satisfy the maximum mass and NM radius constraints when the DM EOS achieves sufficient stiffness to maintain gravitational equilibrium. In these configurations, the NM surface radius meets the NICER threshold for $R_{\rm{NM}}^{1.107 M_\odot}$, while the total mass remains consistent with $M_{\rm max} \geq 2.05 M_\odot$. In contrast, the tidal deformability constraint excludes this halo-dominated regime, as evidenced by the absence of a yellow region in this part of the parameter space. This indicates that configurations with higher $g_\chi/m_{\rm v}$, associated with overly stiff DM EOSs, fail to satisfy $\Lambda_{1.4M_\odot} \leq 580$ due to their extended radii. 
\begin{figure}[tbp]
   \centering
\includegraphics[width=\columnwidth]{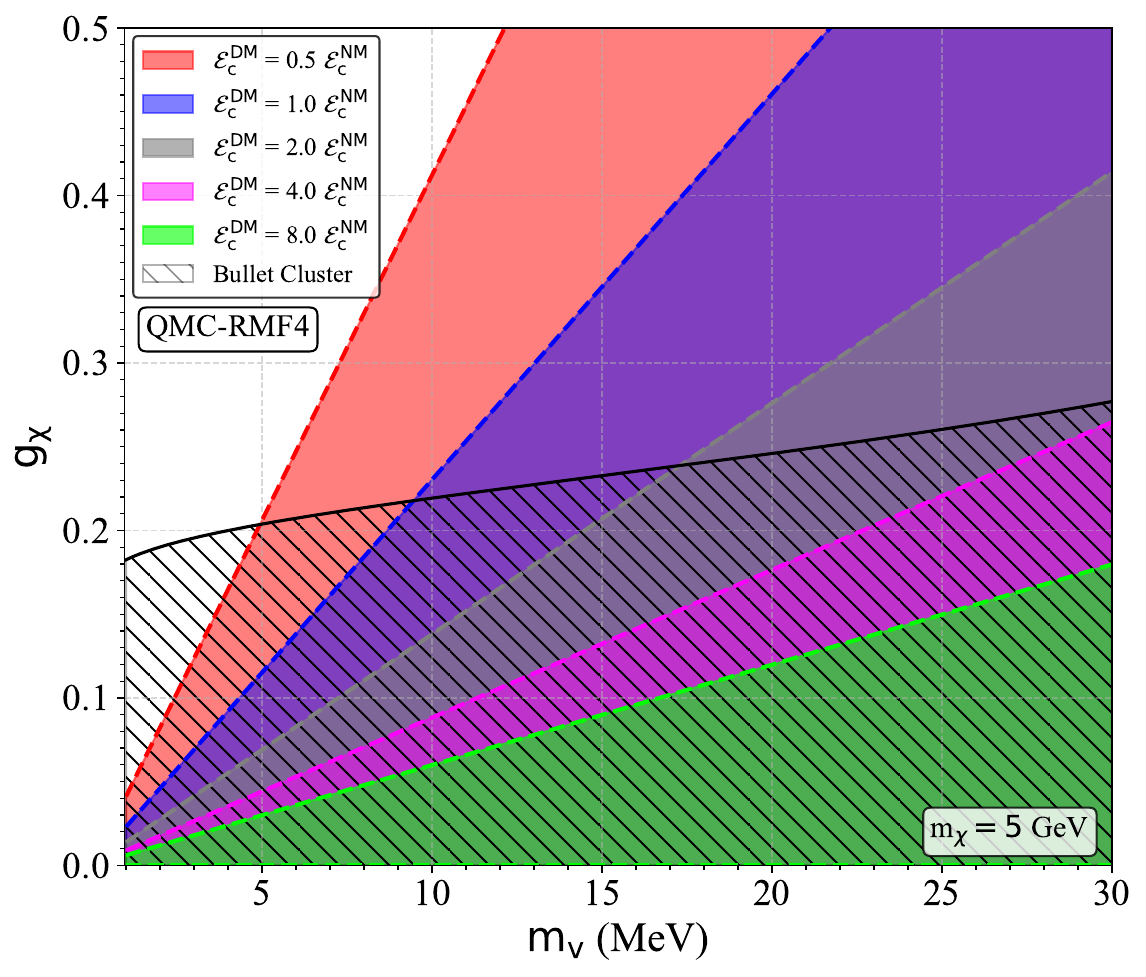} 
   \caption{This plot illustrates the allowed parameter space for the DM self-coupling constant ($g_{\chi}$) as a function of the mediator mass ($m_{\rm v}$) across different central energy density ratios, $\mathcal{E}_{\rm c}^{\rm DM}/\mathcal{E}_{\rm c}^{\rm NM}$. The shaded regions represent the permissible values of $g_{\chi}$ for a given $m_{\rm v}$, translated from the final allowed parameter space identified in Fig. \ref{fig:figure10}, which satisfies all observational constraints, including PSR J0740+6620, GW170817, and NICER analysis for PSR J0030+0451. We note that there is no lower boundary for the allowed region with a fixed value of $\mathcal{E}_{\rm c}^{\rm DM}/\mathcal{E}_{\rm c}^{\rm NM}$.
   The hatched region with slanted lines in this figure represents the allowed region imposed by the Bullet Cluster, constraining the DM self-interaction cross-section. This analysis assumes a DM particle mass of $m_{\chi} = 5$ GeV and is based on the QMC-RMF4 NM EOS.}
    \label{fig:figure11}
\end{figure}
Consequently, the tidal deformability constraint primarily allows core configurations with lower $g_\chi/m_{\rm v}$, where NM plays a dominant role in maintaining the star's gravitational equilibrium. Conversely, the overlap of blue and green regions demonstrate that halo configurations, while not constrained by tidal deformability, can still remain consistent with the maximum mass and NM radius criteria when DM stiffness is appropriately balanced.
\begin{figure*}[tbp]
    \centering
    \includegraphics[width=\textwidth]{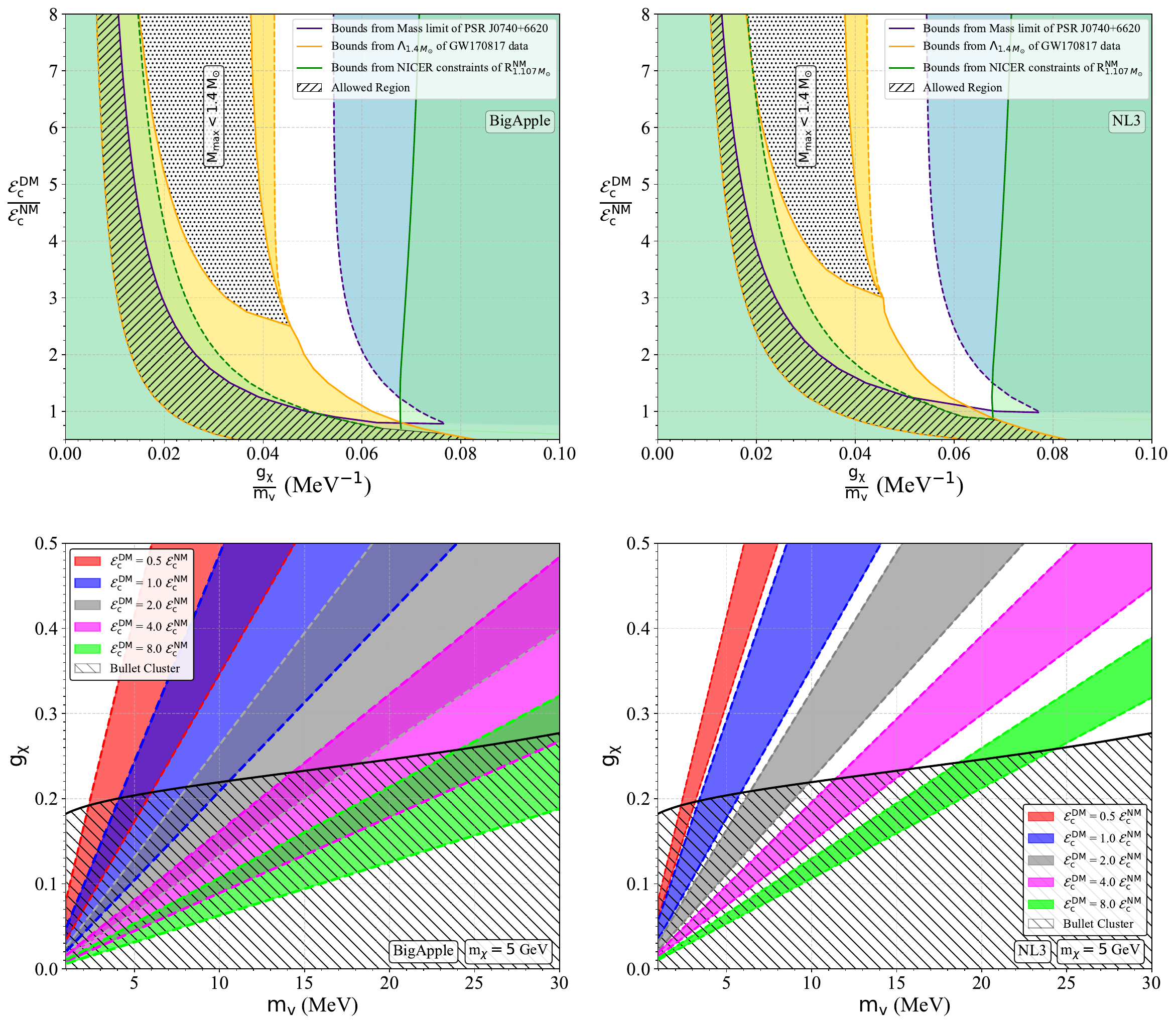} 
    \caption{Top panels: Allowed regions in the parameter space of $g_{\chi}/m_{\rm v}$ and $\mathcal{E}_{\rm c}^{\rm DM}/\mathcal{E}_{\rm c}^{\rm NM}$, derived using the same observational constraints as for QMC-RMF4 (Fig.~\ref{fig:figure10}), but here for BigApple and NL3 RMF parameter sets. Bottom panels: Same as Fig.~\ref{fig:figure11}, but the allowed regions are derived using BigApple and NL3 as NM EOS. The shaded regions represent configurations consistent with observational constraints, while the hatch-filled region shows the Bullet Cluster constraint on $g_{\chi}$ for each value of $m_{\rm v}$}.
    \label{fig:figure12}
\end{figure*}

The hatched region with slanted lines thus represents the intersection of all three constraints and marks the currently allowed parameter space for DM admixed NSs constructed using the QMC-RMF4 EOS for NM. This region serves as a benchmark for theoretical and observational consistency, providing a valuable framework for studying DM in compact stars. However, future observations—such as improved tidal deformability constraints from next-generation gravitational wave detections or refined radius measurements from X-ray telescopes—could further narrow or refine this parameter space.

Figure~\ref{fig:figure11} illustrates the allowed parameter space for the DM self-coupling constant $g_\chi$ as a function of the mediator mass $m_{\rm v}$, translated from the final allowed parameter space of $g_\chi/m_{\rm v}$ satisfying all NS observational constraints (shown as the hatched region with slanted lines //// in Fig.~\ref{fig:figure10}). This plot provides a broader perspective by incorporating constraints from NS observations, while also factoring in cosmological constraint on DM self-interactions from the Bullet Cluster. The analysis assumes a DM particle mass of $m_\chi = 5$ GeV and is based on the QMC-RMF4 NM EOS. 

The color-shaded regions in the plot represent the permissible values of $g_\chi$ for a given $m_{\rm v}$ across various central energy density ratios, ${\cal E}_{\rm c}^{\rm DM}/{\cal E}_{\rm c}^{\rm NM}$, ranging from 0.5 (pink) to 8.0 (green). These regions are derived from the NS observational constraints and reflect the combinations of $g_\chi$ and $m_{\rm v}$ that produce physically viable DM-admixed NS configurations. The hatched space with slanted lines $\backslash\backslash\backslash\backslash$ in Fig.~\ref{fig:figure11} represents the allowed region imposed by the Bullet Cluster, which constrains the DM self-interaction cross-section to $\sigma/m_\chi \lesssim {\cal O} (0.1)$ cm$^{2}$/g. The boundary of this region delineates the maximum $g_\chi$ for a given $m_{\rm v}$, ensuring consistency with the observed dynamics of galaxy clusters while preventing overly strong DM self-interactions on cosmological scales. 

At lower mediator masses, constraints from NS observations, driven by gravitational equilibrium requirements in compact stars, impose tighter restrictions on $g_\chi$. This reflects the impact of gravitational coupling in NSs, where smaller $m_{\rm v}$ enhances the effective DM self-coupling $(g_{\chi}/m_{\rm v})^{2}$, requiring stricter limits to ensure stability and consistency with observations. In contrast, at higher $m_{\rm v}$, cosmological constraints from the Bullet Cluster dominate. The Bullet Cluster imposes an upper bound on the DM self-interaction cross-section, ensuring that DM interactions remain weak enough to preserve large-scale structure. These constraints limit $g_\chi$ at higher $m_{\rm v}$, complementing the restrictions from NS observations at lower $m_{\rm v}$.

The overlap between the shaded regions and the Bullet Cluster boundary delineates the final allowed parameter space, ensuring consistency with both NS and cosmological observations. This interplay demonstrates how constraints from small-scale astrophysical systems like compact stars and large-scale cosmological systems like galaxy clusters work together to define the viable parameter space for DM self-coupling and mediator mass.

Future observational advancements could further refine this parameter space. For instance, improved gravitational wave detections or refined measurements of DM self-interaction cross-sections at the galaxy cluster scale could impose additional restrictions on $g_\chi$ and $m_{\rm v}$. Similarly, direct detection experiments probing the light mediator sector could provide complementary constraints. This combined analysis underscores how the microphysics of DM, as probed by NSs, complements large-scale cosmological observations, offering a comprehensive framework for understanding DM interactions across vastly different physical scales.

Figure~\ref{fig:figure12} extends the analysis by incorporating two additional NM EOSs—BigApple and NL3, providing insights into the sensitivity of DM-admixed NS properties to the underlying nuclear matter interactions. While QMC-RMF4 represents a relatively soft NM EOS, BigApple represents a moderately stiff EOS and NL3 is among the stiffest RMF parameter sets, making it an ideal choice for studying the extremes of NM stiffness. This comparison highlights how uncertainties in NM EOS stiffness influence the allowed parameter space for DM properties and their interplay with NS observational constraints.

The top panels depict the allowed parameter space in terms of $g_\chi/m_{\rm v}$ and ${\cal E}_{\rm c}^{\rm DM}/{\cal E}_{\rm c}^{\rm NM}$, while the bottom panels translate these constraints into the ($g_{\chi}$, $m_{\rm v}$) parameter space for BigApple (left) and NL3 (right). These analyses employ the same observational constraints used for QMC-RMF4 (Figs.~\ref{fig:figure10} and~\ref{fig:figure11}), including the maximum mass from PSR J0740+6620, canonical tidal deformability from GW170817, and NM radius constraints from NICER for PSR J0030+0451. The Bullet Cluster constraint is also incorporated in the bottom panels.

For both BigApple and NL3, the lower boundary of the allowed region arises from the GW170817 constraint for canonical tidal deformability. The upper boundary is dictated by the maximum mass constraint for central density ratios ${\cal E}_{\rm c}^{\rm DM}/{\cal E}_{\rm c}^{\rm NM} \gtrsim 1$, while for ${\cal E}_{\rm c}^{\rm DM}/{\cal E}_{\rm c}^{\rm NM} \lesssim 1$ the upper boundary is initially governed by the NICER radius constraint, but then it shifts to the canonical tidal deformability constraint. Beyond this transition, both the lower and upper boundaries of the allowed region are dictated by the tidal deformability constraint (GW170817) for lower central density ratios, emphasizing its critical role in limiting such configurations. 

A key difference between QMC-RMF4 and the stiffer EOSs (BigApple and NL3) is the presence of a lower boundary for $g_{\chi}/m_{\rm v}$ in the allowed region for the latter. This feature is absent in QMC-RMF4, which satisfies all observational constraints even without DM contributions. For BigApple and NL3, however, this lower boundary emerges because pure NM configurations for these stiffer EOSs yield canonical tidal deformabilities ($\Lambda_{1.4M_{\odot}}$) that exceed observational limits. The inclusion of DM mitigates this effect by softening the EOS, reducing $\Lambda_{1.4M_{\odot}}$, and necessitating a non-zero $g_\chi/m_{\rm v}$ to satisfy the tidal deformability constraint. Additionally, the appearance of a lower boundary in the ($g_{\chi}, m_{\rm v}$) parameter space for each value of ${\cal E}_{\rm c}^{\rm DM}/{\cal E}_{\rm c}^{\rm NM}$ reflects the same trend, arising from the need to constrain the impact of stiffer NM EOSs through DM contributions.

The effect of NM stiffness is also reflected in the allowed ($g_{\chi}, m_{\rm v}$) parameter space, shown as the color-shaded regions for different ${\cal E}_{\rm c}^{\rm DM}/{\cal E}_{\rm c}^{\rm NM}$ in the bottom panels. For both BigApple and NL3, the allowed regions extend to higher $g_\chi$ values compared to QMC-RMF4 due to the increased stiffness of these EOSs. The magnitude of this extension depends on the stiffness of the NM EOS, with NL3 exhibiting narrower permissible regions compared to BigApple. This arises from NL3's inherently larger masses and tidal deformabilities for pure NM configurations, requiring stronger DM self-repulsion to offset the excessive stiffness of the NM EOS and maintain consistency with observational limits. Thus, the comparison demonstrates how stiffer NM EOSs shift and reshape the allowed DM parameter space, emphasizing the interplay between DM and NM contributions in determining the structure of NSs.

The findings in Fig.~\ref{fig:figure12} underscore the profound influence of NM EOS stiffness on the inferred DM parameter space, highlighting the need to account for uncertainties in nuclear interactions at high densities. While QMC-RMF4 provides the most relaxed constraints due to its soft nature, BigApple and NL3 introduce additional restrictions arising from their inherent stiffness. This analysis demonstrates that the choice of NM EOS significantly impacts both the boundaries and extent of the allowed parameter space for DM self-coupling and mediator mass.

\section{Conclusion}
\label{sec:4}
In this study, we have investigated the structural and observational properties of DM admixed NSs within the two-fluid formalism, employing NM and DM EOSs coupled exclusively through gravity. This framework enables a nuanced exploration of how DM influences NS structure while leveraging constraints from gravitational waves, X-ray observations, and observational data from the dynamics of galaxy cluster, such as the Bullet Cluster, to refine the DM parameter space. 

Our results underscore the significant influence of the DM self-coupling parameter ($g_\chi/m_{\rm v}$) and the central energy density ratio (${\cal E}_{\rm c}^{\rm DM}/{\cal E}_{\rm c}^{\rm NM}$) in shaping the structural and compositional properties of DM admixed NSs. The inclusion of DM introduces diverse configurations characterized by core- or halo-dominated structures, governed by the interplay of DM self-repulsion, NM stiffness, and gravitational coupling. This structural diversity manifests in the distinct mass-radius profiles, compositional transitions, and tidal deformability trends.

The comparative analysis of NM EOSs (QMC-RMF4, BigApple, and NL3) reveals the sensitivity of DM parameter constraints to nuclear matter stiffness. Softer EOSs, such as QMC-RMF4, allow for broader DM parameter space, accommodating configurations consistent with all observational constraints without requiring significant DM contributions. In contrast, stiffer EOSs, like NL3, necessitate non-zero $g_\chi/m_{\rm v}$ to mitigate excessive NM stiffness, introducing additional lower boundaries in the allowed DM parameter space. This sensitivity highlights the critical role of uncertainties in high-density nuclear interactions in interpreting DM properties.

Observational constraints play a central role in delineating the DM parameter space. The maximum mass limit from PSR J0740+6620 restricts excessive DM contributions to ensure gravitational stability. Tidal deformability constraints from GW170817 limit the stiffness of both the NM and DM EOSs, preventing configurations with overly extended radii. NICER radius measurements for PSR J0030+0451 provide key benchmarks for NM surface properties in low-mass NSs. Together, these multi-messenger observations define a narrow and observationally consistent DM parameter space, underscoring the synergy between astrophysical and cosmological probes.

A notable finding is the absence of an allowed region from GW170817 tidal deformability data in the halo-dominated regime, suggesting that at least the merging compact stars in this event likely favored core configurations. This result highlights the potential of gravitational wave observations to distinguish between core- and halo-dominated structures. Future detections, particularly those probing the internal structure of merging NSs, could offer critical insights into the relative dominance of core or halo configurations, providing a unique perspective on DM's role in NS structure.

Cosmological constraints from the Bullet Cluster further refine the DM parameter space by limiting the self-interaction cross-section to $\sigma/m_\chi \lesssim \mathcal{O}(0.1)$ cm$^2$/g at relative velocities relevant for the cluster scale. The overlap of allowed regions from NS observations and the Bullet Cluster constraint underscores the importance of reconciling small-scale astrophysical data with large-scale dynamics. While NS observations impose tighter restrictions at lower mediator masses ($m_{\rm v}$) due to gravitational coupling, the Bullet Cluster constraint dominates at higher $m_{\rm v}$, ensuring consistency with galaxy cluster behavior. These findings bridge astrophysical observations and particle physics, indirectly probing the microphysics of DM self-interactions and linking small-scale phenomena within NSs to large-scale cosmological dynamics observed in galaxy clusters and beyond.

In conclusion, this study establishes a robust framework for investigating DM in compact stars, demonstrating the utility of multi-messenger astrophysics and cosmology in probing the microphysics of DM. By systematically linking astrophysical observations to particle physics and cosmology, our results pave the way for deeper insights into DM interactions and their implications for the structure and evolution of NSs. Additionally, direct detection experiments and cosmological studies targeting DM self-interactions and mediator properties could provide valuable complementary insights.

\acknowledgments
This work is supported in part by Japan Society for the Promotion of Science (JSPS) KAKENHI Grant Numbers 
JP23K20848  
and JP24KF0090. \\ 

\appendix
\section{Two fluid Structures of NSs}
\label{sec:appendx1}
\begin{figure*}[tbp]
    \centering
    \includegraphics[width=\textwidth]{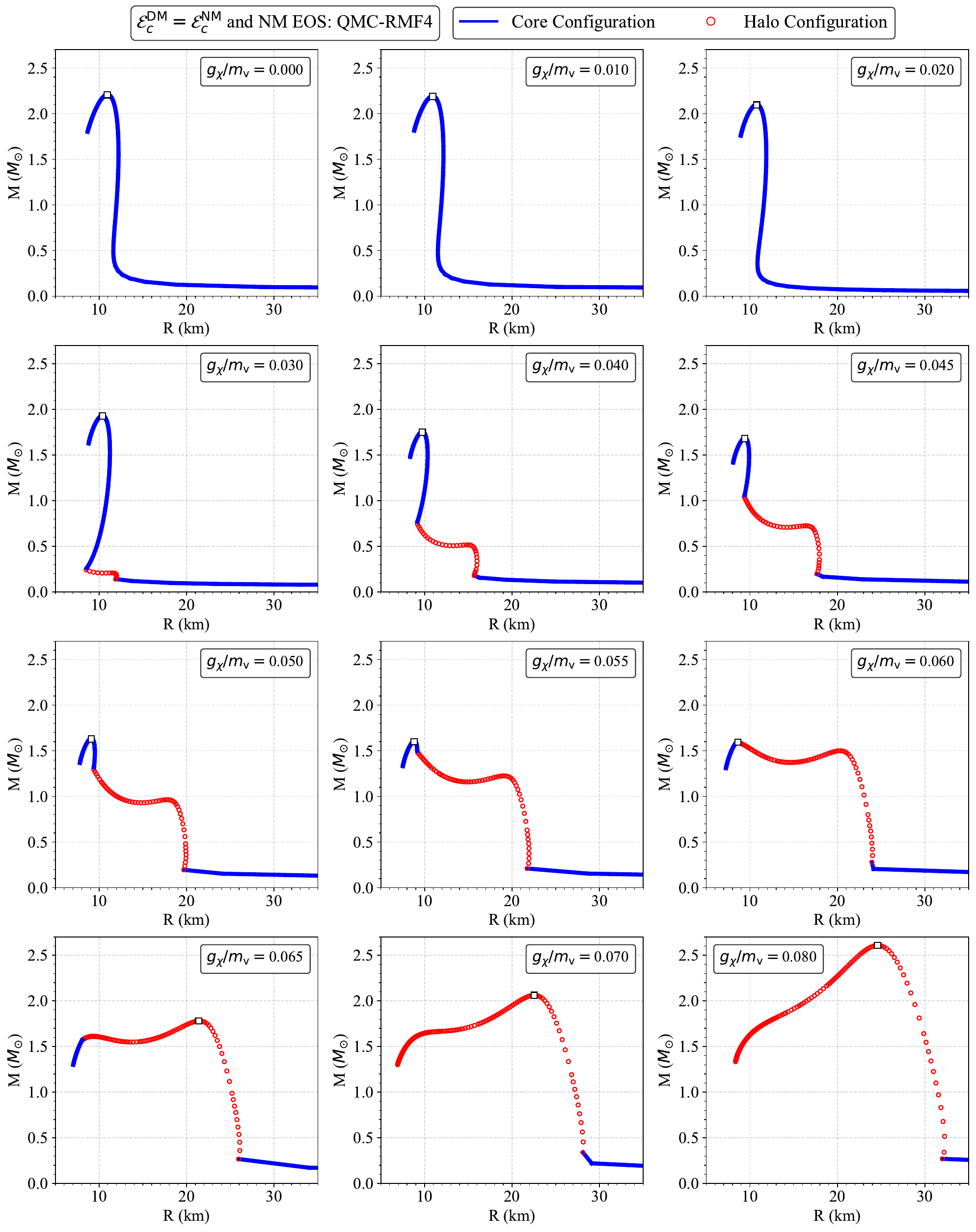} 
    \caption{Mass-radius relations for DM-admixed NSs constructed using the QMC-RMF4 NM EOS, shown for various values of the DM self-coupling parameter $g_{\chi}/m_{\rm v}$, with ${\cal E}^{\rm{DM}}_{\rm c} = {\cal E}^{\rm{NM}}_{\rm c}$. Core configurations ($R_{\rm{NM}} > R_{\rm{DM}}$) are represented by solid blue lines, while halo configurations ($R_{\rm{NM}} < R_{\rm{DM}}$) are denoted by red circles. Maximum mass on each curve is represented by the square marker. This grid highlights the transition between core- and halo-dominated regimes and demonstrates how increasing $g_{\chi}/m_{\rm v}$  affects the structural properties of DM-admixed NSs, including the onset of extended DM halos.}
    \label{fig:figure13}
\end{figure*}

The panels in Fig. \ref{fig:figure13} illustrate the evolution of structural properties in DM-admixed NSs as a function of the DM self-coupling parameter $g_{\chi}/m_{\rm v}$, constructed using the QMC-RMF4 NM EOS under the assumption that ${\cal E}^{\rm{DM}}_{\rm c} = {\cal E}^{\rm{NM}}_{\rm c}$. Core-dominated configurations ($R_{\rm NM} > R_{\rm DM}$) are marked by solid blue lines, while halo-dominated configurations ($R_{\rm NM} < R_{\rm DM}$) are indicated by red circles. The square markers identify the maximum mass for each curve, demonstrating the evolution of maximum mass and radius as the DM self-coupling parameter increases.

For smaller values of $g_{\chi}/m_{\rm v}$, DM contributions to the overall structure are negligible, as evidenced by the absence of red markers. The blue curves, representing core configurations dominate the $M-R$ profiles, indicating that NM primarily dictates the star's structure. In this regime, DM is effectively ``invisible," yet its presence subtly reduces the maximum mass of the star compared to pure NM scenarios. This reduction arises from the softening effect of DM on the overall EOS, even though the DM component does not play a significant structural role at low self-coupling strengths.

As $g_{\chi}/m_{\rm v}$ increases, DM begins to play a more significant role, particularly in lower-mass stars, leading to the emergence of halo-dominated configurations, marked by red circles. This transition reflects the increasing stiffness of the DM EOS due to self-repulsion, which facilitates the formation of extended DM halos. The transition from core- to halo-dominated configurations occurs first in low-mass stars, highlighting the interplay between DM self-repulsion and gravitational coupling.

For larger $g_{\chi}/m_{\rm v}$ values, DM halos become increasingly prominent and dominant across the entire $M-R$ profile, significantly altering their radii. In these cases, the DM component provides sufficient stiffness to support gravitational equilibrium, resulting in an overall increase in the maximum mass of the star compared to pure NM configurations. This is evidenced by the square markers, which denote the maximum mass configurations and demonstrate that at high $g_{\chi}/m_{\rm v}$, DM significantly contributes to the gravitational structure, stabilizing even high-mass stars.

This progression highlights how increasing DM self-coupling parameter transitions NSs from purely NM-dominated structures to configurations where DM dominates, especially in shaping the radii and stabilizing the maximum mass of stars.
\\

\section{Mathematical expressions for tidal response}
\label{sec:appendx2}
To compute $k_{2}$ for the two-fluid system, the following differential equation must be solved with TOV equations as the background, ensuring consistency between the mass, radius, and internal structure of the star:
\begin{widetext}
\begin{align}
    r\frac{dy}{dr} + y(r)^{2} + y(r)F(r) + r^{2} Q(r) = 0
\end{align}
where the functions $F(r)$ and $Q(r)$ are defined as:
\begin{align}
    F(r) = \frac{r-4\pi r^{3}\left[\left({\cal E}_{\rm{NM}} + {\cal E}_{\rm{DM}}\right)-\left(P_{\rm{NM}} + P_{\rm{DM}}\right)\right]}{r-2m},
\end{align}
\begin{align}
    Q(r) = \frac{4\pi r\left[5\left({\cal E}_{\rm{NM}} + {\cal E}_{\rm{DM}}\right) + 9 \left(P_{\rm{NM}} + P_{\rm{DM}}\right) + \frac{{\cal E}_{\rm{NM}}+P_{\rm{NM}}}{\partial P_{\rm{NM}}/{\partial {\cal E}_{\rm{NM}}}} + \frac{{\cal E}_{\rm{DM}}+P_{\rm{DM}}}{\partial P_{\rm{DM}}/{\partial {\cal E}_{\rm{DM}}}} -\frac{6}{4\pi r^{2}}\right]}{r-2m} \nonumber \\
    - 4\left[\frac{m + 4\pi r^{3} \left(P_{\rm{NM}}+P_{\rm{DM}}\right)}{r^{2}\left(1-2m/r\right)}\right]^{2},
\end{align}
where $y(r)$ is a dimensionless quantity related to the metric perturbation function 
$H(r)$, which corresponds to the radial component of the even-parity, quadrupolar metric perturbation in the Regge-Wheeler gauge. Here, $y_{R} = y(R)$ is obtained by solving the above equation from the center ($r=0$) to the star's surface ($r=R$). The total radius $R$ is defined as $R = \rm{max}(R_{\rm{NM}}, R_{\rm{DM}})$, ensuring that the larger of the two radii is used to represent the outer boundary of the star. The Love number $k_{2}$ is then computed using:
\begin{eqnarray}
    k_{2} &=& \frac{8}{5}\left(\frac{M}{R}\right)^{5} \left(1-\frac{2M}{R}\right)^{2} 
    \left[2+\frac{2M}{R}\left(y_{R}-1\right)-y_{R}\right] \times \left\{\frac{2M}{R}\left(6-3y_{R}+\frac{3M}{R}\left(5y_{R}-8\right)\right) \right. \nonumber \\
    && \quad + \left. 4\left(\frac{M}{R}\right)^{3}\left[13-11y_{R}+\frac{M}{R}\left(3y_{R}-2\right)+2\left(\frac{M}{R}\right)^{2}\left(1+y_{R}\right)\right]+3\left(1-\frac{2M}{R}\right)^{2} \right. \nonumber \\
    && \quad \left. \times \left[2-y_{R} + \frac{2M}{R}\left(y_{R}-1\right)\right] \log\left(1-\frac{2M}{R}\right) \right\}^{-1}.
\end{eqnarray}
\end{widetext}
The total tidal deformability $\Lambda$ is computed using the total gravitational mass ($M = M_{\rm{NM}} + M_{\rm{DM}}$) and the total radius of the star ($R = \rm{max}(R_{\rm{NM}}, R_{\rm{DM}})$). This approach ensures that the contributions of both components are consistently incorporated into the calculation of the tidal response.
\bibliographystyle{apsrev4-2}
\bibliography{main.bib}

\begin{thebibliography}{75}%
\makeatletter
\providecommand \@ifxundefined [1]{%
 \@ifx{#1\undefined}
}%
\providecommand \@ifnum [1]{%
 \ifnum #1\expandafter \@firstoftwo
 \else \expandafter \@secondoftwo
 \fi
}%
\providecommand \@ifx [1]{%
 \ifx #1\expandafter \@firstoftwo
 \else \expandafter \@secondoftwo
 \fi
}%
\providecommand \natexlab [1]{#1}%
\providecommand \enquote  [1]{``#1''}%
\providecommand \bibnamefont  [1]{#1}%
\providecommand \bibfnamefont [1]{#1}%
\providecommand \citenamefont [1]{#1}%
\providecommand \href@noop [0]{\@secondoftwo}%
\providecommand \href [0]{\begingroup \@sanitize@url \@href}%
\providecommand \@href[1]{\@@startlink{#1}\@@href}%
\providecommand \@@href[1]{\endgroup#1\@@endlink}%
\providecommand \@sanitize@url [0]{\catcode `\\12\catcode `\$12\catcode `\&12\catcode `\#12\catcode `\^12\catcode `\_12\catcode `\%12\relax}%
\providecommand \@@startlink[1]{}%
\providecommand \@@endlink[0]{}%
\providecommand \url  [0]{\begingroup\@sanitize@url \@url }%
\providecommand \@url [1]{\endgroup\@href {#1}{\urlprefix }}%
\providecommand \urlprefix  [0]{URL }%
\providecommand \Eprint [0]{\href }%
\providecommand \doibase [0]{https://doi.org/}%
\providecommand \selectlanguage [0]{\@gobble}%
\providecommand \bibinfo  [0]{\@secondoftwo}%
\providecommand \bibfield  [0]{\@secondoftwo}%
\providecommand \translation [1]{[#1]}%
\providecommand \BibitemOpen [0]{}%
\providecommand \bibitemStop [0]{}%
\providecommand \bibitemNoStop [0]{.\EOS\space}%
\providecommand \EOS [0]{\spacefactor3000\relax}%
\providecommand \BibitemShut  [1]{\csname bibitem#1\endcsname}%
\let\auto@bib@innerbib\@empty
\bibitem [{\citenamefont {{Planck Collaboration}}\ \emph {et~al.}(2020)\citenamefont {{Planck Collaboration}}, \citenamefont {{Aghanim}}, \citenamefont {{Akrami}}, \citenamefont {{Ashdown}}, \citenamefont {{Aumont}}, \citenamefont {{Baccigalupi}}, \citenamefont {{Ballardini}}, \citenamefont {{Banday}}, \citenamefont {{Barreiro}}, \citenamefont {{Bartolo}}, \citenamefont {{Basak}}, \citenamefont {{Battye}},\ and\ \citenamefont {{et. al.}}}]{2020AA...641A...6P}%
  \BibitemOpen
  \bibfield  {author} {\bibinfo {author} {\bibnamefont {{Planck Collaboration}}}, \bibinfo {author} {\bibfnamefont {N.}~\bibnamefont {{Aghanim}}}, \bibinfo {author} {\bibfnamefont {Y.}~\bibnamefont {{Akrami}}}, \bibinfo {author} {\bibfnamefont {M.}~\bibnamefont {{Ashdown}}}, \bibinfo {author} {\bibfnamefont {J.}~\bibnamefont {{Aumont}}}, \bibinfo {author} {\bibfnamefont {C.}~\bibnamefont {{Baccigalupi}}}, \bibinfo {author} {\bibfnamefont {M.}~\bibnamefont {{Ballardini}}}, \bibinfo {author} {\bibfnamefont {A.~J.}\ \bibnamefont {{Banday}}}, \bibinfo {author} {\bibfnamefont {R.~B.}\ \bibnamefont {{Barreiro}}}, \bibinfo {author} {\bibfnamefont {N.}~\bibnamefont {{Bartolo}}}, \bibinfo {author} {\bibfnamefont {S.}~\bibnamefont {{Basak}}}, \bibinfo {author} {\bibfnamefont {R.}~\bibnamefont {{Battye}}},\ and\ \bibinfo {author} {\bibnamefont {{et. al.}}},\ }\href {https://doi.org/10.1051/0004-6361/201833910} {\bibfield  {journal} {\bibinfo  {journal} {aap}\ }\textbf {\bibinfo {volume} {641}},\ \bibinfo {eid} {A6}
  (\bibinfo {year} {2020})},\ \Eprint {https://arxiv.org/abs/1807.06209} {arXiv:1807.06209 [astro-ph.CO]} \BibitemShut {NoStop}%
\bibitem [{\citenamefont {Group}\ \emph {et~al.}(2022)\citenamefont {Group}, \citenamefont {Workman}, \citenamefont {Burkert}, \citenamefont {Crede}, \citenamefont {Klempt}, \citenamefont {Thoma}, \citenamefont {Tiator}, \citenamefont {Agashe}, \citenamefont {Aielli}, \citenamefont {Allanach}, \citenamefont {Amsler}, \citenamefont {Antonelli}, \citenamefont {Aschenauer},\ and\ \citenamefont {et. al.}}]{10.1093/ptep/ptac097}%
  \BibitemOpen
  \bibfield  {author} {\bibinfo {author} {\bibfnamefont {P.~D.}\ \bibnamefont {Group}}, \bibinfo {author} {\bibfnamefont {R.~L.}\ \bibnamefont {Workman}}, \bibinfo {author} {\bibfnamefont {V.~D.}\ \bibnamefont {Burkert}}, \bibinfo {author} {\bibfnamefont {V.}~\bibnamefont {Crede}}, \bibinfo {author} {\bibfnamefont {E.}~\bibnamefont {Klempt}}, \bibinfo {author} {\bibfnamefont {U.}~\bibnamefont {Thoma}}, \bibinfo {author} {\bibfnamefont {L.}~\bibnamefont {Tiator}}, \bibinfo {author} {\bibfnamefont {K.}~\bibnamefont {Agashe}}, \bibinfo {author} {\bibfnamefont {G.}~\bibnamefont {Aielli}}, \bibinfo {author} {\bibfnamefont {B.~C.}\ \bibnamefont {Allanach}}, \bibinfo {author} {\bibfnamefont {C.}~\bibnamefont {Amsler}}, \bibinfo {author} {\bibfnamefont {M.}~\bibnamefont {Antonelli}}, \bibinfo {author} {\bibfnamefont {E.~C.}\ \bibnamefont {Aschenauer}},\ and\ \bibinfo {author} {\bibnamefont {et. al.}},\ }\href {https://doi.org/10.1093/ptep/ptac097} {\bibfield  {journal} {\bibinfo  {journal} {Progress of Theoretical
  and Experimental Physics}\ }\textbf {\bibinfo {volume} {2022}},\ \bibinfo {pages} {083C01} (\bibinfo {year} {2022})}\BibitemShut {NoStop}%
\bibitem [{\citenamefont {{Davis}}\ \emph {et~al.}(1985)\citenamefont {{Davis}}, \citenamefont {{Efstathiou}}, \citenamefont {{Frenk}},\ and\ \citenamefont {{White}}}]{1985ApJ...292..371D}%
  \BibitemOpen
  \bibfield  {author} {\bibinfo {author} {\bibfnamefont {M.}~\bibnamefont {{Davis}}}, \bibinfo {author} {\bibfnamefont {G.}~\bibnamefont {{Efstathiou}}}, \bibinfo {author} {\bibfnamefont {C.~S.}\ \bibnamefont {{Frenk}}},\ and\ \bibinfo {author} {\bibfnamefont {S.~D.~M.}\ \bibnamefont {{White}}},\ }\href {https://doi.org/10.1086/163168} {\bibfield  {journal} {\bibinfo  {journal} {\apj}\ }\textbf {\bibinfo {volume} {292}},\ \bibinfo {pages} {371} (\bibinfo {year} {1985})}\BibitemShut {NoStop}%
\bibitem [{\citenamefont {Spergel}\ \emph {et~al.}(2003)\citenamefont {Spergel}, \citenamefont {Verde}, \citenamefont {Peiris}, \citenamefont {Komatsu}, \citenamefont {Nolta}, \citenamefont {Bennett}, \citenamefont {Halpern}, \citenamefont {Hinshaw}, \citenamefont {Jarosik}, \citenamefont {Kogut}, \citenamefont {Limon}, \citenamefont {Meyer}, \citenamefont {Page}, \citenamefont {Tucker}, \citenamefont {Weiland}, \citenamefont {Wollack},\ and\ \citenamefont {Wright}}]{Spergel_2003}%
  \BibitemOpen
  \bibfield  {author} {\bibinfo {author} {\bibfnamefont {D.~N.}\ \bibnamefont {Spergel}}, \bibinfo {author} {\bibfnamefont {L.}~\bibnamefont {Verde}}, \bibinfo {author} {\bibfnamefont {H.~V.}\ \bibnamefont {Peiris}}, \bibinfo {author} {\bibfnamefont {E.}~\bibnamefont {Komatsu}}, \bibinfo {author} {\bibfnamefont {M.~R.}\ \bibnamefont {Nolta}}, \bibinfo {author} {\bibfnamefont {C.~L.}\ \bibnamefont {Bennett}}, \bibinfo {author} {\bibfnamefont {M.}~\bibnamefont {Halpern}}, \bibinfo {author} {\bibfnamefont {G.}~\bibnamefont {Hinshaw}}, \bibinfo {author} {\bibfnamefont {N.}~\bibnamefont {Jarosik}}, \bibinfo {author} {\bibfnamefont {A.}~\bibnamefont {Kogut}}, \bibinfo {author} {\bibfnamefont {M.}~\bibnamefont {Limon}}, \bibinfo {author} {\bibfnamefont {S.~S.}\ \bibnamefont {Meyer}}, \bibinfo {author} {\bibfnamefont {L.}~\bibnamefont {Page}}, \bibinfo {author} {\bibfnamefont {G.~S.}\ \bibnamefont {Tucker}}, \bibinfo {author} {\bibfnamefont {J.~L.}\ \bibnamefont {Weiland}}, \bibinfo {author} {\bibfnamefont
  {E.}~\bibnamefont {Wollack}},\ and\ \bibinfo {author} {\bibfnamefont {E.~L.}\ \bibnamefont {Wright}},\ }\href {https://doi.org/10.1086/377226} {\bibfield  {journal} {\bibinfo  {journal} {The Astrophysical Journal Supplement Series}\ }\textbf {\bibinfo {volume} {148}},\ \bibinfo {pages} {175} (\bibinfo {year} {2003})}\BibitemShut {NoStop}%
\bibitem [{\citenamefont {Bertone}\ \emph {et~al.}(2005)\citenamefont {Bertone}, \citenamefont {Hooper},\ and\ \citenamefont {Silk}}]{BERTONE2005279}%
  \BibitemOpen
  \bibfield  {author} {\bibinfo {author} {\bibfnamefont {G.}~\bibnamefont {Bertone}}, \bibinfo {author} {\bibfnamefont {D.}~\bibnamefont {Hooper}},\ and\ \bibinfo {author} {\bibfnamefont {J.}~\bibnamefont {Silk}},\ }\href {https://doi.org/https://doi.org/10.1016/j.physrep.2004.08.031} {\bibfield  {journal} {\bibinfo  {journal} {Physics Reports}\ }\textbf {\bibinfo {volume} {405}},\ \bibinfo {pages} {279} (\bibinfo {year} {2005})}\BibitemShut {NoStop}%
\bibitem [{\citenamefont {Hui}\ \emph {et~al.}(2017)\citenamefont {Hui}, \citenamefont {Ostriker}, \citenamefont {Tremaine},\ and\ \citenamefont {Witten}}]{Hui:2016ltb}%
  \BibitemOpen
  \bibfield  {author} {\bibinfo {author} {\bibfnamefont {L.}~\bibnamefont {Hui}}, \bibinfo {author} {\bibfnamefont {J.~P.}\ \bibnamefont {Ostriker}}, \bibinfo {author} {\bibfnamefont {S.}~\bibnamefont {Tremaine}},\ and\ \bibinfo {author} {\bibfnamefont {E.}~\bibnamefont {Witten}},\ }\href {https://doi.org/10.1103/PhysRevD.95.043541} {\bibfield  {journal} {\bibinfo  {journal} {Phys. Rev. D}\ }\textbf {\bibinfo {volume} {95}},\ \bibinfo {pages} {043541} (\bibinfo {year} {2017})},\ \Eprint {https://arxiv.org/abs/1610.08297} {arXiv:1610.08297 [astro-ph.CO]} \BibitemShut {NoStop}%
\bibitem [{\citenamefont {Carr}\ and\ \citenamefont {Kuhnel}(2020)}]{Carr:2020xqk}%
  \BibitemOpen
  \bibfield  {author} {\bibinfo {author} {\bibfnamefont {B.}~\bibnamefont {Carr}}\ and\ \bibinfo {author} {\bibfnamefont {F.}~\bibnamefont {Kuhnel}},\ }\href {https://doi.org/10.1146/annurev-nucl-050520-125911} {\bibfield  {journal} {\bibinfo  {journal} {Ann. Rev. Nucl. Part. Sci.}\ }\textbf {\bibinfo {volume} {70}},\ \bibinfo {pages} {355} (\bibinfo {year} {2020})},\ \Eprint {https://arxiv.org/abs/2006.02838} {arXiv:2006.02838 [astro-ph.CO]} \BibitemShut {NoStop}%
\bibitem [{\citenamefont {Cirelli}\ \emph {et~al.}(2024)\citenamefont {Cirelli}, \citenamefont {Strumia},\ and\ \citenamefont {Zupan}}]{cirelli2024darkmatter}%
  \BibitemOpen
  \bibfield  {author} {\bibinfo {author} {\bibfnamefont {M.}~\bibnamefont {Cirelli}}, \bibinfo {author} {\bibfnamefont {A.}~\bibnamefont {Strumia}},\ and\ \bibinfo {author} {\bibfnamefont {J.}~\bibnamefont {Zupan}},\ }\href {https://arxiv.org/abs/2406.01705} {\bibinfo {title} {Dark matter}} (\bibinfo {year} {2024}),\ \Eprint {https://arxiv.org/abs/2406.01705} {arXiv:2406.01705 [hep-ph]} \BibitemShut {NoStop}%
\bibitem [{\citenamefont {Kaplan}(1992)}]{Kaplan:1991ah}%
  \BibitemOpen
  \bibfield  {author} {\bibinfo {author} {\bibfnamefont {D.~B.}\ \bibnamefont {Kaplan}},\ }\href {https://doi.org/10.1103/PhysRevLett.68.741} {\bibfield  {journal} {\bibinfo  {journal} {Phys. Rev. Lett.}\ }\textbf {\bibinfo {volume} {68}},\ \bibinfo {pages} {741} (\bibinfo {year} {1992})}\BibitemShut {NoStop}%
\bibitem [{\citenamefont {Zurek}(2014)}]{Zurek:2013wia}%
  \BibitemOpen
  \bibfield  {author} {\bibinfo {author} {\bibfnamefont {K.~M.}\ \bibnamefont {Zurek}},\ }\href {https://doi.org/10.1016/j.physrep.2013.12.001} {\bibfield  {journal} {\bibinfo  {journal} {Phys. Rept.}\ }\textbf {\bibinfo {volume} {537}},\ \bibinfo {pages} {91} (\bibinfo {year} {2014})},\ \Eprint {https://arxiv.org/abs/1308.0338} {arXiv:1308.0338 [hep-ph]} \BibitemShut {NoStop}%
\bibitem [{\citenamefont {Goldman}\ and\ \citenamefont {Nussinov}(1989)}]{PhysRevD.40.3221}%
  \BibitemOpen
  \bibfield  {author} {\bibinfo {author} {\bibfnamefont {I.}~\bibnamefont {Goldman}}\ and\ \bibinfo {author} {\bibfnamefont {S.}~\bibnamefont {Nussinov}},\ }\href {https://doi.org/10.1103/PhysRevD.40.3221} {\bibfield  {journal} {\bibinfo  {journal} {Phys. Rev. D}\ }\textbf {\bibinfo {volume} {40}},\ \bibinfo {pages} {3221} (\bibinfo {year} {1989})}\BibitemShut {NoStop}%
\bibitem [{\citenamefont {Kouvaris}\ and\ \citenamefont {Nielsen}(2015)}]{Kouvaris:2015rea}%
  \BibitemOpen
  \bibfield  {author} {\bibinfo {author} {\bibfnamefont {C.}~\bibnamefont {Kouvaris}}\ and\ \bibinfo {author} {\bibfnamefont {N.~G.}\ \bibnamefont {Nielsen}},\ }\href {https://doi.org/10.1103/PhysRevD.92.063526} {\bibfield  {journal} {\bibinfo  {journal} {Phys. Rev. D}\ }\textbf {\bibinfo {volume} {92}},\ \bibinfo {pages} {063526} (\bibinfo {year} {2015})},\ \Eprint {https://arxiv.org/abs/1507.00959} {arXiv:1507.00959 [hep-ph]} \BibitemShut {NoStop}%
\bibitem [{\citenamefont {Ellis}\ \emph {et~al.}(2018{\natexlab{a}})\citenamefont {Ellis}, \citenamefont {H\"utsi}, \citenamefont {Kannike}, \citenamefont {Marzola}, \citenamefont {Raidal},\ and\ \citenamefont {Vaskonen}}]{Ellis:2018bkr}%
  \BibitemOpen
  \bibfield  {author} {\bibinfo {author} {\bibfnamefont {J.}~\bibnamefont {Ellis}}, \bibinfo {author} {\bibfnamefont {G.}~\bibnamefont {H\"utsi}}, \bibinfo {author} {\bibfnamefont {K.}~\bibnamefont {Kannike}}, \bibinfo {author} {\bibfnamefont {L.}~\bibnamefont {Marzola}}, \bibinfo {author} {\bibfnamefont {M.}~\bibnamefont {Raidal}},\ and\ \bibinfo {author} {\bibfnamefont {V.}~\bibnamefont {Vaskonen}},\ }\href {https://doi.org/10.1103/PhysRevD.97.123007} {\bibfield  {journal} {\bibinfo  {journal} {Phys. Rev. D}\ }\textbf {\bibinfo {volume} {97}},\ \bibinfo {pages} {123007} (\bibinfo {year} {2018}{\natexlab{a}})},\ \Eprint {https://arxiv.org/abs/1804.01418} {arXiv:1804.01418 [astro-ph.CO]} \BibitemShut {NoStop}%
\bibitem [{\citenamefont {Liang}\ and\ \citenamefont {Shao}(2023)}]{Liang_2023}%
  \BibitemOpen
  \bibfield  {author} {\bibinfo {author} {\bibfnamefont {D.}~\bibnamefont {Liang}}\ and\ \bibinfo {author} {\bibfnamefont {L.}~\bibnamefont {Shao}},\ }\href {https://doi.org/10.1088/1475-7516/2023/08/016} {\bibfield  {journal} {\bibinfo  {journal} {Journal of Cosmology and Astroparticle Physics}\ }\textbf {\bibinfo {volume} {2023}}\bibinfo  {number} { (08)},\ \bibinfo {pages} {016}}\BibitemShut {NoStop}%
\bibitem [{\citenamefont {de~Lavallaz}\ and\ \citenamefont {Fairbairn}(2010)}]{PhysRevD.81.123521}%
  \BibitemOpen
\bibfield  {number} {  }\bibfield  {author} {\bibinfo {author} {\bibfnamefont {A.}~\bibnamefont {de~Lavallaz}}\ and\ \bibinfo {author} {\bibfnamefont {M.}~\bibnamefont {Fairbairn}},\ }\href {https://doi.org/10.1103/PhysRevD.81.123521} {\bibfield  {journal} {\bibinfo  {journal} {Phys. Rev. D}\ }\textbf {\bibinfo {volume} {81}},\ \bibinfo {pages} {123521} (\bibinfo {year} {2010})}\BibitemShut {NoStop}%
\bibitem [{\citenamefont {{Miao}}\ \emph {et~al.}(2022)\citenamefont {{Miao}}, \citenamefont {{Zhu}}, \citenamefont {{Li}},\ and\ \citenamefont {{Huang}}}]{2022ApJ...936...69M}%
  \BibitemOpen
  \bibfield  {author} {\bibinfo {author} {\bibfnamefont {Z.}~\bibnamefont {{Miao}}}, \bibinfo {author} {\bibfnamefont {Y.}~\bibnamefont {{Zhu}}}, \bibinfo {author} {\bibfnamefont {A.}~\bibnamefont {{Li}}},\ and\ \bibinfo {author} {\bibfnamefont {F.}~\bibnamefont {{Huang}}},\ }\href {https://doi.org/10.3847/1538-4357/ac8544} {\bibfield  {journal} {\bibinfo  {journal} {\apj}\ }\textbf {\bibinfo {volume} {936}},\ \bibinfo {eid} {69} (\bibinfo {year} {2022})},\ \Eprint {https://arxiv.org/abs/2204.05560} {arXiv:2204.05560 [astro-ph.HE]} \BibitemShut {NoStop}%
\bibitem [{\citenamefont {Ellis}\ \emph {et~al.}(2018{\natexlab{b}})\citenamefont {Ellis}, \citenamefont {Hektor}, \citenamefont {Hütsi}, \citenamefont {Kannike}, \citenamefont {Marzola}, \citenamefont {Raidal},\ and\ \citenamefont {Vaskonen}}]{ELLIS2018607}%
  \BibitemOpen
  \bibfield  {author} {\bibinfo {author} {\bibfnamefont {J.}~\bibnamefont {Ellis}}, \bibinfo {author} {\bibfnamefont {A.}~\bibnamefont {Hektor}}, \bibinfo {author} {\bibfnamefont {G.}~\bibnamefont {Hütsi}}, \bibinfo {author} {\bibfnamefont {K.}~\bibnamefont {Kannike}}, \bibinfo {author} {\bibfnamefont {L.}~\bibnamefont {Marzola}}, \bibinfo {author} {\bibfnamefont {M.}~\bibnamefont {Raidal}},\ and\ \bibinfo {author} {\bibfnamefont {V.}~\bibnamefont {Vaskonen}},\ }\href {https://doi.org/https://doi.org/10.1016/j.physletb.2018.04.048} {\bibfield  {journal} {\bibinfo  {journal} {Physics Letters B}\ }\textbf {\bibinfo {volume} {781}},\ \bibinfo {pages} {607} (\bibinfo {year} {2018}{\natexlab{b}})}\BibitemShut {NoStop}%
\bibitem [{\citenamefont {Raj}\ \emph {et~al.}(2018)\citenamefont {Raj}, \citenamefont {Tanedo},\ and\ \citenamefont {Yu}}]{PhysRevD.97.043006}%
  \BibitemOpen
  \bibfield  {author} {\bibinfo {author} {\bibfnamefont {N.}~\bibnamefont {Raj}}, \bibinfo {author} {\bibfnamefont {P.}~\bibnamefont {Tanedo}},\ and\ \bibinfo {author} {\bibfnamefont {H.-B.}\ \bibnamefont {Yu}},\ }\href {https://doi.org/10.1103/PhysRevD.97.043006} {\bibfield  {journal} {\bibinfo  {journal} {Phys. Rev. D}\ }\textbf {\bibinfo {volume} {97}},\ \bibinfo {pages} {043006} (\bibinfo {year} {2018})}\BibitemShut {NoStop}%
\bibitem [{\citenamefont {Kouvaris}(2008)}]{PhysRevD.77.023006}%
  \BibitemOpen
  \bibfield  {author} {\bibinfo {author} {\bibfnamefont {C.}~\bibnamefont {Kouvaris}},\ }\href {https://doi.org/10.1103/PhysRevD.77.023006} {\bibfield  {journal} {\bibinfo  {journal} {Phys. Rev. D}\ }\textbf {\bibinfo {volume} {77}},\ \bibinfo {pages} {023006} (\bibinfo {year} {2008})}\BibitemShut {NoStop}%
\bibitem [{\citenamefont {Shawqi}\ and\ \citenamefont {Morsink}(2024)}]{Shawqi_2024}%
  \BibitemOpen
  \bibfield  {author} {\bibinfo {author} {\bibfnamefont {S.}~\bibnamefont {Shawqi}}\ and\ \bibinfo {author} {\bibfnamefont {S.~M.}\ \bibnamefont {Morsink}},\ }\href {https://doi.org/10.3847/1538-4357/ad77c1} {\bibfield  {journal} {\bibinfo  {journal} {The Astrophysical Journal}\ }\textbf {\bibinfo {volume} {975}},\ \bibinfo {pages} {123} (\bibinfo {year} {2024})}\BibitemShut {NoStop}%
\bibitem [{\citenamefont {{Gendreau}}\ \emph {et~al.}(2016)\citenamefont {{Gendreau}}, \citenamefont {{Arzoumanian}}, \citenamefont {{Adkins}}, \citenamefont {{Albert}}, \citenamefont {{Anders}}, \citenamefont {{Aylward}}, \citenamefont {{Baker}}, \citenamefont {{Balsamo}},\ and\ \citenamefont {et. al.}}]{2016SPIE.9905E..1HG}%
  \BibitemOpen
  \bibfield  {author} {\bibinfo {author} {\bibfnamefont {K.~C.}\ \bibnamefont {{Gendreau}}}, \bibinfo {author} {\bibfnamefont {Z.}~\bibnamefont {{Arzoumanian}}}, \bibinfo {author} {\bibfnamefont {P.~W.}\ \bibnamefont {{Adkins}}}, \bibinfo {author} {\bibfnamefont {C.~L.}\ \bibnamefont {{Albert}}}, \bibinfo {author} {\bibfnamefont {J.~F.}\ \bibnamefont {{Anders}}}, \bibinfo {author} {\bibfnamefont {A.~T.}\ \bibnamefont {{Aylward}}}, \bibinfo {author} {\bibfnamefont {C.~L.}\ \bibnamefont {{Baker}}}, \bibinfo {author} {\bibfnamefont {E.~R.}\ \bibnamefont {{Balsamo}}},\ and\ \bibinfo {author} {\bibnamefont {et. al.}},\ }in\ \href {https://doi.org/10.1117/12.2231304} {\emph {\bibinfo {booktitle} {Space Telescopes and Instrumentation 2016: Ultraviolet to Gamma Ray}}},\ \bibinfo {series} {Society of Photo-Optical Instrumentation Engineers (SPIE) Conference Series}, Vol.\ \bibinfo {volume} {9905},\ \bibinfo {editor} {edited by\ \bibinfo {editor} {\bibfnamefont {J.-W.~A.}\ \bibnamefont {{den Herder}}}, \bibinfo
  {editor} {\bibfnamefont {T.}~\bibnamefont {{Takahashi}}},\ and\ \bibinfo {editor} {\bibfnamefont {M.}~\bibnamefont {{Bautz}}}}\ (\bibinfo {year} {2016})\ p.\ \bibinfo {pages} {99051H}\BibitemShut {NoStop}%
\bibitem [{\citenamefont {Abbott}\ \emph {et~al.}(2023)\citenamefont {Abbott}, \citenamefont {Abe}, \citenamefont {Acernese}, \citenamefont {Ackley}, \citenamefont {Adhicary}, \citenamefont {Adhikari}, \citenamefont {Adhikari}, \citenamefont {Adkins}, \citenamefont {Adya}, \citenamefont {et.~al. (The LIGO Scientific~Collaboration}, \citenamefont {the Virgo~Collaboration},\ and\ \citenamefont {the KAGRA~Collaboration)}}]{Abbott_2023}%
  \BibitemOpen
  \bibfield  {author} {\bibinfo {author} {\bibfnamefont {R.}~\bibnamefont {Abbott}}, \bibinfo {author} {\bibfnamefont {H.}~\bibnamefont {Abe}}, \bibinfo {author} {\bibfnamefont {F.}~\bibnamefont {Acernese}}, \bibinfo {author} {\bibfnamefont {K.}~\bibnamefont {Ackley}}, \bibinfo {author} {\bibfnamefont {S.}~\bibnamefont {Adhicary}}, \bibinfo {author} {\bibfnamefont {N.}~\bibnamefont {Adhikari}}, \bibinfo {author} {\bibfnamefont {R.~X.}\ \bibnamefont {Adhikari}}, \bibinfo {author} {\bibfnamefont {V.~K.}\ \bibnamefont {Adkins}}, \bibinfo {author} {\bibfnamefont {V.~B.}\ \bibnamefont {Adya}}, \bibinfo {author} {\bibnamefont {et.~al. (The LIGO Scientific~Collaboration}}, \bibinfo {author} {\bibnamefont {the Virgo~Collaboration}},\ and\ \bibinfo {author} {\bibnamefont {the KAGRA~Collaboration)}},\ }\href {https://doi.org/10.3847/1538-4365/acdc9f} {\bibfield  {journal} {\bibinfo  {journal} {The Astrophysical Journal Supplement Series}\ }\textbf {\bibinfo {volume} {267}},\ \bibinfo {pages} {29} (\bibinfo {year}
  {2023})}\BibitemShut {NoStop}%
\bibitem [{\citenamefont {Rutherford}\ \emph {et~al.}(2024)\citenamefont {Rutherford}, \citenamefont {Prescod-Weinstein},\ and\ \citenamefont {Watts}}]{rutherford2024probingfermionicasymmetricdark}%
  \BibitemOpen
  \bibfield  {author} {\bibinfo {author} {\bibfnamefont {N.}~\bibnamefont {Rutherford}}, \bibinfo {author} {\bibfnamefont {C.}~\bibnamefont {Prescod-Weinstein}},\ and\ \bibinfo {author} {\bibfnamefont {A.}~\bibnamefont {Watts}},\ }\href {https://arxiv.org/abs/2410.00140} {\bibinfo {title} {Probing fermionic asymmetric dark matter cores using global neutron star properties}} (\bibinfo {year} {2024}),\ \Eprint {https://arxiv.org/abs/2410.00140} {arXiv:2410.00140 [astro-ph.HE]} \BibitemShut {NoStop}%
\bibitem [{\citenamefont {Nelson}\ \emph {et~al.}(2019)\citenamefont {Nelson}, \citenamefont {Reddy},\ and\ \citenamefont {Zhou}}]{Nelson_2019}%
  \BibitemOpen
  \bibfield  {author} {\bibinfo {author} {\bibfnamefont {A.~E.}\ \bibnamefont {Nelson}}, \bibinfo {author} {\bibfnamefont {S.}~\bibnamefont {Reddy}},\ and\ \bibinfo {author} {\bibfnamefont {D.}~\bibnamefont {Zhou}},\ }\href {https://doi.org/10.1088/1475-7516/2019/07/012} {\bibfield  {journal} {\bibinfo  {journal} {Journal of Cosmology and Astroparticle Physics}\ }\textbf {\bibinfo {volume} {2019}}\bibinfo  {number} { (07)},\ \bibinfo {pages} {012}}\BibitemShut {NoStop}%
\bibitem [{\citenamefont {Ivanytskyi}\ \emph {et~al.}(2020)\citenamefont {Ivanytskyi}, \citenamefont {Sagun},\ and\ \citenamefont {Lopes}}]{PhysRevD.102.063028}%
  \BibitemOpen
\bibfield  {number} {  }\bibfield  {author} {\bibinfo {author} {\bibfnamefont {O.}~\bibnamefont {Ivanytskyi}}, \bibinfo {author} {\bibfnamefont {V.}~\bibnamefont {Sagun}},\ and\ \bibinfo {author} {\bibfnamefont {I.}~\bibnamefont {Lopes}},\ }\href {https://doi.org/10.1103/PhysRevD.102.063028} {\bibfield  {journal} {\bibinfo  {journal} {Phys. Rev. D}\ }\textbf {\bibinfo {volume} {102}},\ \bibinfo {pages} {063028} (\bibinfo {year} {2020})}\BibitemShut {NoStop}%
\bibitem [{\citenamefont {{Biesdorf}}\ \emph {et~al.}(2024)\citenamefont {{Biesdorf}}, \citenamefont {{Schaffner-Bielich}},\ and\ \citenamefont {{Tolos}}}]{2024arXiv241205207B}%
  \BibitemOpen
  \bibfield  {author} {\bibinfo {author} {\bibfnamefont {C.}~\bibnamefont {{Biesdorf}}}, \bibinfo {author} {\bibfnamefont {J.}~\bibnamefont {{Schaffner-Bielich}}},\ and\ \bibinfo {author} {\bibfnamefont {L.}~\bibnamefont {{Tolos}}},\ }\href {https://doi.org/10.48550/arXiv.2412.05207} {\bibfield  {journal} {\bibinfo  {journal} {arXiv e-prints}\ ,\ \bibinfo {eid} {arXiv:2412.05207}} (\bibinfo {year} {2024})},\ \Eprint {https://arxiv.org/abs/2412.05207} {arXiv:2412.05207 [hep-ph]} \BibitemShut {NoStop}%
\bibitem [{\citenamefont {Ellis}\ \emph {et~al.}(2018{\natexlab{c}})\citenamefont {Ellis}, \citenamefont {H\"utsi}, \citenamefont {Kannike}, \citenamefont {Marzola}, \citenamefont {Raidal},\ and\ \citenamefont {Vaskonen}}]{PhysRevD.97.123007}%
  \BibitemOpen
  \bibfield  {author} {\bibinfo {author} {\bibfnamefont {J.}~\bibnamefont {Ellis}}, \bibinfo {author} {\bibfnamefont {G.}~\bibnamefont {H\"utsi}}, \bibinfo {author} {\bibfnamefont {K.}~\bibnamefont {Kannike}}, \bibinfo {author} {\bibfnamefont {L.}~\bibnamefont {Marzola}}, \bibinfo {author} {\bibfnamefont {M.}~\bibnamefont {Raidal}},\ and\ \bibinfo {author} {\bibfnamefont {V.}~\bibnamefont {Vaskonen}},\ }\href {https://doi.org/10.1103/PhysRevD.97.123007} {\bibfield  {journal} {\bibinfo  {journal} {Phys. Rev. D}\ }\textbf {\bibinfo {volume} {97}},\ \bibinfo {pages} {123007} (\bibinfo {year} {2018}{\natexlab{c}})}\BibitemShut {NoStop}%
\bibitem [{\citenamefont {{Fan}}\ \emph {et~al.}(2012)\citenamefont {{Fan}}, \citenamefont {{Yang}},\ and\ \citenamefont {{Chang}}}]{2012arXiv1204.2564F}%
  \BibitemOpen
  \bibfield  {author} {\bibinfo {author} {\bibfnamefont {Y.-z.}\ \bibnamefont {{Fan}}}, \bibinfo {author} {\bibfnamefont {R.-z.}\ \bibnamefont {{Yang}}},\ and\ \bibinfo {author} {\bibfnamefont {J.}~\bibnamefont {{Chang}}},\ }\href {https://doi.org/10.48550/arXiv.1204.2564} {\bibfield  {journal} {\bibinfo  {journal} {arXiv e-prints}\ ,\ \bibinfo {eid} {arXiv:1204.2564}} (\bibinfo {year} {2012})},\ \Eprint {https://arxiv.org/abs/1204.2564} {arXiv:1204.2564 [astro-ph.HE]} \BibitemShut {NoStop}%
\bibitem [{\citenamefont {Rafiei~Karkevandi}\ \emph {et~al.}(2024)\citenamefont {Rafiei~Karkevandi}, \citenamefont {Shahrbaf}, \citenamefont {Shakeri},\ and\ \citenamefont {Typel}}]{particles7010011}%
  \BibitemOpen
  \bibfield  {author} {\bibinfo {author} {\bibfnamefont {D.}~\bibnamefont {Rafiei~Karkevandi}}, \bibinfo {author} {\bibfnamefont {M.}~\bibnamefont {Shahrbaf}}, \bibinfo {author} {\bibfnamefont {S.}~\bibnamefont {Shakeri}},\ and\ \bibinfo {author} {\bibfnamefont {S.}~\bibnamefont {Typel}},\ }\href {https://doi.org/10.3390/particles7010011} {\bibfield  {journal} {\bibinfo  {journal} {Particles}\ }\textbf {\bibinfo {volume} {7}},\ \bibinfo {pages} {201} (\bibinfo {year} {2024})}\BibitemShut {NoStop}%
\bibitem [{\citenamefont {Konstantinou}(2024)}]{Konstantinou_2024}%
  \BibitemOpen
  \bibfield  {author} {\bibinfo {author} {\bibfnamefont {A.}~\bibnamefont {Konstantinou}},\ }\href {https://doi.org/10.3847/1538-4357/ad4701} {\bibfield  {journal} {\bibinfo  {journal} {The Astrophysical Journal}\ }\textbf {\bibinfo {volume} {968}},\ \bibinfo {pages} {83} (\bibinfo {year} {2024})}\BibitemShut {NoStop}%
\bibitem [{\citenamefont {Panotopoulos}\ and\ \citenamefont {Lopes}(2017)}]{PhysRevD.96.083004}%
  \BibitemOpen
  \bibfield  {author} {\bibinfo {author} {\bibfnamefont {G.}~\bibnamefont {Panotopoulos}}\ and\ \bibinfo {author} {\bibfnamefont {I.}~\bibnamefont {Lopes}},\ }\href {https://doi.org/10.1103/PhysRevD.96.083004} {\bibfield  {journal} {\bibinfo  {journal} {Phys. Rev. D}\ }\textbf {\bibinfo {volume} {96}},\ \bibinfo {pages} {083004} (\bibinfo {year} {2017})}\BibitemShut {NoStop}%
\bibitem [{\citenamefont {Das}\ \emph {et~al.}(2022{\natexlab{a}})\citenamefont {Das}, \citenamefont {Kumar}, \citenamefont {Kumar},\ and\ \citenamefont {Patra}}]{galaxies10010014}%
  \BibitemOpen
  \bibfield  {author} {\bibinfo {author} {\bibfnamefont {H.~C.}\ \bibnamefont {Das}}, \bibinfo {author} {\bibfnamefont {A.}~\bibnamefont {Kumar}}, \bibinfo {author} {\bibfnamefont {B.}~\bibnamefont {Kumar}},\ and\ \bibinfo {author} {\bibfnamefont {S.~K.}\ \bibnamefont {Patra}},\ }\bibfield  {journal} {\bibinfo  {journal} {Galaxies}\ }\textbf {\bibinfo {volume} {10}},\ \href {https://doi.org/10.3390/galaxies10010014} {10.3390/galaxies10010014} (\bibinfo {year} {2022}{\natexlab{a}})\BibitemShut {NoStop}%
\bibitem [{\citenamefont {Louren\ifmmode~\mbox{\c{c}}\else \c{c}\fi{}o}\ \emph {et~al.}(2022)\citenamefont {Louren\ifmmode~\mbox{\c{c}}\else \c{c}\fi{}o}, \citenamefont {Lenzi}, \citenamefont {Frederico},\ and\ \citenamefont {Dutra}}]{PhysRevD.106.043010}%
  \BibitemOpen
  \bibfield  {author} {\bibinfo {author} {\bibfnamefont {O.}~\bibnamefont {Louren\ifmmode~\mbox{\c{c}}\else \c{c}\fi{}o}}, \bibinfo {author} {\bibfnamefont {C.~H.}\ \bibnamefont {Lenzi}}, \bibinfo {author} {\bibfnamefont {T.}~\bibnamefont {Frederico}},\ and\ \bibinfo {author} {\bibfnamefont {M.}~\bibnamefont {Dutra}},\ }\href {https://doi.org/10.1103/PhysRevD.106.043010} {\bibfield  {journal} {\bibinfo  {journal} {Phys. Rev. D}\ }\textbf {\bibinfo {volume} {106}},\ \bibinfo {pages} {043010} (\bibinfo {year} {2022})}\BibitemShut {NoStop}%
\bibitem [{\citenamefont {Kumar}\ and\ \citenamefont {Sotani}(2024)}]{PhysRevD.110.063001}%
  \BibitemOpen
  \bibfield  {author} {\bibinfo {author} {\bibfnamefont {A.}~\bibnamefont {Kumar}}\ and\ \bibinfo {author} {\bibfnamefont {H.}~\bibnamefont {Sotani}},\ }\href {https://doi.org/10.1103/PhysRevD.110.063001} {\bibfield  {journal} {\bibinfo  {journal} {Phys. Rev. D}\ }\textbf {\bibinfo {volume} {110}},\ \bibinfo {pages} {063001} (\bibinfo {year} {2024})}\BibitemShut {NoStop}%
\bibitem [{\citenamefont {Fornal}\ and\ \citenamefont {Grinstein}(2018)}]{PhysRevLett.120.191801}%
  \BibitemOpen
  \bibfield  {author} {\bibinfo {author} {\bibfnamefont {B.}~\bibnamefont {Fornal}}\ and\ \bibinfo {author} {\bibfnamefont {B.}~\bibnamefont {Grinstein}},\ }\href {https://doi.org/10.1103/PhysRevLett.120.191801} {\bibfield  {journal} {\bibinfo  {journal} {Phys. Rev. Lett.}\ }\textbf {\bibinfo {volume} {120}},\ \bibinfo {pages} {191801} (\bibinfo {year} {2018})}\BibitemShut {NoStop}%
\bibitem [{\citenamefont {Husain}\ \emph {et~al.}(2022)\citenamefont {Husain}, \citenamefont {Motta},\ and\ \citenamefont {Thomas}}]{Husain_2022}%
  \BibitemOpen
  \bibfield  {author} {\bibinfo {author} {\bibfnamefont {W.}~\bibnamefont {Husain}}, \bibinfo {author} {\bibfnamefont {T.~F.}\ \bibnamefont {Motta}},\ and\ \bibinfo {author} {\bibfnamefont {A.~W.}\ \bibnamefont {Thomas}},\ }\href {https://doi.org/10.1088/1475-7516/2022/10/028} {\bibfield  {journal} {\bibinfo  {journal} {Journal of Cosmology and Astroparticle Physics}\ }\textbf {\bibinfo {volume} {2022}}\bibinfo  {number} { (10)},\ \bibinfo {pages} {028}}\BibitemShut {NoStop}%
\bibitem [{\citenamefont {{Motta}}\ \emph {et~al.}(2018)\citenamefont {{Motta}}, \citenamefont {{Guichon}},\ and\ \citenamefont {{Thomas}}}]{2018IJMPA..3344020M}%
  \BibitemOpen
\bibfield  {number} {  }\bibfield  {author} {\bibinfo {author} {\bibfnamefont {T.~F.}\ \bibnamefont {{Motta}}}, \bibinfo {author} {\bibfnamefont {P.~A.~M.}\ \bibnamefont {{Guichon}}},\ and\ \bibinfo {author} {\bibfnamefont {A.~W.}\ \bibnamefont {{Thomas}}},\ }\href {https://doi.org/10.1142/S0217751X18440207} {\bibfield  {journal} {\bibinfo  {journal} {International Journal of Modern Physics A}\ }\textbf {\bibinfo {volume} {33}},\ \bibinfo {eid} {1844020} (\bibinfo {year} {2018})},\ \Eprint {https://arxiv.org/abs/1806.00903} {arXiv:1806.00903 [nucl-th]} \BibitemShut {NoStop}%
\bibitem [{\citenamefont {Tulin}\ and\ \citenamefont {Yu}(2018)}]{Tulin:2017ara}%
  \BibitemOpen
  \bibfield  {author} {\bibinfo {author} {\bibfnamefont {S.}~\bibnamefont {Tulin}}\ and\ \bibinfo {author} {\bibfnamefont {H.-B.}\ \bibnamefont {Yu}},\ }\href {https://doi.org/10.1016/j.physrep.2017.11.004} {\bibfield  {journal} {\bibinfo  {journal} {Phys. Rept.}\ }\textbf {\bibinfo {volume} {730}},\ \bibinfo {pages} {1} (\bibinfo {year} {2018})},\ \Eprint {https://arxiv.org/abs/1705.02358} {arXiv:1705.02358 [hep-ph]} \BibitemShut {NoStop}%
\bibitem [{\citenamefont {Randall}\ \emph {et~al.}(2008)\citenamefont {Randall}, \citenamefont {Markevitch}, \citenamefont {Clowe}, \citenamefont {Gonzalez},\ and\ \citenamefont {Bradac}}]{Randall:2008ppe}%
  \BibitemOpen
  \bibfield  {author} {\bibinfo {author} {\bibfnamefont {S.~W.}\ \bibnamefont {Randall}}, \bibinfo {author} {\bibfnamefont {M.}~\bibnamefont {Markevitch}}, \bibinfo {author} {\bibfnamefont {D.}~\bibnamefont {Clowe}}, \bibinfo {author} {\bibfnamefont {A.~H.}\ \bibnamefont {Gonzalez}},\ and\ \bibinfo {author} {\bibfnamefont {M.}~\bibnamefont {Bradac}},\ }\href {https://doi.org/10.1086/587859} {\bibfield  {journal} {\bibinfo  {journal} {Astrophys. J.}\ }\textbf {\bibinfo {volume} {679}},\ \bibinfo {pages} {1173} (\bibinfo {year} {2008})},\ \Eprint {https://arxiv.org/abs/0704.0261} {arXiv:0704.0261 [astro-ph]} \BibitemShut {NoStop}%
\bibitem [{\citenamefont {Ray}\ \emph {et~al.}(2019)\citenamefont {Ray}, \citenamefont {Arzoumanian}, \citenamefont {Ballantyne}, \citenamefont {Bozzo}, \citenamefont {Brandt}, \citenamefont {Brenneman} \emph {et~al.}}]{ray2019strobexxraytimingspectroscopy}%
  \BibitemOpen
  \bibfield  {author} {\bibinfo {author} {\bibfnamefont {P.~S.}\ \bibnamefont {Ray}}, \bibinfo {author} {\bibfnamefont {Z.}~\bibnamefont {Arzoumanian}}, \bibinfo {author} {\bibfnamefont {D.}~\bibnamefont {Ballantyne}}, \bibinfo {author} {\bibfnamefont {E.}~\bibnamefont {Bozzo}}, \bibinfo {author} {\bibfnamefont {S.}~\bibnamefont {Brandt}}, \bibinfo {author} {\bibfnamefont {L.}~\bibnamefont {Brenneman}}, \emph {et~al.},\ }\href {https://arxiv.org/abs/1903.03035} {\bibinfo {title} {Strobe-x: X-ray timing and spectroscopy on dynamical timescales from microseconds to years}} (\bibinfo {year} {2019}),\ \Eprint {https://arxiv.org/abs/1903.03035} {arXiv:1903.03035 [astro-ph.IM]} \BibitemShut {NoStop}%
\bibitem [{\citenamefont {Müller}\ and\ \citenamefont {Serot}(1996)}]{MULLER1996508}%
  \BibitemOpen
  \bibfield  {author} {\bibinfo {author} {\bibfnamefont {H.}~\bibnamefont {Müller}}\ and\ \bibinfo {author} {\bibfnamefont {B.~D.}\ \bibnamefont {Serot}},\ }\href {https://doi.org/https://doi.org/10.1016/0375-9474(96)00187-X} {\bibfield  {journal} {\bibinfo  {journal} {Nuclear Physics A}\ }\textbf {\bibinfo {volume} {606}},\ \bibinfo {pages} {508} (\bibinfo {year} {1996})}\BibitemShut {NoStop}%
\bibitem [{\citenamefont {Boguta}\ and\ \citenamefont {Bodmer}(1977)}]{BOGUTA1977413}%
  \BibitemOpen
  \bibfield  {author} {\bibinfo {author} {\bibfnamefont {J.}~\bibnamefont {Boguta}}\ and\ \bibinfo {author} {\bibfnamefont {A.}~\bibnamefont {Bodmer}},\ }\href {https://doi.org/https://doi.org/10.1016/0375-9474(77)90626-1} {\bibfield  {journal} {\bibinfo  {journal} {Nuclear Physics A}\ }\textbf {\bibinfo {volume} {292}},\ \bibinfo {pages} {413} (\bibinfo {year} {1977})}\BibitemShut {NoStop}%
\bibitem [{\citenamefont {Kubis}\ and\ \citenamefont {Kutschera}(1997)}]{KUBIS1997191}%
  \BibitemOpen
  \bibfield  {author} {\bibinfo {author} {\bibfnamefont {S.}~\bibnamefont {Kubis}}\ and\ \bibinfo {author} {\bibfnamefont {M.}~\bibnamefont {Kutschera}},\ }\href {https://doi.org/https://doi.org/10.1016/S0370-2693(97)00306-7} {\bibfield  {journal} {\bibinfo  {journal} {Physics Letters B}\ }\textbf {\bibinfo {volume} {399}},\ \bibinfo {pages} {191} (\bibinfo {year} {1997})}\BibitemShut {NoStop}%
\bibitem [{\citenamefont {Lalazissis}\ \emph {et~al.}(1997)\citenamefont {Lalazissis}, \citenamefont {K\"onig},\ and\ \citenamefont {Ring}}]{PhysRevC.55.540}%
  \BibitemOpen
  \bibfield  {author} {\bibinfo {author} {\bibfnamefont {G.~A.}\ \bibnamefont {Lalazissis}}, \bibinfo {author} {\bibfnamefont {J.}~\bibnamefont {K\"onig}},\ and\ \bibinfo {author} {\bibfnamefont {P.}~\bibnamefont {Ring}},\ }\href {https://doi.org/10.1103/PhysRevC.55.540} {\bibfield  {journal} {\bibinfo  {journal} {Phys. Rev. C}\ }\textbf {\bibinfo {volume} {55}},\ \bibinfo {pages} {540} (\bibinfo {year} {1997})}\BibitemShut {NoStop}%
\bibitem [{\citenamefont {Ma\ifmmode~\acute{n}\else \'{n}\fi{}ka}\ \emph {et~al.}(2000)\citenamefont {Ma\ifmmode~\acute{n}\else \'{n}\fi{}ka}, \citenamefont {Bednarek},\ and\ \citenamefont {Przyby\l{}a}}]{PhysRevC.62.015802}%
  \BibitemOpen
  \bibfield  {author} {\bibinfo {author} {\bibfnamefont {R.}~\bibnamefont {Ma\ifmmode~\acute{n}\else \'{n}\fi{}ka}}, \bibinfo {author} {\bibfnamefont {I.}~\bibnamefont {Bednarek}},\ and\ \bibinfo {author} {\bibfnamefont {G.}~\bibnamefont {Przyby\l{}a}},\ }\href {https://doi.org/10.1103/PhysRevC.62.015802} {\bibfield  {journal} {\bibinfo  {journal} {Phys. Rev. C}\ }\textbf {\bibinfo {volume} {62}},\ \bibinfo {pages} {015802} (\bibinfo {year} {2000})}\BibitemShut {NoStop}%
\bibitem [{\citenamefont {Chen}\ and\ \citenamefont {Piekarewicz}(2014)}]{PhysRevC.90.044305}%
  \BibitemOpen
  \bibfield  {author} {\bibinfo {author} {\bibfnamefont {W.-C.}\ \bibnamefont {Chen}}\ and\ \bibinfo {author} {\bibfnamefont {J.}~\bibnamefont {Piekarewicz}},\ }\href {https://doi.org/10.1103/PhysRevC.90.044305} {\bibfield  {journal} {\bibinfo  {journal} {Phys. Rev. C}\ }\textbf {\bibinfo {volume} {90}},\ \bibinfo {pages} {044305} (\bibinfo {year} {2014})}\BibitemShut {NoStop}%
\bibitem [{\citenamefont {Kumar}\ \emph {et~al.}(2020)\citenamefont {Kumar}, \citenamefont {Das}, \citenamefont {Biswal}, \citenamefont {Kumar},\ and\ \citenamefont {Patra}}]{Kumar2020-zh}%
  \BibitemOpen
  \bibfield  {author} {\bibinfo {author} {\bibfnamefont {A.}~\bibnamefont {Kumar}}, \bibinfo {author} {\bibfnamefont {H.~C.}\ \bibnamefont {Das}}, \bibinfo {author} {\bibfnamefont {S.~K.}\ \bibnamefont {Biswal}}, \bibinfo {author} {\bibfnamefont {B.}~\bibnamefont {Kumar}},\ and\ \bibinfo {author} {\bibfnamefont {S.~K.}\ \bibnamefont {Patra}},\ }\href@noop {} {\bibfield  {journal} {\bibinfo  {journal} {The European Physical Journal C}\ }\textbf {\bibinfo {volume} {80}},\ \bibinfo {pages} {775} (\bibinfo {year} {2020})}\BibitemShut {NoStop}%
\bibitem [{\citenamefont {Alford}\ \emph {et~al.}(2022)\citenamefont {Alford}, \citenamefont {Brodie}, \citenamefont {Haber},\ and\ \citenamefont {Tews}}]{PhysRevC.106.055804}%
  \BibitemOpen
  \bibfield  {author} {\bibinfo {author} {\bibfnamefont {M.~G.}\ \bibnamefont {Alford}}, \bibinfo {author} {\bibfnamefont {L.}~\bibnamefont {Brodie}}, \bibinfo {author} {\bibfnamefont {A.}~\bibnamefont {Haber}},\ and\ \bibinfo {author} {\bibfnamefont {I.}~\bibnamefont {Tews}},\ }\href {https://doi.org/10.1103/PhysRevC.106.055804} {\bibfield  {journal} {\bibinfo  {journal} {Phys. Rev. C}\ }\textbf {\bibinfo {volume} {106}},\ \bibinfo {pages} {055804} (\bibinfo {year} {2022})}\BibitemShut {NoStop}%
\bibitem [{\citenamefont {Fattoyev}\ \emph {et~al.}(2020)\citenamefont {Fattoyev}, \citenamefont {Horowitz}, \citenamefont {Piekarewicz},\ and\ \citenamefont {Reed}}]{PhysRevC.102.065805}%
  \BibitemOpen
  \bibfield  {author} {\bibinfo {author} {\bibfnamefont {F.~J.}\ \bibnamefont {Fattoyev}}, \bibinfo {author} {\bibfnamefont {C.~J.}\ \bibnamefont {Horowitz}}, \bibinfo {author} {\bibfnamefont {J.}~\bibnamefont {Piekarewicz}},\ and\ \bibinfo {author} {\bibfnamefont {B.}~\bibnamefont {Reed}},\ }\href {https://doi.org/10.1103/PhysRevC.102.065805} {\bibfield  {journal} {\bibinfo  {journal} {Phys. Rev. C}\ }\textbf {\bibinfo {volume} {102}},\ \bibinfo {pages} {065805} (\bibinfo {year} {2020})}\BibitemShut {NoStop}%
\bibitem [{\citenamefont {Tolman}(1939)}]{PhysRev.55.364}%
  \BibitemOpen
  \bibfield  {author} {\bibinfo {author} {\bibfnamefont {R.~C.}\ \bibnamefont {Tolman}},\ }\href {https://doi.org/10.1103/PhysRev.55.364} {\bibfield  {journal} {\bibinfo  {journal} {Phys. Rev.}\ }\textbf {\bibinfo {volume} {55}},\ \bibinfo {pages} {364} (\bibinfo {year} {1939})}\BibitemShut {NoStop}%
\bibitem [{\citenamefont {Oppenheimer}\ and\ \citenamefont {Volkoff}(1939)}]{PhysRev.55.374}%
  \BibitemOpen
  \bibfield  {author} {\bibinfo {author} {\bibfnamefont {J.~R.}\ \bibnamefont {Oppenheimer}}\ and\ \bibinfo {author} {\bibfnamefont {G.~M.}\ \bibnamefont {Volkoff}},\ }\href {https://doi.org/10.1103/PhysRev.55.374} {\bibfield  {journal} {\bibinfo  {journal} {Phys. Rev.}\ }\textbf {\bibinfo {volume} {55}},\ \bibinfo {pages} {374} (\bibinfo {year} {1939})}\BibitemShut {NoStop}%
\bibitem [{\citenamefont {{Douchin, F.}}\ and\ \citenamefont {{Haensel, P.}}(2001)}]{refId0}%
  \BibitemOpen
  \bibfield  {author} {\bibinfo {author} {\bibnamefont {{Douchin, F.}}}\ and\ \bibinfo {author} {\bibnamefont {{Haensel, P.}}},\ }\href {https://doi.org/10.1051/0004-6361:20011402} {\bibfield  {journal} {\bibinfo  {journal} {A\&A}\ }\textbf {\bibinfo {volume} {380}},\ \bibinfo {pages} {151} (\bibinfo {year} {2001})}\BibitemShut {NoStop}%
\bibitem [{\citenamefont {{Cromartie}}\ \emph {et~al.}(2020)\citenamefont {{Cromartie}}, \citenamefont {{Fonseca}}, \citenamefont {{Ransom}}, \citenamefont {{Demorest}}, \citenamefont {{Arzoumanian}}, \citenamefont {{Blumer}}, \citenamefont {{Brook}}, \citenamefont {{DeCesar}}, \citenamefont {{Dolch}}, \citenamefont {{Ellis}}, \citenamefont {{Ferdman}}, \citenamefont {{Ferrara}}, \citenamefont {{Garver-Daniels}}, \citenamefont {{Gentile}}, \citenamefont {{Jones}}, \citenamefont {{Lam}}, \citenamefont {{Lorimer}}, \citenamefont {{Lynch}}, \citenamefont {{McLaughlin}}, \citenamefont {{Ng}}, \citenamefont {{Nice}}, \citenamefont {{Pennucci}}, \citenamefont {{Spiewak}}, \citenamefont {{Stairs}}, \citenamefont {{Stovall}}, \citenamefont {{Swiggum}},\ and\ \citenamefont {{Zhu}}}]{2020NatAs...4...72C}%
  \BibitemOpen
  \bibfield  {author} {\bibinfo {author} {\bibfnamefont {H.~T.}\ \bibnamefont {{Cromartie}}}, \bibinfo {author} {\bibfnamefont {E.}~\bibnamefont {{Fonseca}}}, \bibinfo {author} {\bibfnamefont {S.~M.}\ \bibnamefont {{Ransom}}}, \bibinfo {author} {\bibfnamefont {P.~B.}\ \bibnamefont {{Demorest}}}, \bibinfo {author} {\bibfnamefont {Z.}~\bibnamefont {{Arzoumanian}}}, \bibinfo {author} {\bibfnamefont {H.}~\bibnamefont {{Blumer}}}, \bibinfo {author} {\bibfnamefont {P.~R.}\ \bibnamefont {{Brook}}}, \bibinfo {author} {\bibfnamefont {M.~E.}\ \bibnamefont {{DeCesar}}}, \bibinfo {author} {\bibfnamefont {T.}~\bibnamefont {{Dolch}}}, \bibinfo {author} {\bibfnamefont {J.~A.}\ \bibnamefont {{Ellis}}}, \bibinfo {author} {\bibfnamefont {R.~D.}\ \bibnamefont {{Ferdman}}}, \bibinfo {author} {\bibfnamefont {E.~C.}\ \bibnamefont {{Ferrara}}}, \bibinfo {author} {\bibfnamefont {N.}~\bibnamefont {{Garver-Daniels}}}, \bibinfo {author} {\bibfnamefont {P.~A.}\ \bibnamefont {{Gentile}}}, \bibinfo {author} {\bibfnamefont {M.~L.}\
  \bibnamefont {{Jones}}}, \bibinfo {author} {\bibfnamefont {M.~T.}\ \bibnamefont {{Lam}}}, \bibinfo {author} {\bibfnamefont {D.~R.}\ \bibnamefont {{Lorimer}}}, \bibinfo {author} {\bibfnamefont {R.~S.}\ \bibnamefont {{Lynch}}}, \bibinfo {author} {\bibfnamefont {M.~A.}\ \bibnamefont {{McLaughlin}}}, \bibinfo {author} {\bibfnamefont {C.}~\bibnamefont {{Ng}}}, \bibinfo {author} {\bibfnamefont {D.~J.}\ \bibnamefont {{Nice}}}, \bibinfo {author} {\bibfnamefont {T.~T.}\ \bibnamefont {{Pennucci}}}, \bibinfo {author} {\bibfnamefont {R.}~\bibnamefont {{Spiewak}}}, \bibinfo {author} {\bibfnamefont {I.~H.}\ \bibnamefont {{Stairs}}}, \bibinfo {author} {\bibfnamefont {K.}~\bibnamefont {{Stovall}}}, \bibinfo {author} {\bibfnamefont {J.~K.}\ \bibnamefont {{Swiggum}}},\ and\ \bibinfo {author} {\bibfnamefont {W.~W.}\ \bibnamefont {{Zhu}}},\ }\href {https://doi.org/10.1038/s41550-019-0880-2} {\bibfield  {journal} {\bibinfo  {journal} {Nature Astronomy}\ }\textbf {\bibinfo {volume} {4}},\ \bibinfo {pages} {72} (\bibinfo {year}
  {2020})},\ \Eprint {https://arxiv.org/abs/1904.06759} {arXiv:1904.06759 [astro-ph.HE]} \BibitemShut {NoStop}%
\bibitem [{\citenamefont {Abbott}\ \emph {et~al.}(2018)\citenamefont {Abbott}, \citenamefont {Abbott}, \citenamefont {Abbott}, \citenamefont {Acernese}, \citenamefont {Ackley}, \citenamefont {Adams}, \citenamefont {Adams}, \citenamefont {Addesso}, \citenamefont {Adhikari}, \citenamefont {Adya},\ and\ \citenamefont {et. al.}}]{PhysRevLett.121.161101}%
  \BibitemOpen
  \bibfield  {author} {\bibinfo {author} {\bibfnamefont {B.~P.}\ \bibnamefont {Abbott}}, \bibinfo {author} {\bibfnamefont {R.}~\bibnamefont {Abbott}}, \bibinfo {author} {\bibfnamefont {T.~D.}\ \bibnamefont {Abbott}}, \bibinfo {author} {\bibfnamefont {F.}~\bibnamefont {Acernese}}, \bibinfo {author} {\bibfnamefont {K.}~\bibnamefont {Ackley}}, \bibinfo {author} {\bibfnamefont {C.}~\bibnamefont {Adams}}, \bibinfo {author} {\bibfnamefont {T.}~\bibnamefont {Adams}}, \bibinfo {author} {\bibfnamefont {P.}~\bibnamefont {Addesso}}, \bibinfo {author} {\bibfnamefont {R.~X.}\ \bibnamefont {Adhikari}}, \bibinfo {author} {\bibfnamefont {V.~B.}\ \bibnamefont {Adya}},\ and\ \bibinfo {author} {\bibnamefont {et. al.}} (\bibinfo {collaboration} {The LIGO Scientific Collaboration and the Virgo Collaboration}),\ }\href {https://doi.org/10.1103/PhysRevLett.121.161101} {\bibfield  {journal} {\bibinfo  {journal} {Phys. Rev. Lett.}\ }\textbf {\bibinfo {volume} {121}},\ \bibinfo {pages} {161101} (\bibinfo {year} {2018})}\BibitemShut
  {NoStop}%
\bibitem [{\citenamefont {Roberts}\ \emph {et~al.}(2024)\citenamefont {Roberts}, \citenamefont {Kaplinghat}, \citenamefont {Valli},\ and\ \citenamefont {Yu}}]{Roberts:2024uyw}%
  \BibitemOpen
  \bibfield  {author} {\bibinfo {author} {\bibfnamefont {M.~G.}\ \bibnamefont {Roberts}}, \bibinfo {author} {\bibfnamefont {M.}~\bibnamefont {Kaplinghat}}, \bibinfo {author} {\bibfnamefont {M.}~\bibnamefont {Valli}},\ and\ \bibinfo {author} {\bibfnamefont {H.-B.}\ \bibnamefont {Yu}},\ }\href@noop {} {\  (\bibinfo {year} {2024})},\ \Eprint {https://arxiv.org/abs/2407.15005} {arXiv:2407.15005 [astro-ph.GA]} \BibitemShut {NoStop}%
\bibitem [{\citenamefont {Correa}\ \emph {et~al.}(2022)\citenamefont {Correa}, \citenamefont {Schaller}, \citenamefont {Ploeckinger}, \citenamefont {Anau~Montel}, \citenamefont {Weniger},\ and\ \citenamefont {Ando}}]{Correa:2022dey}%
  \BibitemOpen
  \bibfield  {author} {\bibinfo {author} {\bibfnamefont {C.~A.}\ \bibnamefont {Correa}}, \bibinfo {author} {\bibfnamefont {M.}~\bibnamefont {Schaller}}, \bibinfo {author} {\bibfnamefont {S.}~\bibnamefont {Ploeckinger}}, \bibinfo {author} {\bibfnamefont {N.}~\bibnamefont {Anau~Montel}}, \bibinfo {author} {\bibfnamefont {C.}~\bibnamefont {Weniger}},\ and\ \bibinfo {author} {\bibfnamefont {S.}~\bibnamefont {Ando}},\ }\href {https://doi.org/10.1093/mnras/stac2830} {\bibfield  {journal} {\bibinfo  {journal} {Mon. Not. Roy. Astron. Soc.}\ }\textbf {\bibinfo {volume} {517}},\ \bibinfo {pages} {3045} (\bibinfo {year} {2022})},\ \Eprint {https://arxiv.org/abs/2206.11298} {arXiv:2206.11298 [astro-ph.GA]} \BibitemShut {NoStop}%
\bibitem [{\citenamefont {Kaplinghat}\ \emph {et~al.}(2014)\citenamefont {Kaplinghat}, \citenamefont {Tulin},\ and\ \citenamefont {Yu}}]{Kaplinghat:2013yxa}%
  \BibitemOpen
  \bibfield  {author} {\bibinfo {author} {\bibfnamefont {M.}~\bibnamefont {Kaplinghat}}, \bibinfo {author} {\bibfnamefont {S.}~\bibnamefont {Tulin}},\ and\ \bibinfo {author} {\bibfnamefont {H.-B.}\ \bibnamefont {Yu}},\ }\href {https://doi.org/10.1103/PhysRevD.89.035009} {\bibfield  {journal} {\bibinfo  {journal} {Phys. Rev. D}\ }\textbf {\bibinfo {volume} {89}},\ \bibinfo {pages} {035009} (\bibinfo {year} {2014})},\ \Eprint {https://arxiv.org/abs/1310.7945} {arXiv:1310.7945 [hep-ph]} \BibitemShut {NoStop}%
\bibitem [{\citenamefont {Fujikura}\ \emph {et~al.}(2024)\citenamefont {Fujikura}, \citenamefont {Girmohanta}, \citenamefont {Nakai},\ and\ \citenamefont {Zhang}}]{Fujikura:2024jto}%
  \BibitemOpen
  \bibfield  {author} {\bibinfo {author} {\bibfnamefont {K.}~\bibnamefont {Fujikura}}, \bibinfo {author} {\bibfnamefont {S.}~\bibnamefont {Girmohanta}}, \bibinfo {author} {\bibfnamefont {Y.}~\bibnamefont {Nakai}},\ and\ \bibinfo {author} {\bibfnamefont {Z.}~\bibnamefont {Zhang}},\ }\href {https://doi.org/10.1016/j.physletb.2024.139045} {\bibfield  {journal} {\bibinfo  {journal} {Phys. Lett. B}\ }\textbf {\bibinfo {volume} {858}},\ \bibinfo {pages} {139045} (\bibinfo {year} {2024})},\ \Eprint {https://arxiv.org/abs/2406.12956} {arXiv:2406.12956 [hep-ph]} \BibitemShut {NoStop}%
\bibitem [{\citenamefont {Gresham}\ and\ \citenamefont {Zurek}(2019)}]{PhysRevD.99.083008}%
  \BibitemOpen
  \bibfield  {author} {\bibinfo {author} {\bibfnamefont {M.~I.}\ \bibnamefont {Gresham}}\ and\ \bibinfo {author} {\bibfnamefont {K.~M.}\ \bibnamefont {Zurek}},\ }\href {https://doi.org/10.1103/PhysRevD.99.083008} {\bibfield  {journal} {\bibinfo  {journal} {Phys. Rev. D}\ }\textbf {\bibinfo {volume} {99}},\ \bibinfo {pages} {083008} (\bibinfo {year} {2019})}\BibitemShut {NoStop}%
\bibitem [{\citenamefont {Girmohanta}\ and\ \citenamefont {Shrock}(2022)}]{Girmohanta:2022dog}%
  \BibitemOpen
  \bibfield  {author} {\bibinfo {author} {\bibfnamefont {S.}~\bibnamefont {Girmohanta}}\ and\ \bibinfo {author} {\bibfnamefont {R.}~\bibnamefont {Shrock}},\ }\href {https://doi.org/10.1103/PhysRevD.106.063013} {\bibfield  {journal} {\bibinfo  {journal} {Phys. Rev. D}\ }\textbf {\bibinfo {volume} {106}},\ \bibinfo {pages} {063013} (\bibinfo {year} {2022})},\ \Eprint {https://arxiv.org/abs/2206.14395} {arXiv:2206.14395 [hep-ph]} \BibitemShut {NoStop}%
\bibitem [{\citenamefont {Girmohanta}\ and\ \citenamefont {Shrock}(2023)}]{Girmohanta:2022izb}%
  \BibitemOpen
  \bibfield  {author} {\bibinfo {author} {\bibfnamefont {S.}~\bibnamefont {Girmohanta}}\ and\ \bibinfo {author} {\bibfnamefont {R.}~\bibnamefont {Shrock}},\ }\href {https://doi.org/10.1103/PhysRevD.107.063006} {\bibfield  {journal} {\bibinfo  {journal} {Phys. Rev. D}\ }\textbf {\bibinfo {volume} {107}},\ \bibinfo {pages} {063006} (\bibinfo {year} {2023})},\ \Eprint {https://arxiv.org/abs/2210.01132} {arXiv:2210.01132 [hep-ph]} \BibitemShut {NoStop}%
\bibitem [{\citenamefont {Khrapak}\ \emph {et~al.}(2003)\citenamefont {Khrapak}, \citenamefont {Ivlev}, \citenamefont {Morfill},\ and\ \citenamefont {Zhdanov}}]{Khrapak:2003kjw}%
  \BibitemOpen
  \bibfield  {author} {\bibinfo {author} {\bibfnamefont {S.~A.}\ \bibnamefont {Khrapak}}, \bibinfo {author} {\bibfnamefont {A.~V.}\ \bibnamefont {Ivlev}}, \bibinfo {author} {\bibfnamefont {G.~E.}\ \bibnamefont {Morfill}},\ and\ \bibinfo {author} {\bibfnamefont {S.~K.}\ \bibnamefont {Zhdanov}},\ }\href {https://doi.org/10.1103/PhysRevLett.90.225002} {\bibfield  {journal} {\bibinfo  {journal} {Phys. Rev. Lett.}\ }\textbf {\bibinfo {volume} {90}},\ \bibinfo {pages} {225002} (\bibinfo {year} {2003})}\BibitemShut {NoStop}%
\bibitem [{\citenamefont {Tulin}\ \emph {et~al.}(2013)\citenamefont {Tulin}, \citenamefont {Yu},\ and\ \citenamefont {Zurek}}]{Tulin:2013teo}%
  \BibitemOpen
  \bibfield  {author} {\bibinfo {author} {\bibfnamefont {S.}~\bibnamefont {Tulin}}, \bibinfo {author} {\bibfnamefont {H.-B.}\ \bibnamefont {Yu}},\ and\ \bibinfo {author} {\bibfnamefont {K.~M.}\ \bibnamefont {Zurek}},\ }\href {https://doi.org/10.1103/PhysRevD.87.115007} {\bibfield  {journal} {\bibinfo  {journal} {Phys. Rev. D}\ }\textbf {\bibinfo {volume} {87}},\ \bibinfo {pages} {115007} (\bibinfo {year} {2013})},\ \Eprint {https://arxiv.org/abs/1302.3898} {arXiv:1302.3898 [hep-ph]} \BibitemShut {NoStop}%
\bibitem [{\citenamefont {Ciarcelluti}\ and\ \citenamefont {Sandin}(2011)}]{CIARCELLUTI201119}%
  \BibitemOpen
  \bibfield  {author} {\bibinfo {author} {\bibfnamefont {P.}~\bibnamefont {Ciarcelluti}}\ and\ \bibinfo {author} {\bibfnamefont {F.}~\bibnamefont {Sandin}},\ }\href {https://doi.org/https://doi.org/10.1016/j.physletb.2010.11.021} {\bibfield  {journal} {\bibinfo  {journal} {Physics Letters B}\ }\textbf {\bibinfo {volume} {695}},\ \bibinfo {pages} {19} (\bibinfo {year} {2011})}\BibitemShut {NoStop}%
\bibitem [{\citenamefont {{Sandin}}\ and\ \citenamefont {{Ciarcelluti}}(2009)}]{2009APh....32..278S}%
  \BibitemOpen
  \bibfield  {author} {\bibinfo {author} {\bibfnamefont {F.}~\bibnamefont {{Sandin}}}\ and\ \bibinfo {author} {\bibfnamefont {P.}~\bibnamefont {{Ciarcelluti}}},\ }\href {https://doi.org/10.1016/j.astropartphys.2009.09.005} {\bibfield  {journal} {\bibinfo  {journal} {Astroparticle Physics}\ }\textbf {\bibinfo {volume} {32}},\ \bibinfo {pages} {278} (\bibinfo {year} {2009})},\ \Eprint {https://arxiv.org/abs/0809.2942} {arXiv:0809.2942 [astro-ph]} \BibitemShut {NoStop}%
\bibitem [{\citenamefont {Goldman}\ \emph {et~al.}(2013)\citenamefont {Goldman}, \citenamefont {Mohapatra}, \citenamefont {Nussinov}, \citenamefont {Rosenbaum},\ and\ \citenamefont {Teplitz}}]{GOLDMAN2013200}%
  \BibitemOpen
  \bibfield  {author} {\bibinfo {author} {\bibfnamefont {I.}~\bibnamefont {Goldman}}, \bibinfo {author} {\bibfnamefont {R.}~\bibnamefont {Mohapatra}}, \bibinfo {author} {\bibfnamefont {S.}~\bibnamefont {Nussinov}}, \bibinfo {author} {\bibfnamefont {D.}~\bibnamefont {Rosenbaum}},\ and\ \bibinfo {author} {\bibfnamefont {V.}~\bibnamefont {Teplitz}},\ }\href {https://doi.org/https://doi.org/10.1016/j.physletb.2013.07.017} {\bibfield  {journal} {\bibinfo  {journal} {Physics Letters B}\ }\textbf {\bibinfo {volume} {725}},\ \bibinfo {pages} {200} (\bibinfo {year} {2013})}\BibitemShut {NoStop}%
\bibitem [{\citenamefont {Sagun}\ \emph {et~al.}(2023)\citenamefont {Sagun}, \citenamefont {Giangrandi}, \citenamefont {Dietrich}, \citenamefont {Ivanytskyi}, \citenamefont {Negreiros},\ and\ \citenamefont {Providência}}]{Sagun_2023}%
  \BibitemOpen
  \bibfield  {author} {\bibinfo {author} {\bibfnamefont {V.}~\bibnamefont {Sagun}}, \bibinfo {author} {\bibfnamefont {E.}~\bibnamefont {Giangrandi}}, \bibinfo {author} {\bibfnamefont {T.}~\bibnamefont {Dietrich}}, \bibinfo {author} {\bibfnamefont {O.}~\bibnamefont {Ivanytskyi}}, \bibinfo {author} {\bibfnamefont {R.}~\bibnamefont {Negreiros}},\ and\ \bibinfo {author} {\bibfnamefont {C.}~\bibnamefont {Providência}},\ }\href {https://doi.org/10.3847/1538-4357/acfc9e} {\bibfield  {journal} {\bibinfo  {journal} {The Astrophysical Journal}\ }\textbf {\bibinfo {volume} {958}},\ \bibinfo {pages} {49} (\bibinfo {year} {2023})}\BibitemShut {NoStop}%
\bibitem [{\citenamefont {Hinderer}(2008)}]{Hinderer_2008}%
  \BibitemOpen
  \bibfield  {author} {\bibinfo {author} {\bibfnamefont {T.}~\bibnamefont {Hinderer}},\ }\href {https://doi.org/10.1086/533487} {\bibfield  {journal} {\bibinfo  {journal} {The Astrophysical Journal}\ }\textbf {\bibinfo {volume} {677}},\ \bibinfo {pages} {1216} (\bibinfo {year} {2008})}\BibitemShut {NoStop}%
\bibitem [{\citenamefont {Hippert}\ \emph {et~al.}(2023)\citenamefont {Hippert}, \citenamefont {Dillingham}, \citenamefont {Tan}, \citenamefont {Curtin}, \citenamefont {Noronha-Hostler},\ and\ \citenamefont {Yunes}}]{PhysRevD.107.115028}%
  \BibitemOpen
  \bibfield  {author} {\bibinfo {author} {\bibfnamefont {M.}~\bibnamefont {Hippert}}, \bibinfo {author} {\bibfnamefont {E.}~\bibnamefont {Dillingham}}, \bibinfo {author} {\bibfnamefont {H.}~\bibnamefont {Tan}}, \bibinfo {author} {\bibfnamefont {D.}~\bibnamefont {Curtin}}, \bibinfo {author} {\bibfnamefont {J.}~\bibnamefont {Noronha-Hostler}},\ and\ \bibinfo {author} {\bibfnamefont {N.}~\bibnamefont {Yunes}},\ }\href {https://doi.org/10.1103/PhysRevD.107.115028} {\bibfield  {journal} {\bibinfo  {journal} {Phys. Rev. D}\ }\textbf {\bibinfo {volume} {107}},\ \bibinfo {pages} {115028} (\bibinfo {year} {2023})}\BibitemShut {NoStop}%
\bibitem [{\citenamefont {Hinderer}\ \emph {et~al.}(2010)\citenamefont {Hinderer}, \citenamefont {Lackey}, \citenamefont {Lang},\ and\ \citenamefont {Read}}]{PhysRevD.81.123016}%
  \BibitemOpen
  \bibfield  {author} {\bibinfo {author} {\bibfnamefont {T.}~\bibnamefont {Hinderer}}, \bibinfo {author} {\bibfnamefont {B.~D.}\ \bibnamefont {Lackey}}, \bibinfo {author} {\bibfnamefont {R.~N.}\ \bibnamefont {Lang}},\ and\ \bibinfo {author} {\bibfnamefont {J.~S.}\ \bibnamefont {Read}},\ }\href {https://doi.org/10.1103/PhysRevD.81.123016} {\bibfield  {journal} {\bibinfo  {journal} {Phys. Rev. D}\ }\textbf {\bibinfo {volume} {81}},\ \bibinfo {pages} {123016} (\bibinfo {year} {2010})}\BibitemShut {NoStop}%
\bibitem [{\citenamefont {Leung}\ \emph {et~al.}(2022)\citenamefont {Leung}, \citenamefont {Chu},\ and\ \citenamefont {Lin}}]{PhysRevD.105.123010}%
  \BibitemOpen
  \bibfield  {author} {\bibinfo {author} {\bibfnamefont {K.-L.}\ \bibnamefont {Leung}}, \bibinfo {author} {\bibfnamefont {M.-c.}\ \bibnamefont {Chu}},\ and\ \bibinfo {author} {\bibfnamefont {L.-M.}\ \bibnamefont {Lin}},\ }\href {https://doi.org/10.1103/PhysRevD.105.123010} {\bibfield  {journal} {\bibinfo  {journal} {Phys. Rev. D}\ }\textbf {\bibinfo {volume} {105}},\ \bibinfo {pages} {123010} (\bibinfo {year} {2022})}\BibitemShut {NoStop}%
\bibitem [{\citenamefont {Das}\ \emph {et~al.}(2022{\natexlab{b}})\citenamefont {Das}, \citenamefont {Malik},\ and\ \citenamefont {Nayak}}]{PhysRevD.105.123034}%
  \BibitemOpen
  \bibfield  {author} {\bibinfo {author} {\bibfnamefont {A.}~\bibnamefont {Das}}, \bibinfo {author} {\bibfnamefont {T.}~\bibnamefont {Malik}},\ and\ \bibinfo {author} {\bibfnamefont {A.~C.}\ \bibnamefont {Nayak}},\ }\href {https://doi.org/10.1103/PhysRevD.105.123034} {\bibfield  {journal} {\bibinfo  {journal} {Phys. Rev. D}\ }\textbf {\bibinfo {volume} {105}},\ \bibinfo {pages} {123034} (\bibinfo {year} {2022}{\natexlab{b}})}\BibitemShut {NoStop}%
\bibitem [{\citenamefont {Fonseca}\ \emph {et~al.}(2021)\citenamefont {Fonseca}, \citenamefont {Cromartie}, \citenamefont {Pennucci}, \citenamefont {Ray}, \citenamefont {Kirichenko}, \citenamefont {Ransom}, \citenamefont {Demorest}, \citenamefont {Stairs}, \citenamefont {Arzoumanian}, \citenamefont {Guillemot}, \citenamefont {Parthasarathy}, \citenamefont {Kerr}, \citenamefont {Cognard}, \citenamefont {Baker}, \citenamefont {Blumer}, \citenamefont {Brook}, \citenamefont {DeCesar}, \citenamefont {Dolch}, \citenamefont {Dong}, \citenamefont {Ferrara}, \citenamefont {Fiore}, \citenamefont {Garver-Daniels}, \citenamefont {Good}, \citenamefont {Jennings}, \citenamefont {Jones}, \citenamefont {Kaspi}, \citenamefont {Lam}, \citenamefont {Lorimer}, \citenamefont {Luo}, \citenamefont {McEwen}, \citenamefont {McKee}, \citenamefont {McLaughlin}, \citenamefont {McMann}, \citenamefont {Meyers}, \citenamefont {Naidu}, \citenamefont {Ng}, \citenamefont {Nice}, \citenamefont {Pol}, \citenamefont {Radovan}, \citenamefont
  {Shapiro-Albert}, \citenamefont {Tan}, \citenamefont {Tendulkar}, \citenamefont {Swiggum}, \citenamefont {Wahl},\ and\ \citenamefont {Zhu}}]{Fonseca_2021}%
  \BibitemOpen
  \bibfield  {author} {\bibinfo {author} {\bibfnamefont {E.}~\bibnamefont {Fonseca}}, \bibinfo {author} {\bibfnamefont {H.~T.}\ \bibnamefont {Cromartie}}, \bibinfo {author} {\bibfnamefont {T.~T.}\ \bibnamefont {Pennucci}}, \bibinfo {author} {\bibfnamefont {P.~S.}\ \bibnamefont {Ray}}, \bibinfo {author} {\bibfnamefont {A.~Y.}\ \bibnamefont {Kirichenko}}, \bibinfo {author} {\bibfnamefont {S.~M.}\ \bibnamefont {Ransom}}, \bibinfo {author} {\bibfnamefont {P.~B.}\ \bibnamefont {Demorest}}, \bibinfo {author} {\bibfnamefont {I.~H.}\ \bibnamefont {Stairs}}, \bibinfo {author} {\bibfnamefont {Z.}~\bibnamefont {Arzoumanian}}, \bibinfo {author} {\bibfnamefont {L.}~\bibnamefont {Guillemot}}, \bibinfo {author} {\bibfnamefont {A.}~\bibnamefont {Parthasarathy}}, \bibinfo {author} {\bibfnamefont {M.}~\bibnamefont {Kerr}}, \bibinfo {author} {\bibfnamefont {I.}~\bibnamefont {Cognard}}, \bibinfo {author} {\bibfnamefont {P.~T.}\ \bibnamefont {Baker}}, \bibinfo {author} {\bibfnamefont {H.}~\bibnamefont {Blumer}}, \bibinfo {author}
  {\bibfnamefont {P.~R.}\ \bibnamefont {Brook}}, \bibinfo {author} {\bibfnamefont {M.}~\bibnamefont {DeCesar}}, \bibinfo {author} {\bibfnamefont {T.}~\bibnamefont {Dolch}}, \bibinfo {author} {\bibfnamefont {F.~A.}\ \bibnamefont {Dong}}, \bibinfo {author} {\bibfnamefont {E.~C.}\ \bibnamefont {Ferrara}}, \bibinfo {author} {\bibfnamefont {W.}~\bibnamefont {Fiore}}, \bibinfo {author} {\bibfnamefont {N.}~\bibnamefont {Garver-Daniels}}, \bibinfo {author} {\bibfnamefont {D.~C.}\ \bibnamefont {Good}}, \bibinfo {author} {\bibfnamefont {R.}~\bibnamefont {Jennings}}, \bibinfo {author} {\bibfnamefont {M.~L.}\ \bibnamefont {Jones}}, \bibinfo {author} {\bibfnamefont {V.~M.}\ \bibnamefont {Kaspi}}, \bibinfo {author} {\bibfnamefont {M.~T.}\ \bibnamefont {Lam}}, \bibinfo {author} {\bibfnamefont {D.~R.}\ \bibnamefont {Lorimer}}, \bibinfo {author} {\bibfnamefont {J.}~\bibnamefont {Luo}}, \bibinfo {author} {\bibfnamefont {A.}~\bibnamefont {McEwen}}, \bibinfo {author} {\bibfnamefont {J.~W.}\ \bibnamefont {McKee}}, \bibinfo
  {author} {\bibfnamefont {M.~A.}\ \bibnamefont {McLaughlin}}, \bibinfo {author} {\bibfnamefont {N.}~\bibnamefont {McMann}}, \bibinfo {author} {\bibfnamefont {B.~W.}\ \bibnamefont {Meyers}}, \bibinfo {author} {\bibfnamefont {A.}~\bibnamefont {Naidu}}, \bibinfo {author} {\bibfnamefont {C.}~\bibnamefont {Ng}}, \bibinfo {author} {\bibfnamefont {D.~J.}\ \bibnamefont {Nice}}, \bibinfo {author} {\bibfnamefont {N.}~\bibnamefont {Pol}}, \bibinfo {author} {\bibfnamefont {H.~A.}\ \bibnamefont {Radovan}}, \bibinfo {author} {\bibfnamefont {B.}~\bibnamefont {Shapiro-Albert}}, \bibinfo {author} {\bibfnamefont {C.~M.}\ \bibnamefont {Tan}}, \bibinfo {author} {\bibfnamefont {S.~P.}\ \bibnamefont {Tendulkar}}, \bibinfo {author} {\bibfnamefont {J.~K.}\ \bibnamefont {Swiggum}}, \bibinfo {author} {\bibfnamefont {H.~M.}\ \bibnamefont {Wahl}},\ and\ \bibinfo {author} {\bibfnamefont {W.~W.}\ \bibnamefont {Zhu}},\ }\href {https://doi.org/10.3847/2041-8213/ac03b8} {\bibfield  {journal} {\bibinfo  {journal} {The Astrophysical Journal
  Letters}\ }\textbf {\bibinfo {volume} {915}},\ \bibinfo {pages} {L12} (\bibinfo {year} {2021})}\BibitemShut {NoStop}%
\bibitem [{\citenamefont {Miller}\ \emph {et~al.}(2019)\citenamefont {Miller}, \citenamefont {Lamb}, \citenamefont {Dittmann}, \citenamefont {Bogdanov}, \citenamefont {Arzoumanian}, \citenamefont {Gendreau}, \citenamefont {Guillot}, \citenamefont {Harding}, \citenamefont {Ho}, \citenamefont {Lattimer}, \citenamefont {Ludlam}, \citenamefont {Mahmoodifar}, \citenamefont {Morsink}, \citenamefont {Ray}, \citenamefont {Strohmayer}, \citenamefont {Wood}, \citenamefont {Enoto}, \citenamefont {Foster}, \citenamefont {Okajima}, \citenamefont {Prigozhin},\ and\ \citenamefont {Soong}}]{Miller_2019}%
  \BibitemOpen
  \bibfield  {author} {\bibinfo {author} {\bibfnamefont {M.~C.}\ \bibnamefont {Miller}}, \bibinfo {author} {\bibfnamefont {F.~K.}\ \bibnamefont {Lamb}}, \bibinfo {author} {\bibfnamefont {A.~J.}\ \bibnamefont {Dittmann}}, \bibinfo {author} {\bibfnamefont {S.}~\bibnamefont {Bogdanov}}, \bibinfo {author} {\bibfnamefont {Z.}~\bibnamefont {Arzoumanian}}, \bibinfo {author} {\bibfnamefont {K.~C.}\ \bibnamefont {Gendreau}}, \bibinfo {author} {\bibfnamefont {S.}~\bibnamefont {Guillot}}, \bibinfo {author} {\bibfnamefont {A.~K.}\ \bibnamefont {Harding}}, \bibinfo {author} {\bibfnamefont {W.~C.~G.}\ \bibnamefont {Ho}}, \bibinfo {author} {\bibfnamefont {J.~M.}\ \bibnamefont {Lattimer}}, \bibinfo {author} {\bibfnamefont {R.~M.}\ \bibnamefont {Ludlam}}, \bibinfo {author} {\bibfnamefont {S.}~\bibnamefont {Mahmoodifar}}, \bibinfo {author} {\bibfnamefont {S.~M.}\ \bibnamefont {Morsink}}, \bibinfo {author} {\bibfnamefont {P.~S.}\ \bibnamefont {Ray}}, \bibinfo {author} {\bibfnamefont {T.~E.}\ \bibnamefont {Strohmayer}}, \bibinfo
  {author} {\bibfnamefont {K.~S.}\ \bibnamefont {Wood}}, \bibinfo {author} {\bibfnamefont {T.}~\bibnamefont {Enoto}}, \bibinfo {author} {\bibfnamefont {R.}~\bibnamefont {Foster}}, \bibinfo {author} {\bibfnamefont {T.}~\bibnamefont {Okajima}}, \bibinfo {author} {\bibfnamefont {G.}~\bibnamefont {Prigozhin}},\ and\ \bibinfo {author} {\bibfnamefont {Y.}~\bibnamefont {Soong}},\ }\href {https://doi.org/10.3847/2041-8213/ab50c5} {\bibfield  {journal} {\bibinfo  {journal} {The Astrophysical Journal Letters}\ }\textbf {\bibinfo {volume} {887}},\ \bibinfo {pages} {L24} (\bibinfo {year} {2019})}\BibitemShut {NoStop}%
\bibitem [{\citenamefont {Riley}\ \emph {et~al.}(2019)\citenamefont {Riley}, \citenamefont {Watts}, \citenamefont {Bogdanov}, \citenamefont {Ray}, \citenamefont {Ludlam}, \citenamefont {Guillot}, \citenamefont {Arzoumanian}, \citenamefont {Baker}, \citenamefont {Bilous}, \citenamefont {Chakrabarty}, \citenamefont {Gendreau}, \citenamefont {Harding}, \citenamefont {Ho}, \citenamefont {Lattimer}, \citenamefont {Morsink},\ and\ \citenamefont {Strohmayer}}]{Riley_2019}%
  \BibitemOpen
  \bibfield  {author} {\bibinfo {author} {\bibfnamefont {T.~E.}\ \bibnamefont {Riley}}, \bibinfo {author} {\bibfnamefont {A.~L.}\ \bibnamefont {Watts}}, \bibinfo {author} {\bibfnamefont {S.}~\bibnamefont {Bogdanov}}, \bibinfo {author} {\bibfnamefont {P.~S.}\ \bibnamefont {Ray}}, \bibinfo {author} {\bibfnamefont {R.~M.}\ \bibnamefont {Ludlam}}, \bibinfo {author} {\bibfnamefont {S.}~\bibnamefont {Guillot}}, \bibinfo {author} {\bibfnamefont {Z.}~\bibnamefont {Arzoumanian}}, \bibinfo {author} {\bibfnamefont {C.~L.}\ \bibnamefont {Baker}}, \bibinfo {author} {\bibfnamefont {A.~V.}\ \bibnamefont {Bilous}}, \bibinfo {author} {\bibfnamefont {D.}~\bibnamefont {Chakrabarty}}, \bibinfo {author} {\bibfnamefont {K.~C.}\ \bibnamefont {Gendreau}}, \bibinfo {author} {\bibfnamefont {A.~K.}\ \bibnamefont {Harding}}, \bibinfo {author} {\bibfnamefont {W.~C.~G.}\ \bibnamefont {Ho}}, \bibinfo {author} {\bibfnamefont {J.~M.}\ \bibnamefont {Lattimer}}, \bibinfo {author} {\bibfnamefont {S.~M.}\ \bibnamefont {Morsink}},\ and\ \bibinfo
  {author} {\bibfnamefont {T.~E.}\ \bibnamefont {Strohmayer}},\ }\href {https://doi.org/10.3847/2041-8213/ab481c} {\bibfield  {journal} {\bibinfo  {journal} {The Astrophysical Journal Letters}\ }\textbf {\bibinfo {volume} {887}},\ \bibinfo {pages} {L21} (\bibinfo {year} {2019})}\BibitemShut {NoStop}%
\end{thebibliography}%

\end{document}